\documentclass[12pt]{article}
\usepackage{amsmath,amsfonts}
\usepackage{amscd,amsthm}
\usepackage{geometry}
\usepackage{graphicx}
\usepackage{color}
\usepackage[applemac]{inputenc}
\usepackage{tikz}
\usepackage{graphicx}
\usepackage{color}
\usepackage[applemac]{inputenc}
\usepackage{cases}
\usepackage{tikz}
\usepackage{setspace}
\usepackage{float}

\newcommand{\keyw}[1]{{\bf #1}}

\begin{document}

\title{Emotional Strategies as Catalysts for Cooperation in Signed Networks}

\author{Simone Righi \thanks{MTA TK "Lend\"{u}let" Research Center for Educational and Network Studies (RECENS),
Hungarian Academy of Sciences. Mailing address:
Orsz\'{a}gh\'{a}z utca 30, 
1014 Budapest, Hungary.
Email: simone.righi@tk.mta.hu}\and K\'{a}roly Tak\'{a}cs\thanks{MTA TK "Lend\"{u}let" Research Center for Educational and Network Studies (RECENS),
Hungarian Academy of Sciences. Mailing address:
Orsz\'{a}gh\'{a}z utca 30, 
1014 Budapest, Hungary.
Email: takacs.karoly@tk.mta.hu}}
\maketitle

\begin{abstract}
The evolution of unconditional cooperation is one of the fundamental problems in science. A new solution is proposed to solve this puzzle. We treat this issue with an evolutionary model in which agents play the Prisoner's Dilemma on signed networks. The topology is allowed to co-evolve with relational signs as well as with agent strategies. We introduce a strategy that is conditional on the emotional content embedded in network signs. We show that this strategy acts as a catalyst and creates favorable conditions for the spread of unconditional cooperation. In line with the literature, we found evidence that the evolution of cooperation most likely occurs in networks with relatively high chances of rewiring and with low likelihood of strategy adoption. While a low likelihood of rewiring enhances cooperation, a very high likelihood seems to limit its diffusion. Furthermore, unlike in non-signed networks, cooperation becomes more prevalent in denser topologies.
\vspace{1cm}


\end{abstract}

\section{Introduction}\label{Introduction}
The reasons and the conditions for the evolution of cooperative behavior in human communities remain one of the most actively targeted puzzles in science \cite{axelrod2006evolution,axelrod1997complexity,axelrod1981evolution,burtsev2006evolution,fehr2003nature,mcnamara2008coevolution,nowak2006evolutionary,nowak2006five,wedekind2000cooperation}. The topic has been extensively studied by evolutionary game theory, which focuses on the dynamic change of strategies in populations of interacting agents subject to natural selection pressures. 
The Prisoner's Dilemma (PD) game is frequently used to model the problem of cooperation as it very clearly depicts the paradox concerned \cite{flood1958some,poundstone1992prisoner}. The PD describes a symmetric setup in which two players can choose between two alternative strategies: to cooperate or to defect. Defection provides players with higher payoffs than cooperation, regardless of the other individual's choice. Mutual defection, however, is suboptimal to mutual cooperation. 
In this game, defection is the dominant strategy, but it results in a Pareto-inefficient equilibrium. Moreover, from an evolutionary perspective, defection is the sole evolutionary stable stategy (ESS).

It was shown previously that cooperation can be sustained if social dilemma games (such as the PD) are played in networks, with interactions taking place only between immediate neighbors \cite{Virtuallabs,hauert2004spatial,lieberman2005evolutionary,ohtsuki2006simple,santos2006cooperation,wang2008learning}. 
The sparseness of interactions allows groups of cooperators to invade populations of defectors \cite{nowak2006five,ohtsuki2006simple,santos2006cooperation}. The network structure is important because persistent social ties (both direct and indirect) can be used to control behavior, which thus contributes to the establishment of cooperation through reputation mechanisms, such as image scoring \cite{wedekind2000cooperation}.
The introduction of structured interactions opens the question of which network topologies sustain or foster cooperation. The literature has frequently adopted a numerical approach (e.g., \cite{Virtuallabs,masuda2003spatial}) to this problem, as the evolution of cooperation is beyond analytical reach even when behavior evolves in a fixed interaction topology  \cite{lieberman2005evolutionary,ohtsuki2006simple,vukov}. Further results have been obtained with the aid of agent-based simulations for evolving networks \cite{santos2006cooperation,wang2013int,yamagishi1996selective,yamagishi1994prisoner}. In this context, it has been shown that the possibility of exclusion, exit, and partner selection favor the emergence of cooperation \cite{schuessler1989exit,vanberg1992rationality,yamagishi1996selective}.  

Virtually all studies that analyze the co-evolution of network structure and cooperation assume only positive relations between players. In this paper, we introduce negative ties with the assumption that they may act as accelerators of cooperation as suggested by both the literature on the efficiency of altruistic punishment \cite{bowles2004evolution,dreber2008winners,ernst2005human,fowler2005altruistic,fowler2005egalitarian}, and on stigmatization and social exclusion \cite{kerr2008detection,kurzban2001evolutionary}. 
Including negative ties allows us to capture the dyadic social-psychological mechanisms of vengeance and anger, which are generally modeled with trigger strategies in the context of repeated PD games \cite{axelrod2006evolution,axelrod1997complexity,axelrod1981evolution,trivers1971evolution}. However, the main enhancement the model can offer with the integration of negative ties is that they allow for effective representation of emotions that encode relevant reputational information. Speculations about why emotions have evolved in humans are in fact largely linked to their function in human interaction \cite{darwin1965expression,frank1988passions,keltner2006social,trivers1971evolution}. Individuals do not keep exact records of all their interactions, but use affection, sentiments (e.g., like, dislike) and emotions as simplified tools of recall. Relational signs are simple information that is easy to remember and costless to maintain between potential partners, and can thus act as effective signals of cooperative intentions.

Feelings shape and are shaped by the interaction experience with partners. On the one hand, relations in everyday life are filled with emotions that trigger appropriate behavioral responses. On the other hand, sentiments partly replace factual bookkeeping of interactions and, at the same time, constitute effective reputational signals, as they change as a consequence of partners' behavior. In our setup, relational signs make the model more realistic, as they represent the emotions towards interacting partners.

In this paper we focus on the role of emotions in a setting where the single shot PD is played and where relational signs, network topology and the agents' strategies co-evolve. We show that a simple strategy that is conditional on the direct relational sign with the partner, but does not have a memory of past choices, catalyzes the chances of cooperation to become dominant. 
Unlike universal cooperation (UC) and universal defection (UD), the emotional conditional strategy (COND) triggers different responses of the same individual towards different partners. COND prescribes cooperation with those who are liked (positive relational sign) and defection with those who are disliked (negative relational sign).~\footnote{COND is similar to Tit-for-Tat (TFT), with the difference that the former does not recall previous choices and is therefore a strategy for the single-shot PD.} 
In our simulations, COND played against other strategies without memory (UC and UD) and we explored how the inclusion of this strategy affected the prevalence of cooperative behavior under different parametric conditions. Different strategies induce different choices in the game and thus create asymmetric emotional experiences that can result in emotional tension. We assume that such tension is resolved by a probabilistic and non-strategic sign update or by elimination and rewiring of the relationship. 

First, we manipulated the proportion of the agents playing the emotional strategy at the outset and the likelihood of network updates, showing that the presence of conditional strategies facilitates the evolution of cooperation. We show that rewiring is necessary for cooperation to emerge but it hampers its diffusion when it becomes the dominant force.
Moreover, we determine the parametric conditions under which the presence of emotional strategies facilitates the evolution of cooperation in signed networks. We compare our findings with those observed in the literature on games in unsigned networks and note both similarities and differences. We observe that the presence of negative ties reverses the results of \cite{ohtsuki2006simple}, as we show that network density increases the chances of cooperation to endure.
Noting that the frequency of sign updates has a relatively minor effect on results, we focus on the exploration of the relative importance of network updates compared to strategy updates. In line with the literature on positive ties \cite{santos2006cooperation} we find that more frequent network updates provide favorable conditions for cooperation, as they allow cooperators to be matched with cooperators, while evolutionary updates support the strategy with a higher payoff, i.e., defection. Most importantly, the presence of emotional strategies enlarges the parameter set under which cooperation survives. 
  
We study these questions in networks the sizes of which are calibrated to resemble that of human ancestors' communities \cite{dunbar1992neocortex}. In line with most of the literature, we provide results for a model in which the average payoff from all dyadic interactions determines individual fitness.

In the subsequent sections of the paper, we introduce our model (Section \ref{TheModel}), present our results in detail (Section \ref{Results}), and discuss their importance and limitations (Section \ref{Conclusions}). 

\section{The Model}\label{TheModel}
Consider a set of agents $N$. Each agent  $i \in N$ is placed on a node in a non-weighted non-directed signed network. Agents are characterized by a type and by a set of connections with a subset of the whole population $\mathcal{F}_i \subset N$. Network ties, which constrain the possibility of playing a two-person single-shot PD, are signed and each of them is either \textit{negative} or \textit{positive}. Strategy types define the behavior of the agents in the PD. Specifically, there are three types of strategies:
\begin{itemize}
\item Unconditional Defector. This type of agent always defects regardless of the sign of his relationships. Both the strategy and the agents playing it are labeled UD.
\item Unconditional Cooperator (UC). This type of agent always cooperates, regardless of the sign of the tie with the partner.
\item Conditional Player (COND). The behavior of this agent is determined by the sign of the tie he shares with the current partner in the PD game. Specifically, he cooperates with agents he shares a positive tie with and defects with agents he has a negative tie with. \footnote{We do not consider strategies that are conditional on the valence of the tie in the opposite way: cooperate in the case of a negative tie and defect in the case of a positive tie.}
\end{itemize}

The agents play the Prisoner's Dilemma, characterized by the classical payoff structure, with each of their neighbors (see Table 1). During every dyadic interaction, agents observe the link sign and play according to their type.  

\begin{table}[h!]
\centering
\label{PDPayff}
$\begin{array}{|p{0.7cm}|c|c|}
\hline
& \textrm{C} & \textrm{D} \\
\hline \textrm{C} &  (R=3,R=3) & (S=0,T=5) \\
\hline \textrm{D} & (T=5,S=0) & (P=1,P=1) \\
\hline
\end{array}$

\caption{The Prisoner's Dilemma payoff matrix ($T>R>P>S$). The numerical payoffs used here are the same as in \cite{axelrod2006evolution}.}
\end{table}

The dynamics of our model allows for the co-evolution of relational signs, agent strategies, and network structure. Time is divided in discrete periods and simulations continue until equilibrium is reached.~\footnote{To consider a state as equilibrium, two strict conditions must be met. First, the simulation cannot end before a transitory period of 150 steps has expired. Second, the configuration of relational signs, network topology, and agent types need to be precisely the same in five randomly determined periods of time. Each $t$ had a probability 0.1 to be selected for this end rule. When at least two periods were selected, they were compared. If their configuration was not exactly the same, then simulation continued and new $t$s were randomly selected for the end rule. Robustness checks have been performed and the results are stable with alternative parameterizations of the end rule.}
The intra-step dynamics is summarized in Algorithm 1. At each time step $t$, each agent $i$ contemporaneously plays the Prisoner's Dilemma with all agents in his current first order social neighborhood, i.e., with each $j\in\mathcal{F}_i^{t-1}$. After each dyadic interaction, emotional tension may emerge as a consequence of the behavior of the agents. Emotional tension emerges if and only if the interacting partners opted for a different action (one cooperates and the other defects).
This can be solved either by deleting the problematic relationship and substituting it with a new one, or with a sign update of the current relationship. Both cases require a more detailed discussion provided in the following. Subsequently, each agent computes his average payoff and compares it with that of peers in the direct neighborhood (i.e., with all those he has played with). In case one or more of the neighbors have a payoff higher than his own, he adopts (with probability $P_{adopt}$) the strategy played by one of them, selected uniformly at random.
As the order in which the agents are selected for update should not make a difference in evolution, update is done in parallel.
The literature on the PD played in networks has consistently showed that when individual fitness is determined on the basis of the average payoff in all individual encounters, the proportion of cooperators increases with respect to the case in which updates are made after each dyadic interaction \cite{wang2013int}.~\footnote{The choice of using the average payoff instead of sum relies on the fact that we want a measure of performance independent of the degree.} This is also the case in our model (see \cite{righi2014parallel}). We will therefore limit our analysis to the situation in which individual fitness is determined as the average payoff from all dyadic interactions with neighbors. 

\begin{tabbing}
\quad \=\quad \=\quad \=\quad \=\quad\kill
\keyw{For} each agent $i$ \\
\> Compute its social neighborhood $\mathcal{F}_i^{t-1} \in N$\\
\> \keyw{For} each agent $j \in \mathcal{F}^{t-1}_i$ \\
\> \> Let $i$ and $j$ play the PD according to their types in $t-1$ and determine outcome\\
\> \> Update relational signs between $i$ and $j$ for $t$\\
\> \> \keyw{If} $i$ cooperated and $j$ defected, \keyw{Then} \\
\> \> \>  with probability $P_{rew}$ delete the link between $i$ and $j$ ... \\
\> \> \> ... and connect $i$ to somebody in its second order neighborhood \\
\> \> \keyw{end}\\
\> \keyw{end}\\
\> Compute the average payoff in $t$ of agent $i$\\
\keyw{end}\\
\keyw{For} each agent $i$ \\
\> Observe the average payoffs of period $t$ for each agent $j \in \mathcal{F}_i^{t-1}$\\
\> Adopt a random (strictly) better strategy in $j \in \mathcal{F}_i^{t-1}$ (with probability $P_{adopt}$)\\
\keyw{end}\\
All rewirings are observed. \\
All relational signs become effective.\\
New agent types become effective.\\

\keyw{Algorithm 1.} Intra-step dynamics, repeated at each time step $t$.\\
\end{tabbing}\label{parallelalgo}%

There are three main mechanisms that drive the evolution of our model. The first is the relational sign update, which allows the agents' behavior to influence the nature of their relationship. The second is the rewiring of tense links, through which the agents' behavior influence the structure of social interactions. The third is the adoption of a better strategy, through which the agents update their behavior in order to improve their welfare. 
For each of these mechanisms, we selected simple and yet empirically founded updating rules. Let us discuss each of these building blocks in detail. 

{\bf Relational sign updates} model the consequences of behavior in the PD on the emotional content of relations with peers. When two players cooperate they are both satisfied with the outcome. An existing positive link thus remains positive, and a negative one becomes positive. For similar reasons, when both agents defect, a positive relation becomes negative and a negative one remains so. More problematic is the situation in which behaviors differ: one agent cooperates while the other defects. In this case the emotional content of the relationship is subject to tension. The defecting partner is content to remain friends with the cooperator. In fact, a positive relationship to the cooperator provides him with a strictly higher payoff if he is paired with a COND player, whose action is sensitive to the sign of their relationship. If the link is negative, the defector might be interested in turning it into a positive tie. We assume that it happens with probability $P_{pos}$.
The cooperating agent is, however, frustrated by the partner's behavior and may want to change a positive link to negative with some probability $P_{neg}$. It is logical to assume that the frustration of the cooperator in the latter case is stronger than the will of the defector to turn the troubled relationship positive. Therefore, we assume that $P_{neg}>>P_{pos}$.~\footnote{Note that tension in a particular dyad is always asymmetric. The probabilities $P_{pos}$ and $P_{neg}$ are therefore never applied at the same time to any particular dyad. In most simulations reported here, we fixed $P_{neg}=0.2$ and $P_{pos}=0.1$. These values are assumed equal for all agents. We have run simulations also for other parameter values and observed that all results are robust to these changes. An analysis of alternative parameter values is reported in the Appendix.}

It should be noted that, while all links are signed and evolve as a consequence of the behavior adopted in interactions between partners, only COND acts on them conditionally. Hence, COND is the only truly emotional strategy in our model. Meanwhile, also UCs and UDs update their network signs in certain situations. This way, we separated the automatic update of network signs from the evolution and spread of strategies. Consequently, similarly to other evolutionary models, we are able to make fair comparisons about the evolutionary success of conditional cooperation and other simple strategies in the PD.

The second update mechanism in our model is {\bf  rewiring}, through which network topology changes endogenously as a consequence of the agents' behavior. A similar mechanism has been studied by \cite{santos2006cooperation}. A cooperator, frustrated by the behavior of a defecting partner, may decide to delete the social connection altogether. We assume that such a choice is made non-strategically with probability $P_{rew}$ (or rewiring probability). When rewiring takes place, a new link  is created with another agent. In line with the sociological literature on partnership formation, we assume that new links are created more frequently with players at social distance two. In other words, there is a certain tendency to transitive closure \cite{granovetter1973strength}. More precisely, we assume that a new link connects the frustrated cooperator with someone who is a friend of his friends. The new link is initially {\it positive}. This restriction excludes the possibility of the formation of ties with the friend of an enemy or with the enemy of a friend, and naturally increases clustering in the network. In order to introduce some noise into this process, with a probability $P_{rand}$ rewiring is done with a random agent in the population. \footnote{This parameter was fixed at $P_{rand}=0.01$ in the main simulations of this paper. However, as shown in the Appendix, the main results have been qualitatively preserved even for the extreme case of purely random relinking ($P_{rand}=1$).} 

The third element required for an evolutionary model as ours is a {\bf strategy update} rule that determines how successful strategies are adopted by others. 
At each time step and for each agent $i$, the average payoff across all his interactions is calculated.  The agent compares his payoff with the ones of all his peers in his first order social neighborhood (i.e., all those he has played with). If a subset of agents in $\mathcal{F}_i$ has a payoff higher than his own, agent $i$ adopts the strategy played by one of the agents in this pool of better performing peers, selected uniformly at random. Strategy update happens, for each agent, with probability $P_{adopt}$.

Finally, in order to focus on some key parameters, we made a few additional assumptions about other variables and {\bf calibrated our simulations} consequently.  For the sake of simplicity, we initialized relations according to an Erd\H{o}s-R\'{e}nyi random graph \cite{erdHos1959random}. \footnote{We address alternative network initializations in another paper \cite{righi2014degree}.} Each pair of nodes was connected with an independent probability $P_{link}$ (i.e., each possible link existed with probability $P_{link}$). Moreover, at the outset, the agents were randomly assigned with one of the three strategies in proportions $(\mu_{UC}, \mu_{UD} , \mu_{COND}) \in [0;1]$ s.t. $(\mu_{UC}+\mu_{UD}+\mu_{COND}) \equiv 1$. In all simulations reported we assumed $\mu_{UC} = \mu_{UD}$. Finally, relational signs were randomly distributed and initialized so that each link had the probability $1/2$ of being either negative or positive. \footnote{Results from alternative parametrizations are available upon request.}

\section{Results}\label{Results}

{\bf Fixed networks.} In the absence of emotional strategies (COND) and rewiring, the dynamics of our model converges to a state dominated by unconditional defection and negative signs. In this baseline setup, UCs are systematically cheated on by UDs. Sign changes do not influence the outcome as cooperators do not act upon them. Thus, the baseline dynamics leads to a Hobbesian state dominated by defection and dislike.

\begin{figure}[ht!]
\centering
\includegraphics[width=1\textwidth]{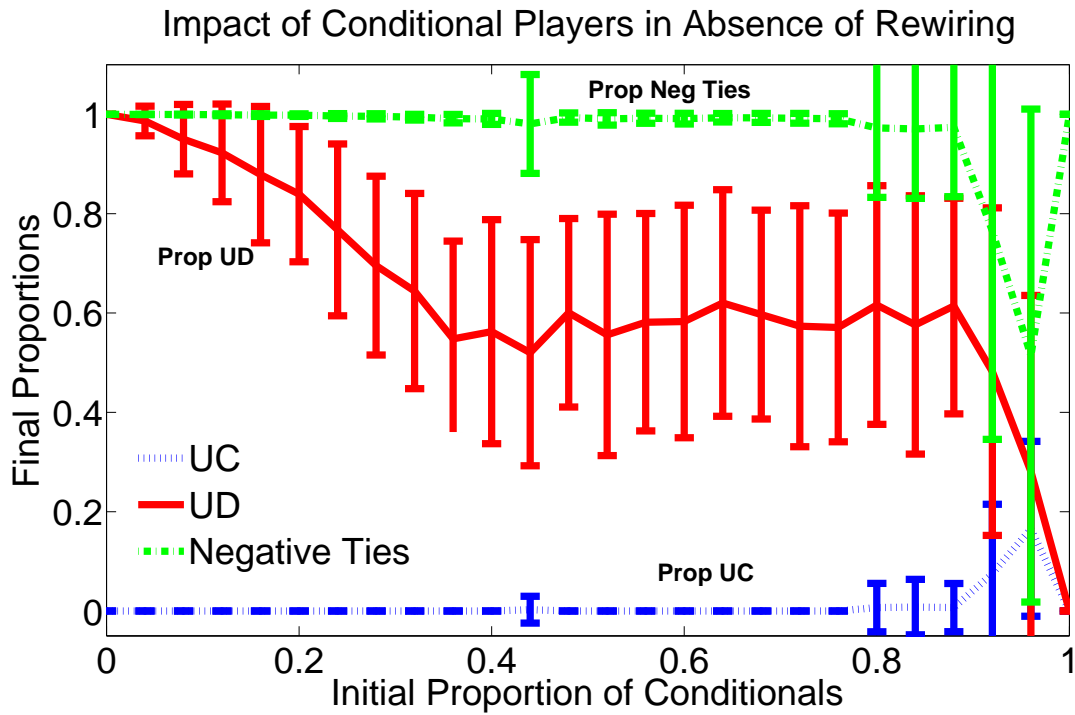}
\caption{The final proportion of UCs, UDs and of negative signs in the network as the initial proportion of CONDs increases and the network topology is fixed (i.e., $P_{rew}=0$). Each data point represents the average and the standard deviation of values obtained from 100 simulations. In each simulation $N=200$, $P_{link}=0.05$,  $P_{neg}=0.2$, $P_{neg}=0.1$ and $P_{rand}=0.01$. The presence of COND affects positively the prevalence of cooperative behavior only if more than 80\% of the population starts with this strategy.In this range, however, the high proportion of CONDs makes any result possible, hence the high variability of results.}
\label{Norewire_cond}
\end{figure}

Moreover, if the network topology is fixed, the introduction of conditional players is not sufficient to guarantee the dissemination of cooperation. To study the impact of CONDs, at the outset we manipulated the initial proportion of the population that plays this strategy and we observed the proportions of agent types and that of network signs at the end of the simulation (Figure \ref{Norewire_cond}). The conditional strategy did not disappear (the sum of proportions of UC and UD was persistently smaller than one), and remained  more likely in the final population in case its initial proportion was higher. Conditional agents, however, became functionally undistinguishable from unconditional defectors since all ties were negative. 

Since the presence of some evolution of network topology (rewiring) is a necessary condition in order to observe cooperation, in the subsequent simulations we assume $P_{rew}>0$. Moreover, when it is not stated otherwise, the initial proportion of each agent's type in the population is assumed to be the same. 
As it is noted in Table 2 and shown in the next sections, cooperation can emerge as a dominant behavior in our setup, if the network topology is dynamic and conditional players are present. In the following, we thus concentrate on this sub-case. 

\begin{table}[h!]
\centering

\begin{tabular}{| p{2cm} | p{4cm} | p{4cm} |}
\hline 
   & $P_{rew}=0$ & $P_{rew}>0$  \\
\hline 
$\mu_{COND}=0$ & No cooperation & Cooperation through clustering of strategies \\
\hline 
$\mu_{COND}>0$ & Some cooperation, only if $\mu_{COND}\rightarrow1$  & {\bf Cooperation survival/diffusion} \\  
\hline 
\end{tabular}}
\caption{Summary of the results. A positive rewiring probability and the presence of conditional strategies are both required for the spread of unconditional cooperation}{
\label{SummaryResults}
\end{table}

\subsection{Emotional strategies as catalysts for cooperation}
Our main goal is to uncover the impact of emotional strategies, and hence negative relational signs, on the evolution of cooperation. Our results produce convincing evidence that the COND strategy largely facilitates the emergence and the diffusion of {\it unconditional cooperation} in a mixed population. 

\begin{figure}[ht!]
\centering
\includegraphics[width=0.49\textwidth]{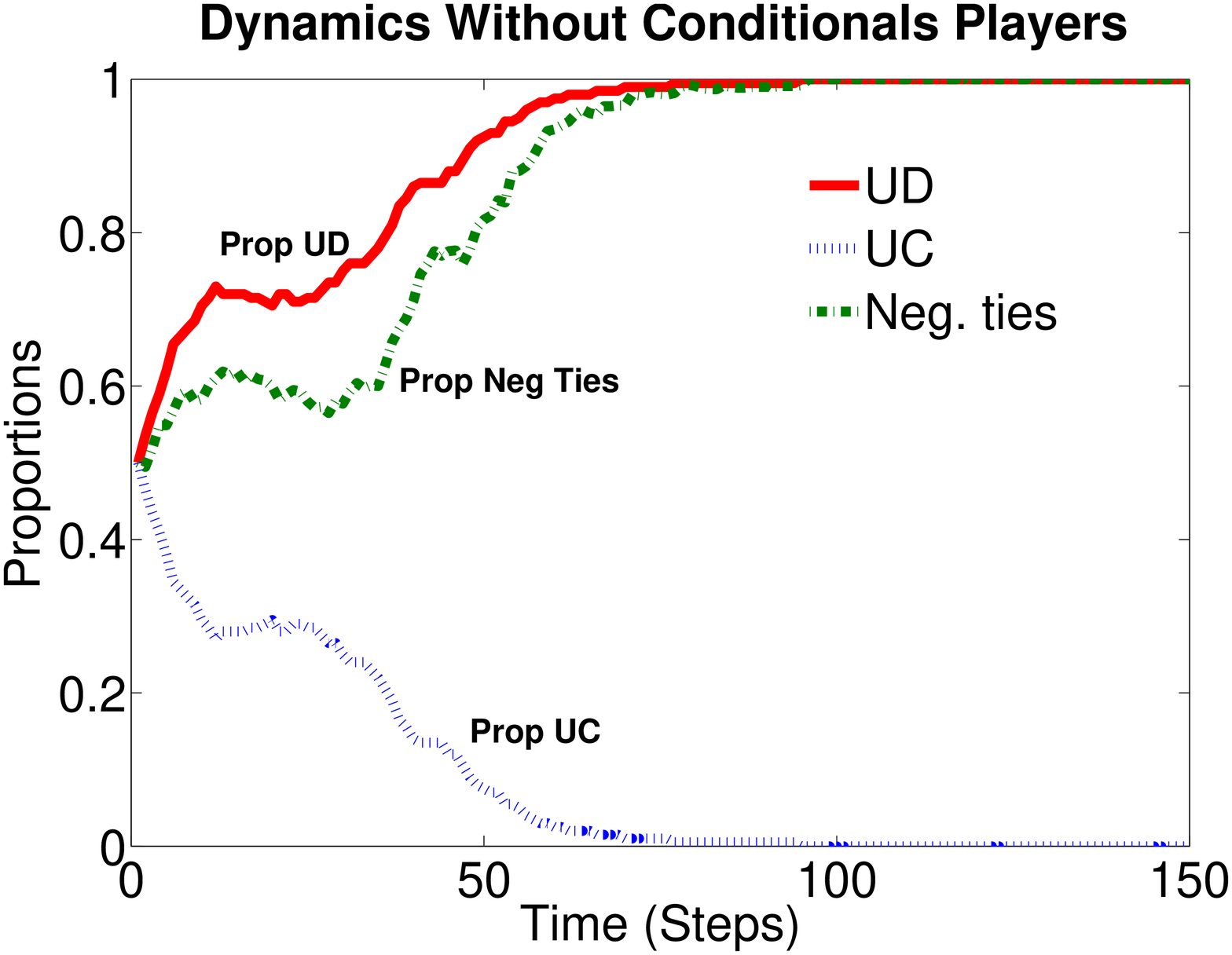}
\includegraphics[width=0.49\textwidth]{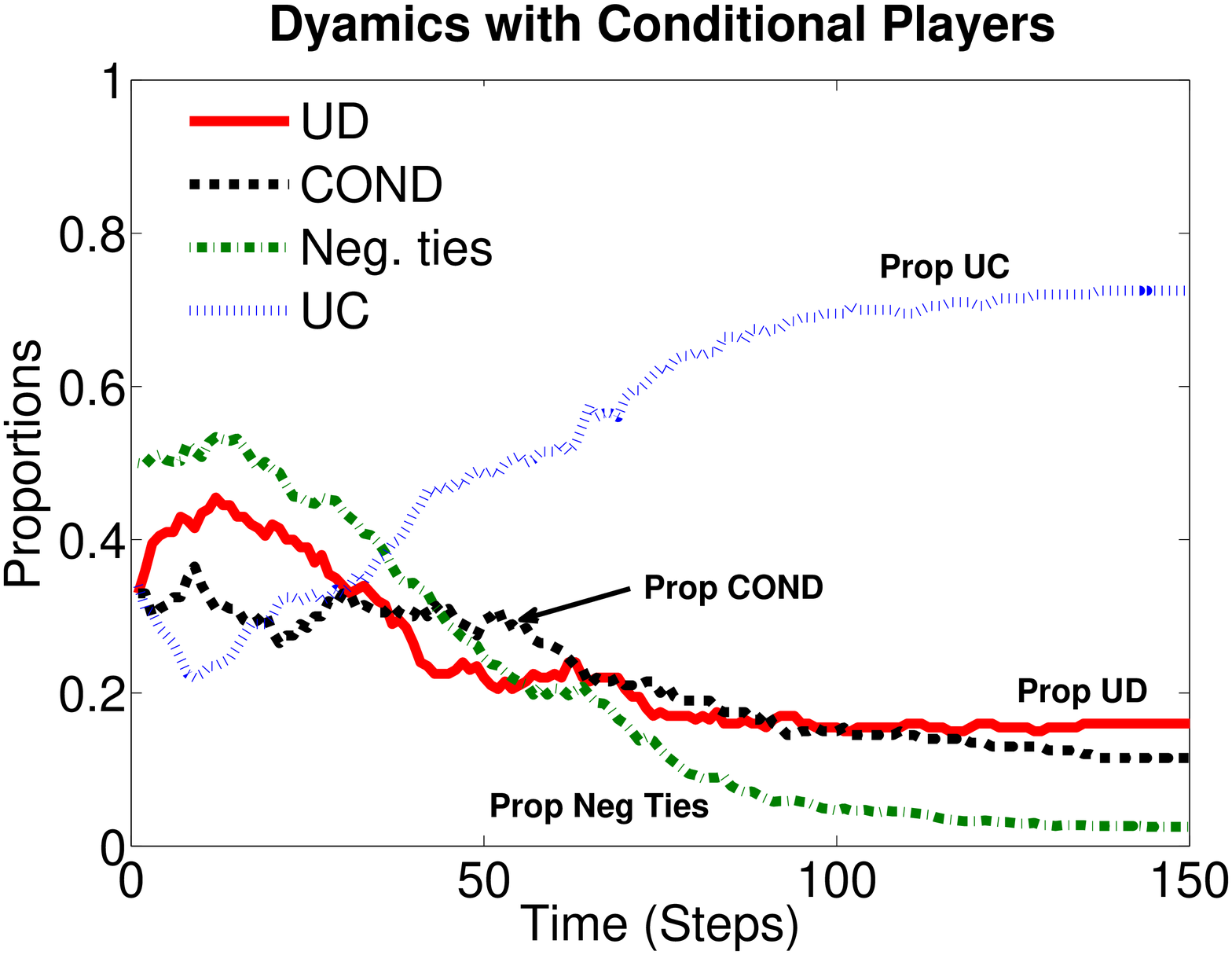}\\
\includegraphics[width=0.49\textwidth]{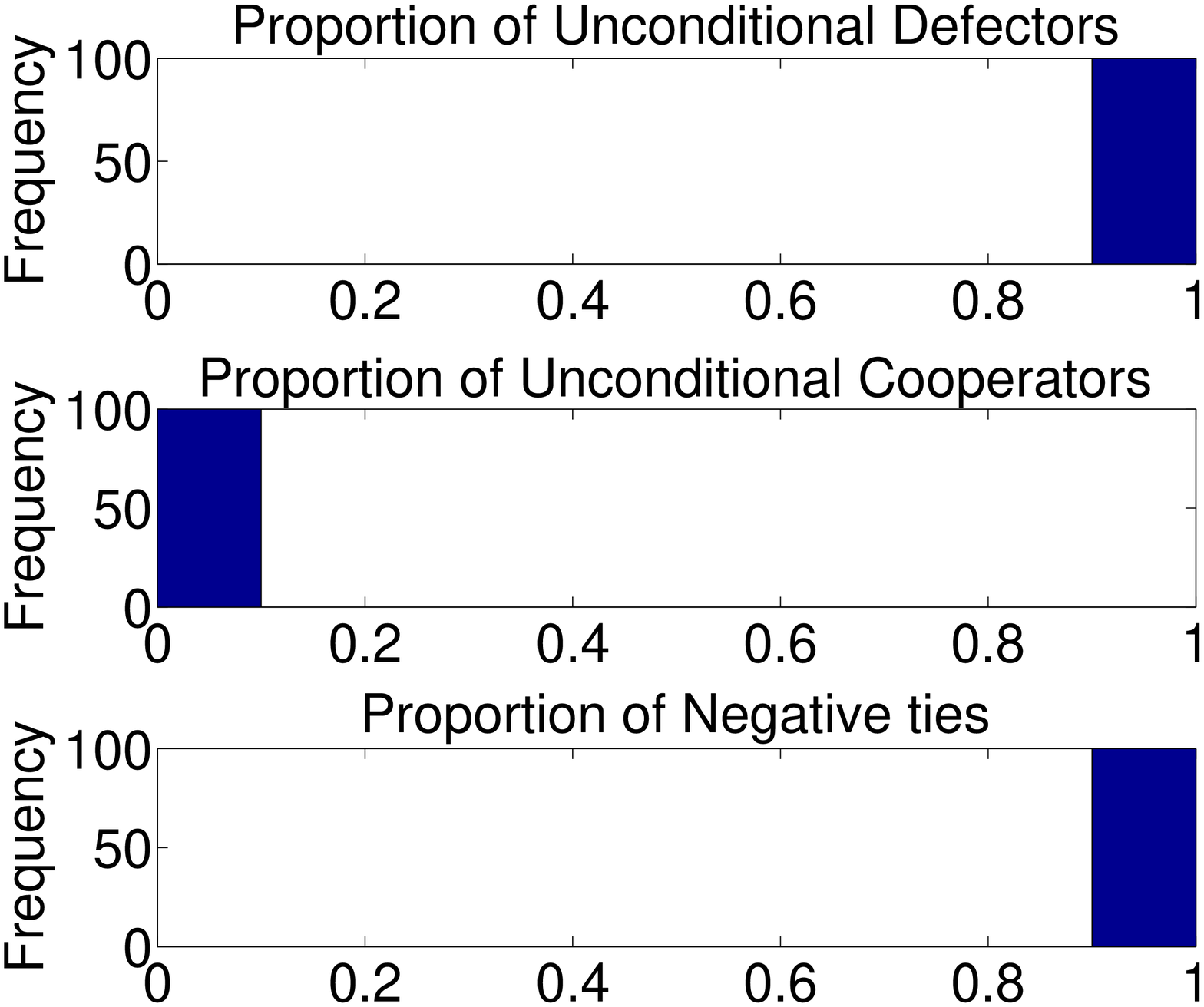}
\includegraphics[width=0.49\textwidth]{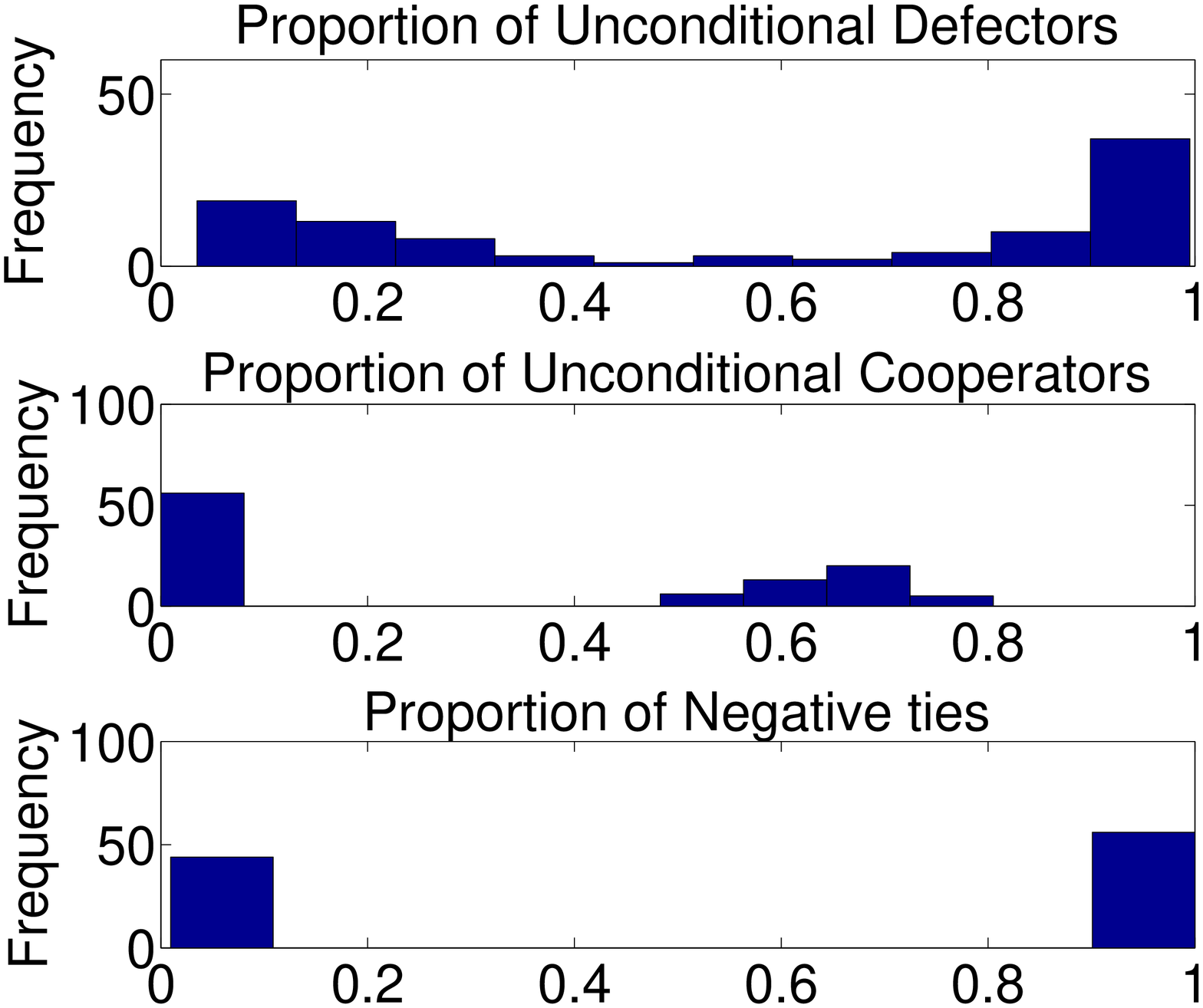}
\caption{Upper Panels: Dynamics of the proportions of agent types and network signs in typical simulations. In the Left Panel, the population is initialized as equally divided between UCs and UDs. In the Right Panel, the population is divided equally among UC, UD, and COND strategies.  Lower Panels: The distribution (calculated on 100 simulations) of the final proportions of UDs, UCs, and negative ties without (Left Panel) and with (Right Panel) conditional players. For all simulations: N=200, $P_{rew}=P_{adopt}=0.1$, $P_{neg}=0.2$, $P_{neg}=0.1$ and $P_{rand}=0.01$. The network signs were randomly initialized as positive or negative with equal probability and the probability of existence for each tie was $P_{link}=0.05$.}
\label{examples_and_distros}
\end{figure}

The Upper Panels of Figure \ref{examples_and_distros} illustrate the co-evolution of proportions of strategy types and of network signs in the absence and in the presence of emotional conditional strategies in two simulation runs, where all other parameter values were left identical. Without conditional players, universal defectors always gain dominance and the population reaches the Hobbesian state with everyone defecting and all ties being negative. The population dynamics when conditional strategies are present can be different. The share of agents using the COND strategy decreases over time and thus this strategy does not become prevalent in the population. Its presence, however, assists universal cooperators to gain dominance. At the same time, the network reaches a state with an overwhelming proportion of positive relations.
The Lower Panels of Figure \ref{examples_and_distros} report on the distribution of outcomes in a set of 100 simulations. This confirms that without CONDs, defection dominates and the Hobbesian state is the only outcome. With the introduction of the emotional strategy, however, the distribution of the results becomes bimodal for the parameters chosen, showing the existence of two clearly distinct cases. Besides the dynamics where defection becomes dominant, a second case emerges where cooperative behavior and positive signs dominate. 

The COND strategy facilitates the spread of UCs, but does not become dominant itself in the population. This is a remarkable result in itself, which originates from two distinct micro-mechanisms. First, COND efficiently copes with defectors, and prevents the exploitation of UCs by UDs partly. Second, when the COND strategy is played in mixed neighborhoods, it gains higher payoffs than UDs and lower payoffs than UCs on average. Hence, CONDs tend to conquer neighboring sites occupied by unconditional defectors, while their original positions are taken over by unconditional cooperators. This mechanism extends the area of cooperation in the network progressively.  Rewiring emotionally tense dyads reinforces this process, contributing to less exposure of UCs to UDs (through the elimination of the links between agents playing strategies which induces different behaviors).
In the absence of either emotional strategies or rewiring, the UCs are exposed to direct contact to UDs more often. In a direct interaction, the former is always payoff-dominated by the latter and thus tends to disappear.

\begin{figure}[ht!]
\centering
\includegraphics[width=0.49\textwidth]{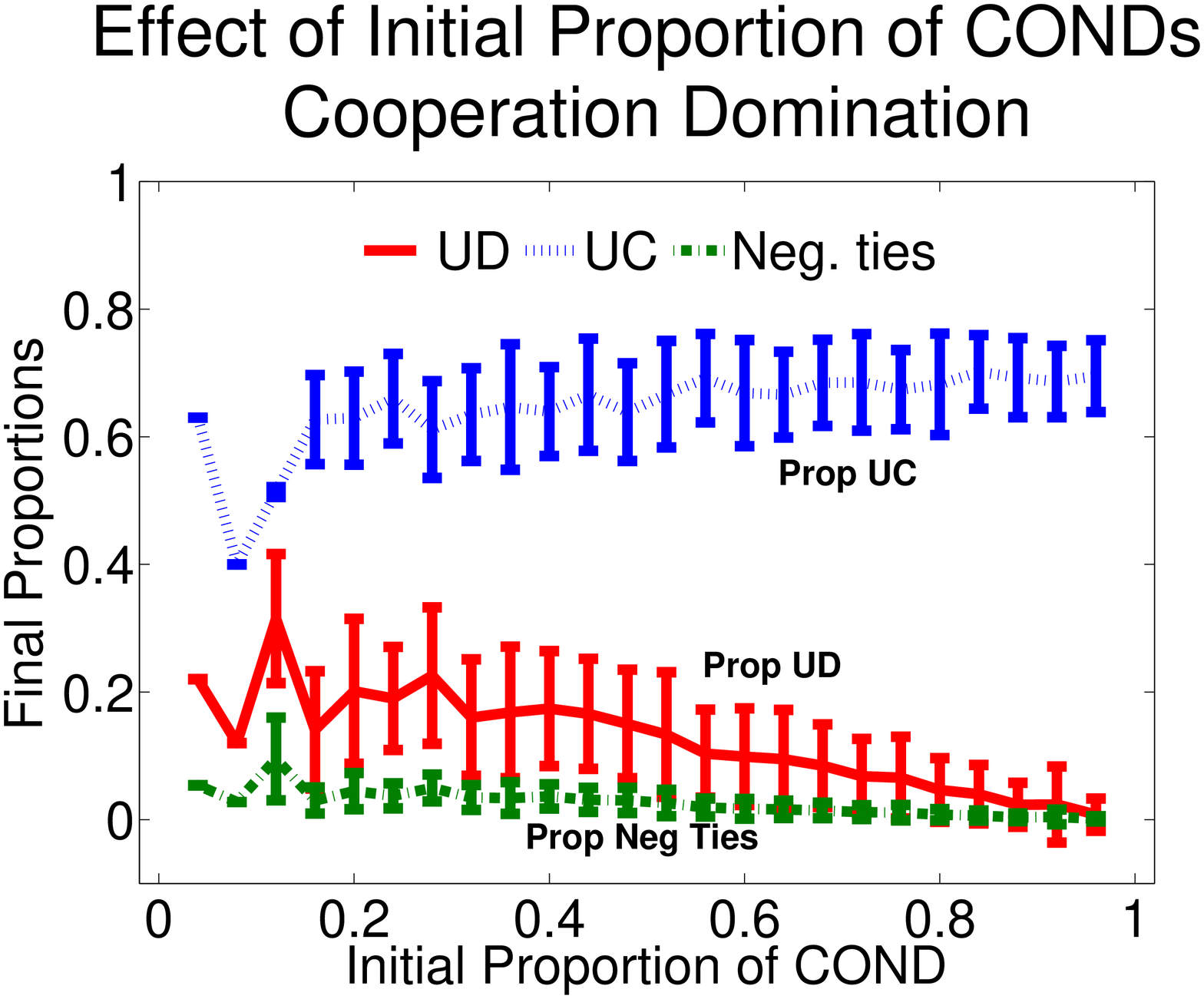}
\includegraphics[width=0.49\textwidth]{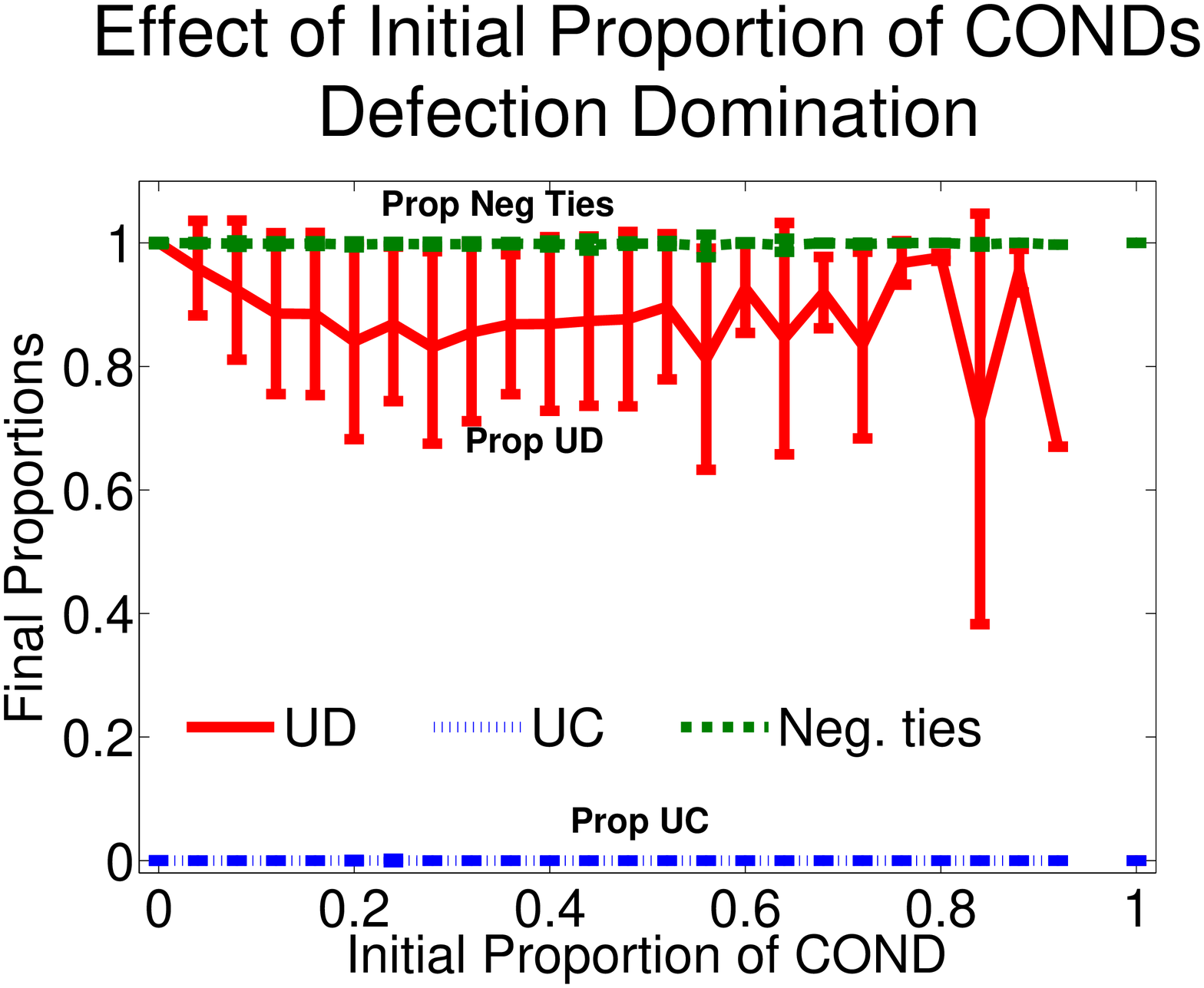}\\
\includegraphics[width=0.49\textwidth]{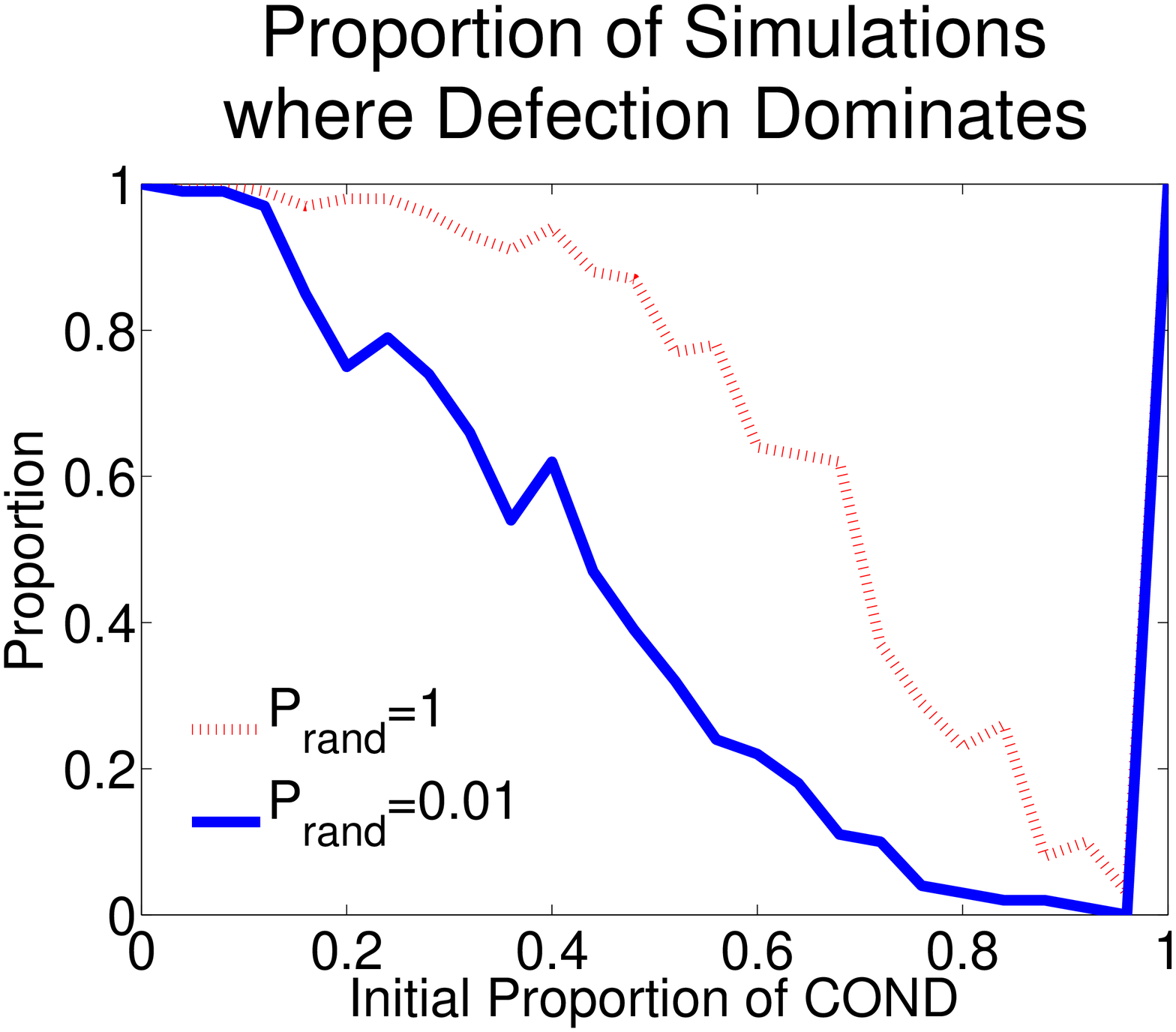}
\includegraphics[width=0.49\textwidth]{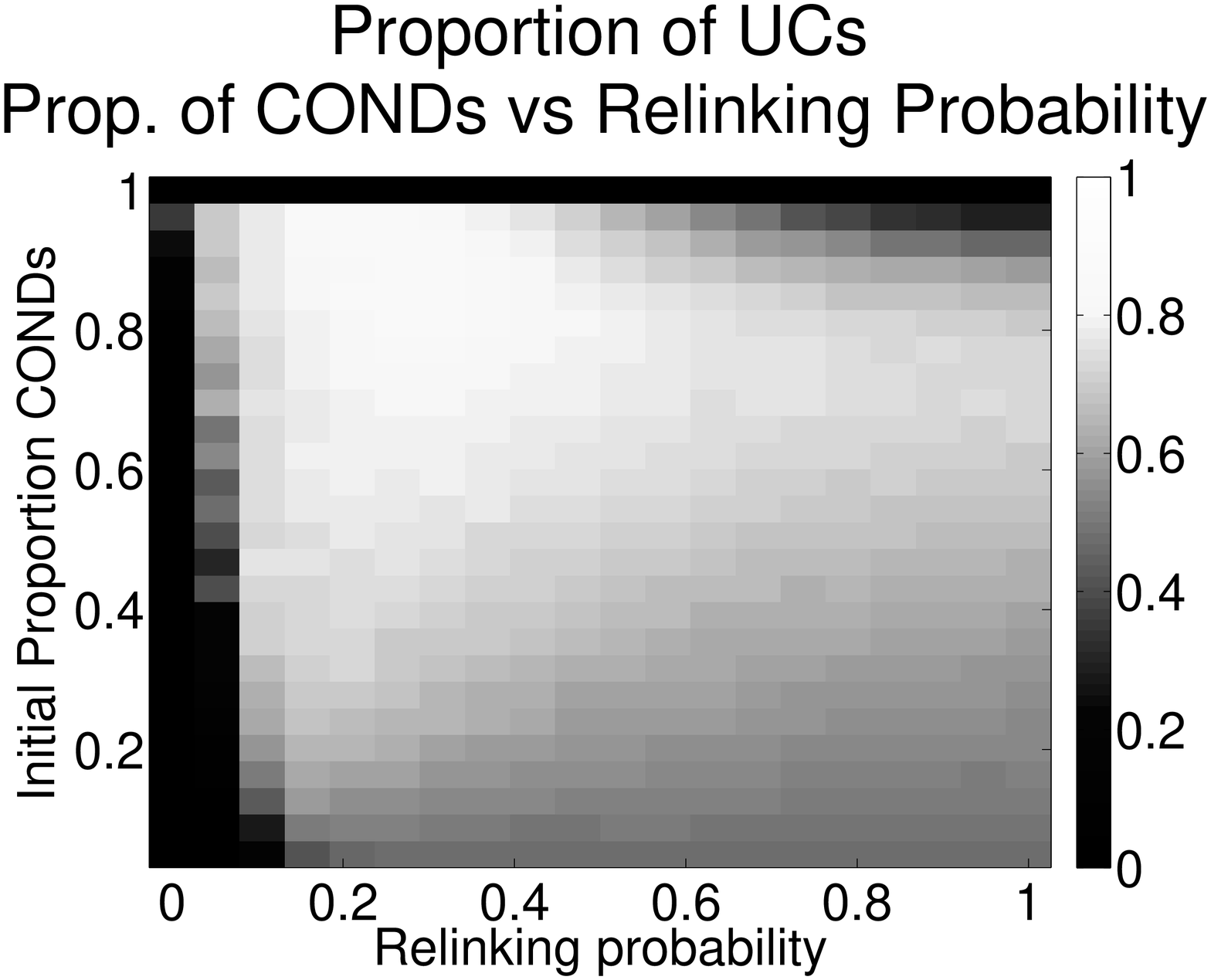}
\caption{Top Panels: mean and standard deviation of the proportion of negative signs, UDs and UCs at the end of the simulations as a function of the initial proportion of CONDs. Two types of equilibrium states exist: one in which cooperative behavior dominates (Left Panel) and one in which defection dominates (Right Panel). The proportion of the simulations in which defection dominates is reported in the Bottom Left Panel, for the cases $P_{rand}=0.01$ and $P_{rand}=1$. In these simulations $P_{rew}=P_{adopt}=0.1$. Results reported in Top Panels are similar for both $P_rand=0.01$ and $P_rand=1$. Bottom Right Panel: final proportion of UCs as the initial proportion of CONDs (vertical axis) and the rewiring probabilities (horizontal axis) are changed. In all simulations: $P_{adopt}=0.1$, $N=200$, $P_{neg}=0.2$, $P_{neg}=0.1$ and $P_{rand}=0.01$. Network signs are initially randomly positive or negative with equal probability and the probability of existence for each tie is $P_{link}=0.05$.}
\label{prop_cond}
\end{figure}

Next, we assess the contribution of emotional strategies to the evolution of cooperation more systematically. We manipulated the initial proportion of CONDs, progressively increasing it from zero to one, while dividing the rest of the population between UCs and UDs in identical shares. The results, reported in Figure \ref{prop_cond}, confirm the existence of two different types of equilibrium, the characteristics of which are quite similar across different initial shares of CONDs. In the first type of equilibrium, cooperation dominates and relational signs become mostly positive (Top Left Panel). In this case, it is interesting to note that unconditional cooperation becomes the most diffused form of cooperation regardless of the initial share of CONDs. 
The second type of equilibrium (Top Right Panel) is dominated by defection. In this case, unconditional cooperation disappears completely, while the conditional players, albeit marginal in number, become all defectors.
The share of CONDs in the population determines which of the two types of equilibrium appears more often. The Bottom Left Panel of Figure \ref{prop_cond} clearly indicates a negative relation between the initial proportion of CONDs and the share of simulations in which defection dominates. In this sense, increasing the share of agents making use of relational signs makes cooperation more likely to emerge. 
The co-existence of the two types of equilibrium here depends on the fact that we studied a situation in which the two main drivers of the dynamics (probability of rewiring and that of strategy update) have equal strength, that is, $P_{rew}=P_{adopt}$. As we will show in the insets of Figure \ref{FixedPFlip} and in Figure \ref{stds} (in Appendix), by manipulating the relative strengths of these two evolutionary forces, we can characterize the situation in which the number of equilibrium states is reduced to one and defection never dominates. Indeed, as we will note in Section \ref{rewvsadopt}, the multiple possible equilibrium states are reduced to one when the two dynamics differ significantly. Moreover, the still significant variance in the results displayed here reduces as network size increases (see the discussion in Section \ref{ConnandSize}).

The Bottom Right Panel of Figure \ref{prop_cond} presents a more general experiment where both the initial proportion of conditional agents and the rewiring probability are manipulated. \footnote{Only the proportion of UCs is reported as all relevant information can be deduced from it. The final proportion of CONDs is homogeneous across the parameter space at a value around $0.25 \pm 0.01$. The proportion of UDs decreases progressively and monotonically as the initial proportion of CONDs increases, regardless of the rewiring probability. The proportion of negative signs is complementary to the proportion of UCs; with the maximum value where the UCs are at a minimum and vice-versa. Graphs and data are available upon request.} In order to correctly interpret this study, it is important to note that only the two leftmost columns report on  simulations which ended with defection dominating. This is due to the relative strength of  $P_{rew}$ and that of $P_{adopt}$, the role of which in driving the results is studied systematically in Section \ref{rewvsadopt}. Shifting our attention to the rest of the Panel, we can notice that while the presence of some rewiring allows cooperation to become prevalent, too much of it can also limit its diffusion. For any given initial proportion of CONDs, as $P_{rew}$ increases, the proportion of UCs decreases and the proportion of negative signs increases at the end of the simulations. This interesting result is a by-product of the progressive diffusion of unconditional cooperation through the catalyzation effect produced by the emotional strategy. As soon as there is some rewiring, some tense links are deleted, and this leads to an increase in the final proportion of cooperators. When rewiring is relatively low, the ties that are deleted are those that are more likely to produce tension, i.e., those that link agents to UDs.  When the rewiring probability increases, however, the links between CONDs and UCs, which can be initially negative, are also deleted before having the opportunity of becoming positive. When this happens, the emotional strategy tends to become segregated from unconditional cooperation thus reducing the effectiveness of the synergy between these two strategies, which is at the core of the effective diffusion of cooperation in our model. One obvious way of compensating for this phenomenon is to increase the proportion of CONDs, as this reduces the impact of the segregating force of rewiring. 

With this first set of results, we showed that the presence of emotional strategies at the outset is necessary for the evolution of cooperation in the single-shot Prisoner's Dilemma played in signed networks. Emotional strategies did not gain dominance themselves, but acted as catalysts for cooperation. The higher their initial proportion was, the better the chances were, in the aggregate, for universal cooperators to outperform, and therefore eliminate, unconditional defectors.

\subsection{Low adoption probability and high rewiring support cooperation}\label{rewvsadopt}

\begin{figure}[ht!]
\centering

\includegraphics[width=0.32\textwidth]{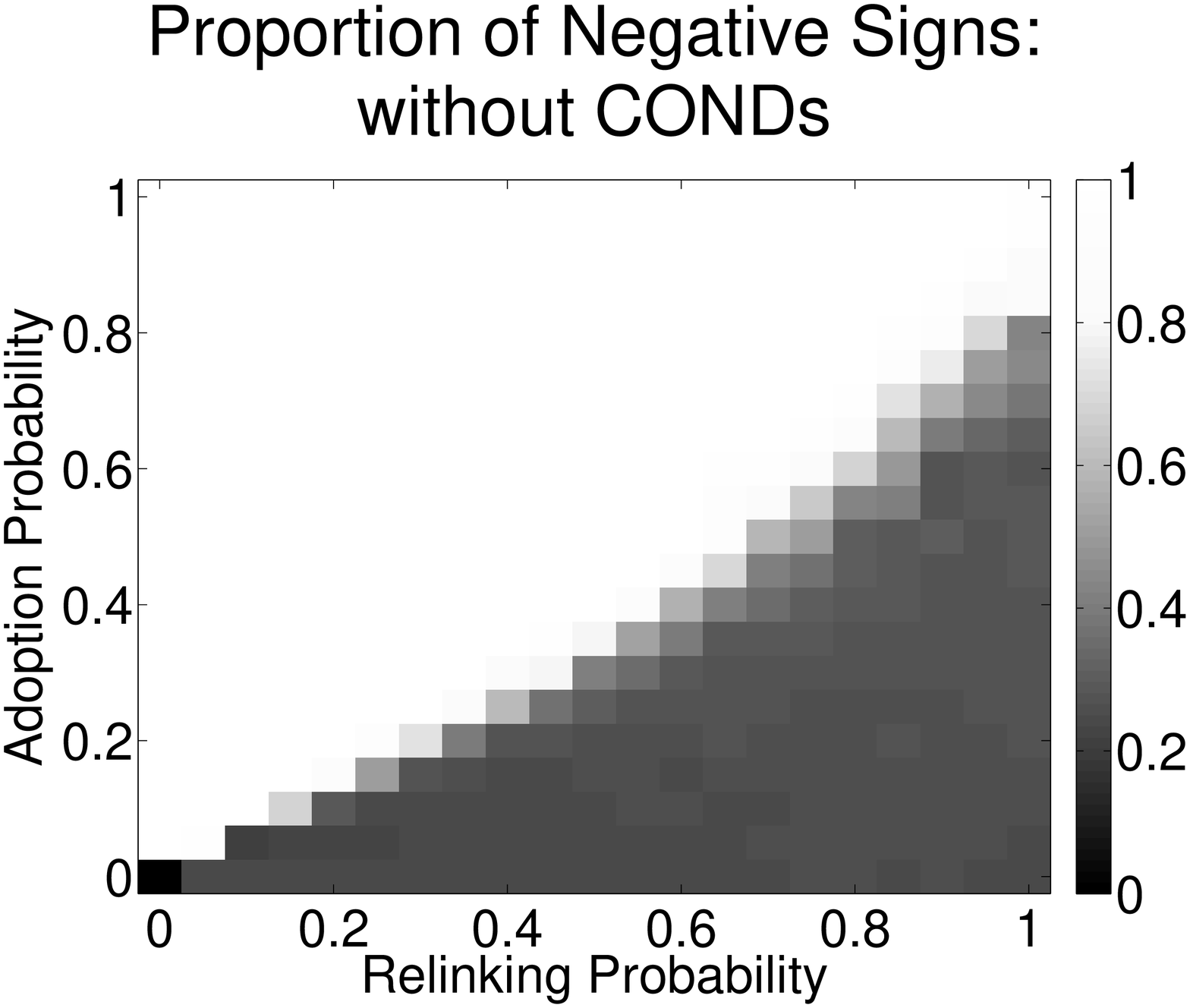}
\includegraphics[width=0.32\textwidth]{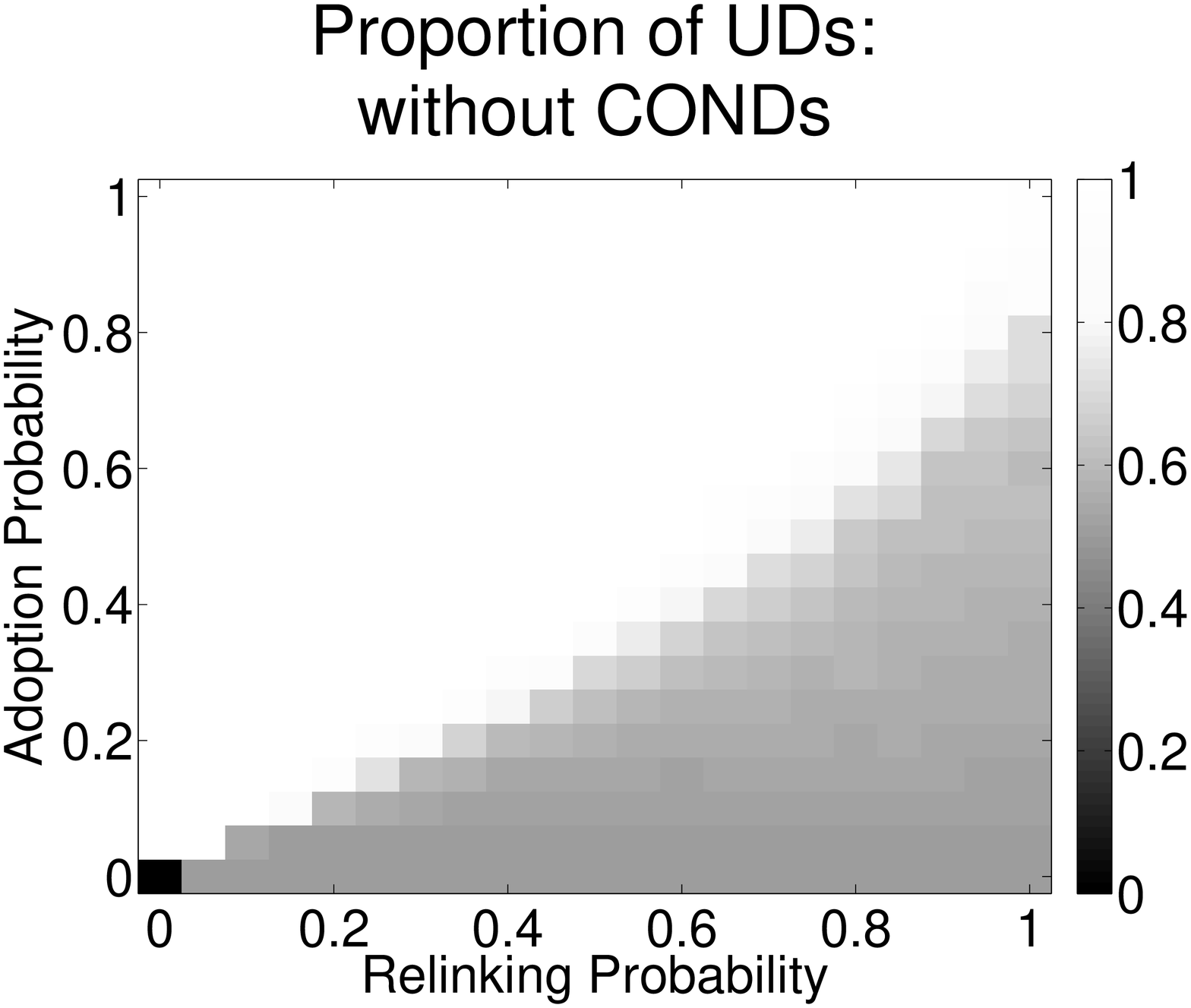}
\includegraphics[width=0.32\textwidth]{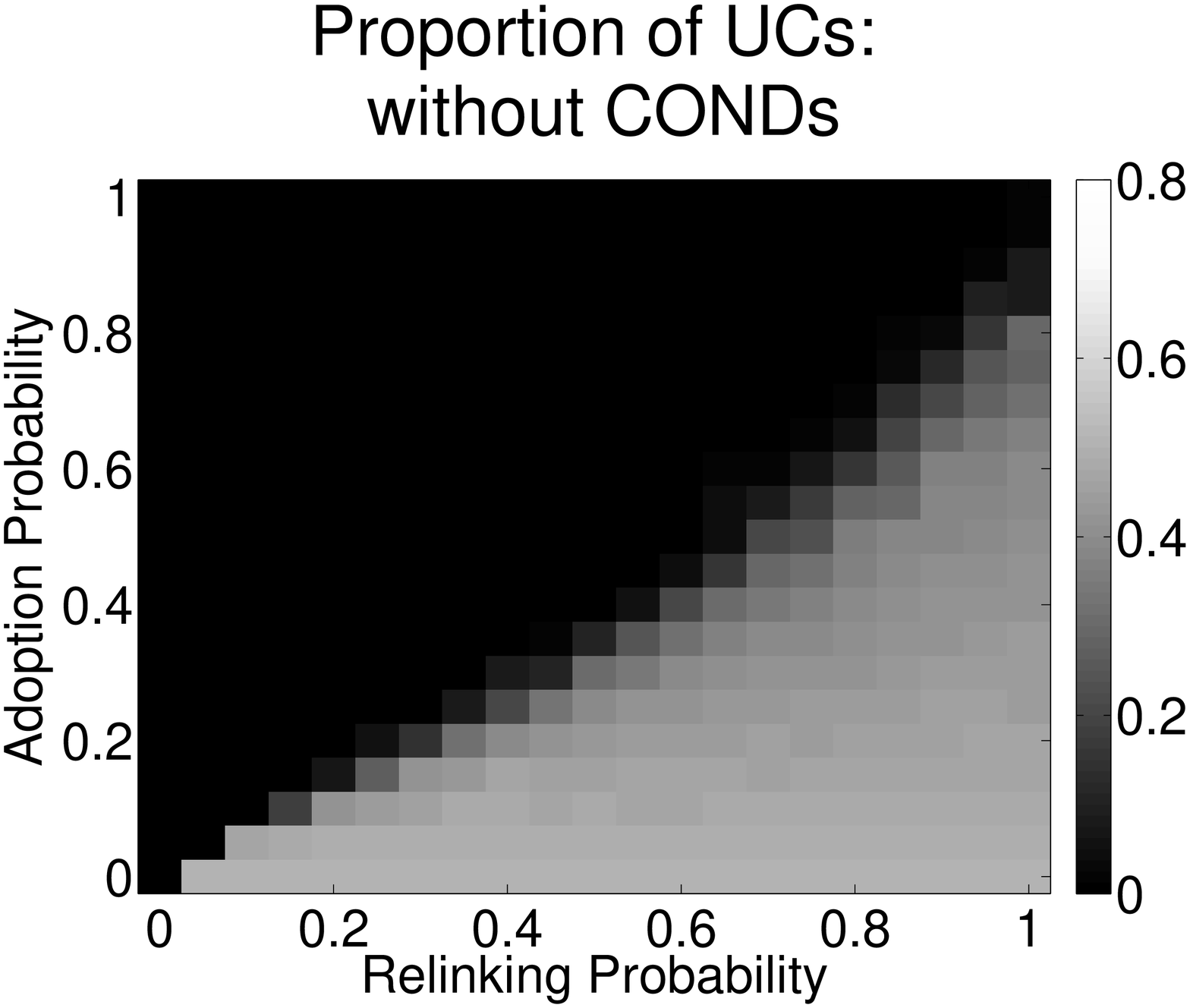}\\
\includegraphics[width=0.32\textwidth]{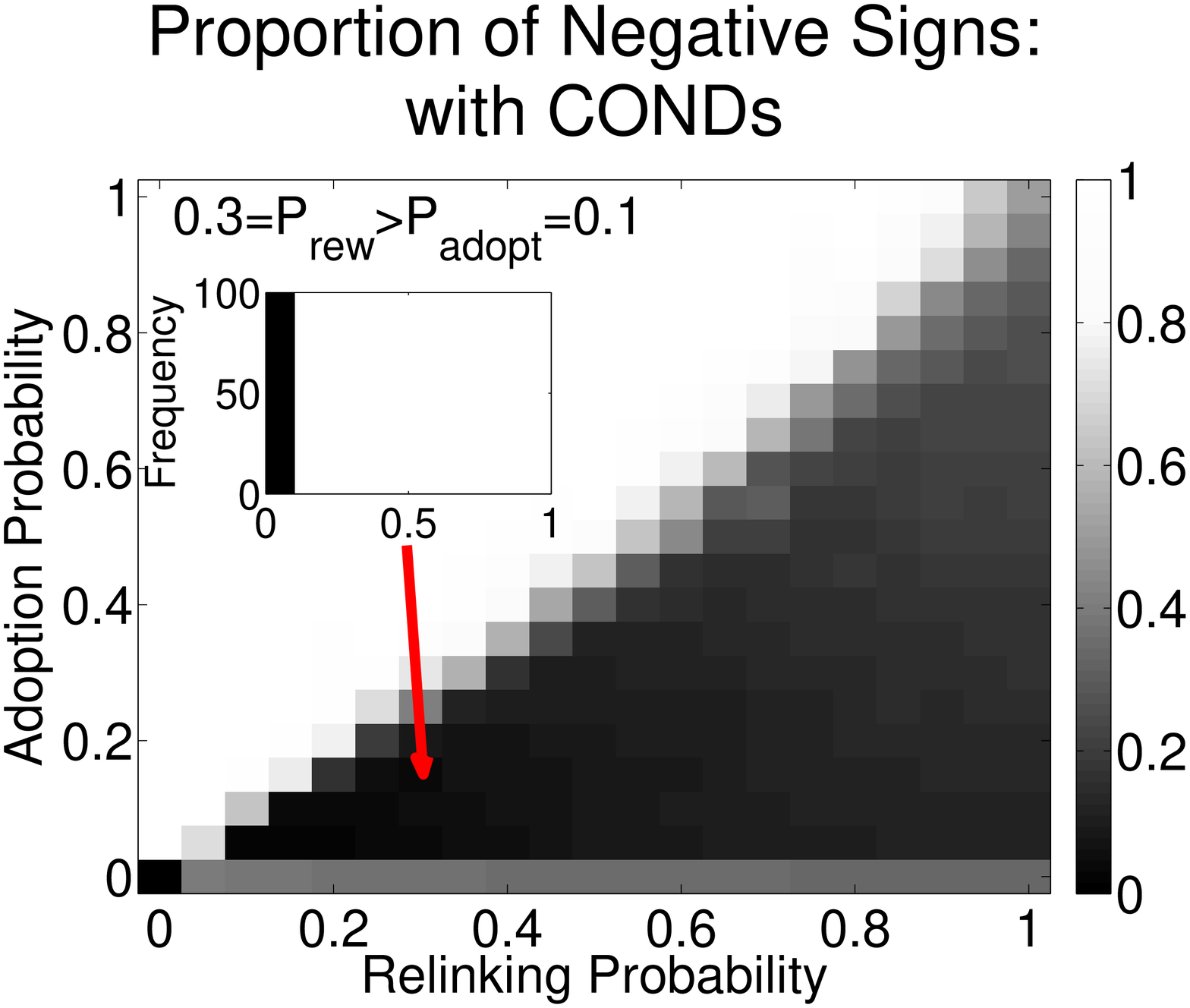}
\includegraphics[width=0.32\textwidth]{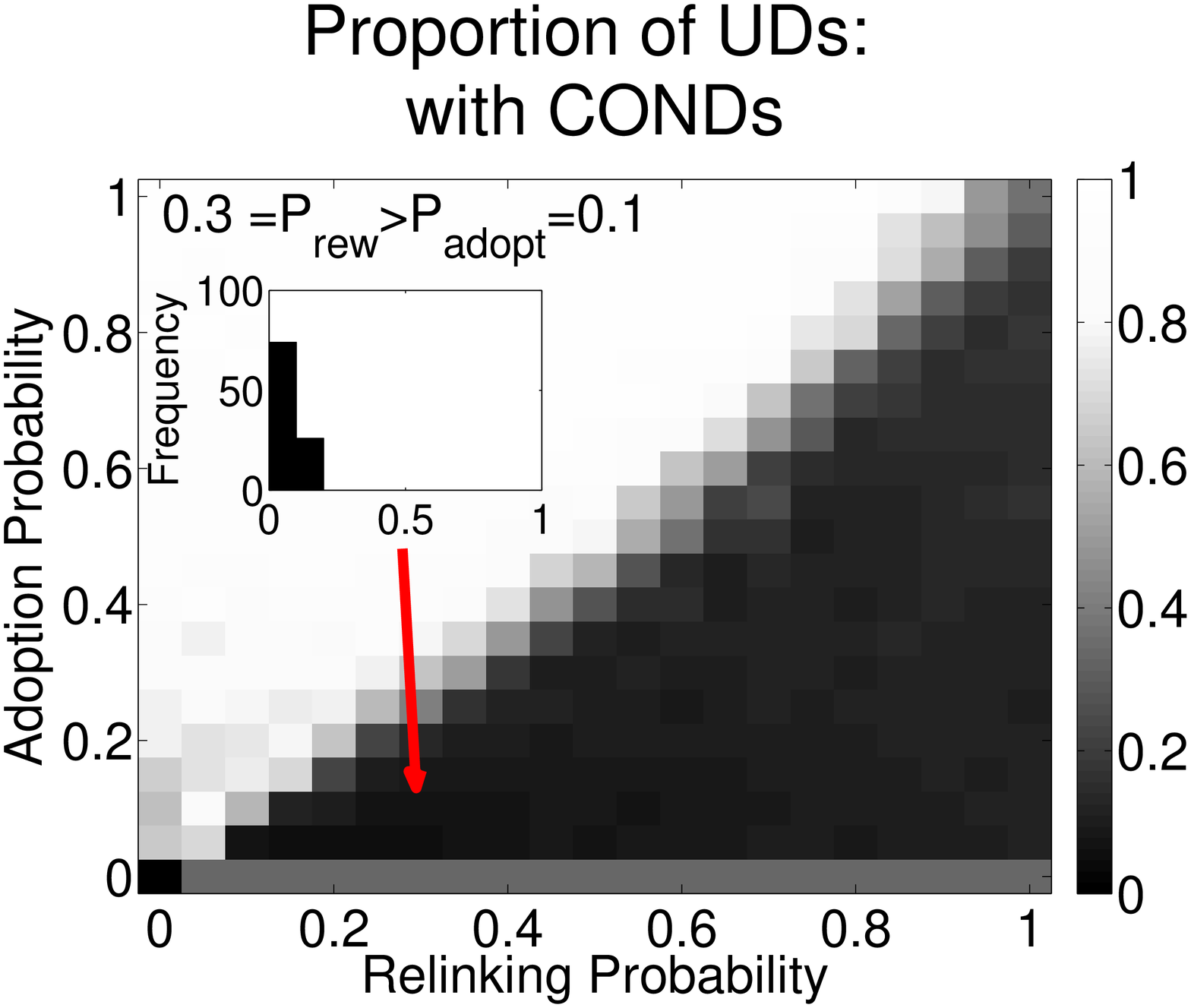}
\includegraphics[width=0.32\textwidth]{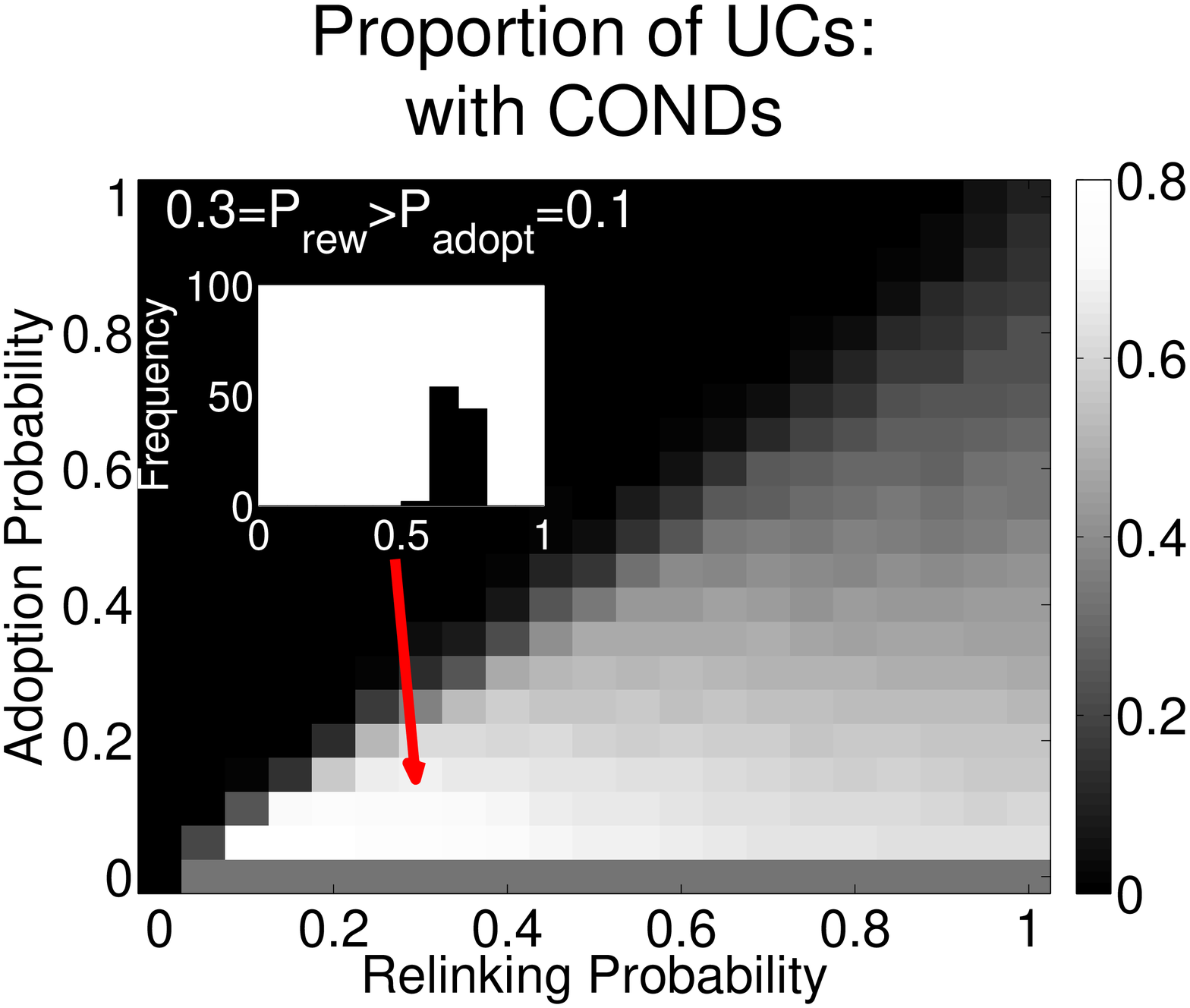}\\
\includegraphics[width=0.32\textwidth]{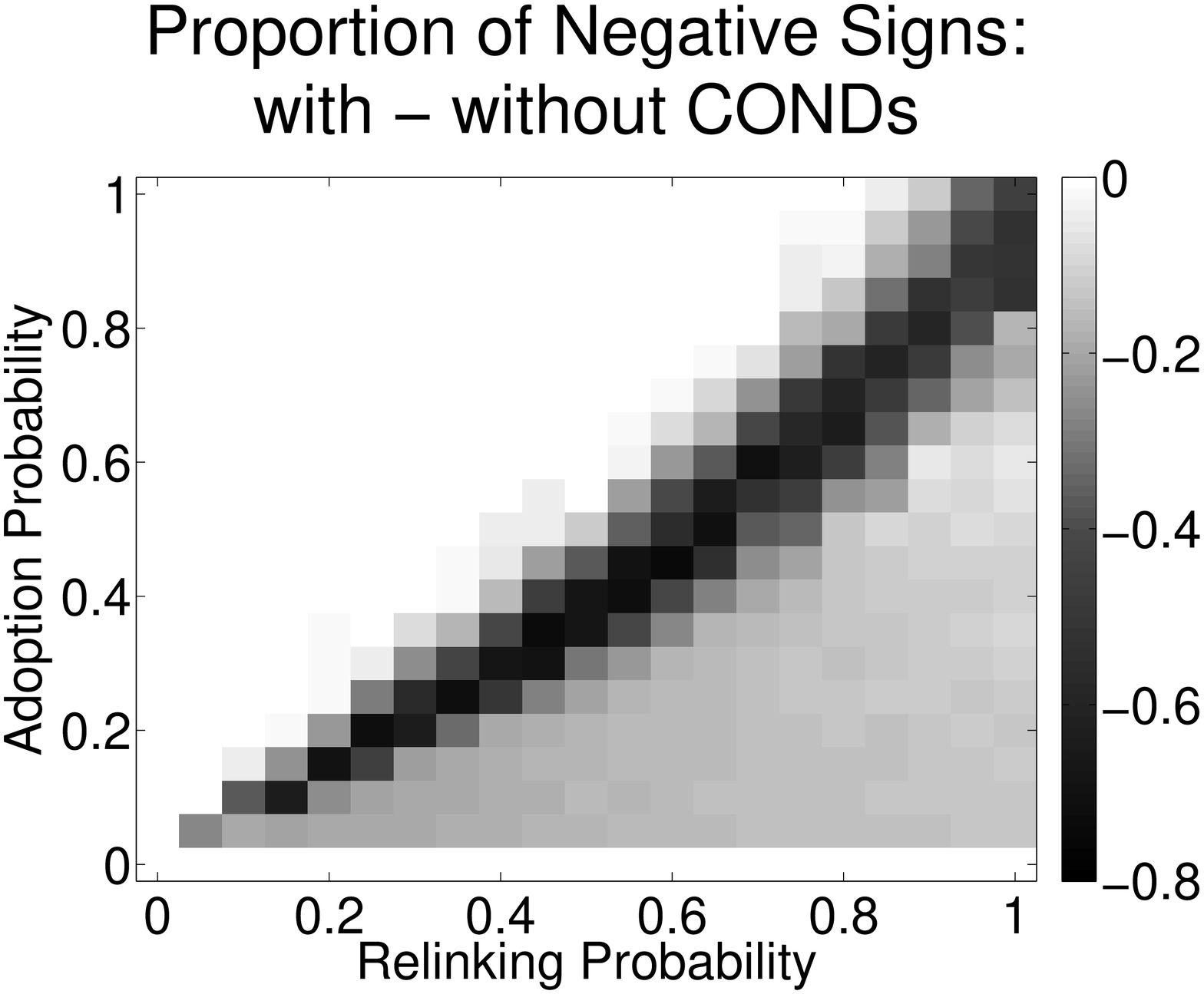}
\includegraphics[width=0.32\textwidth]{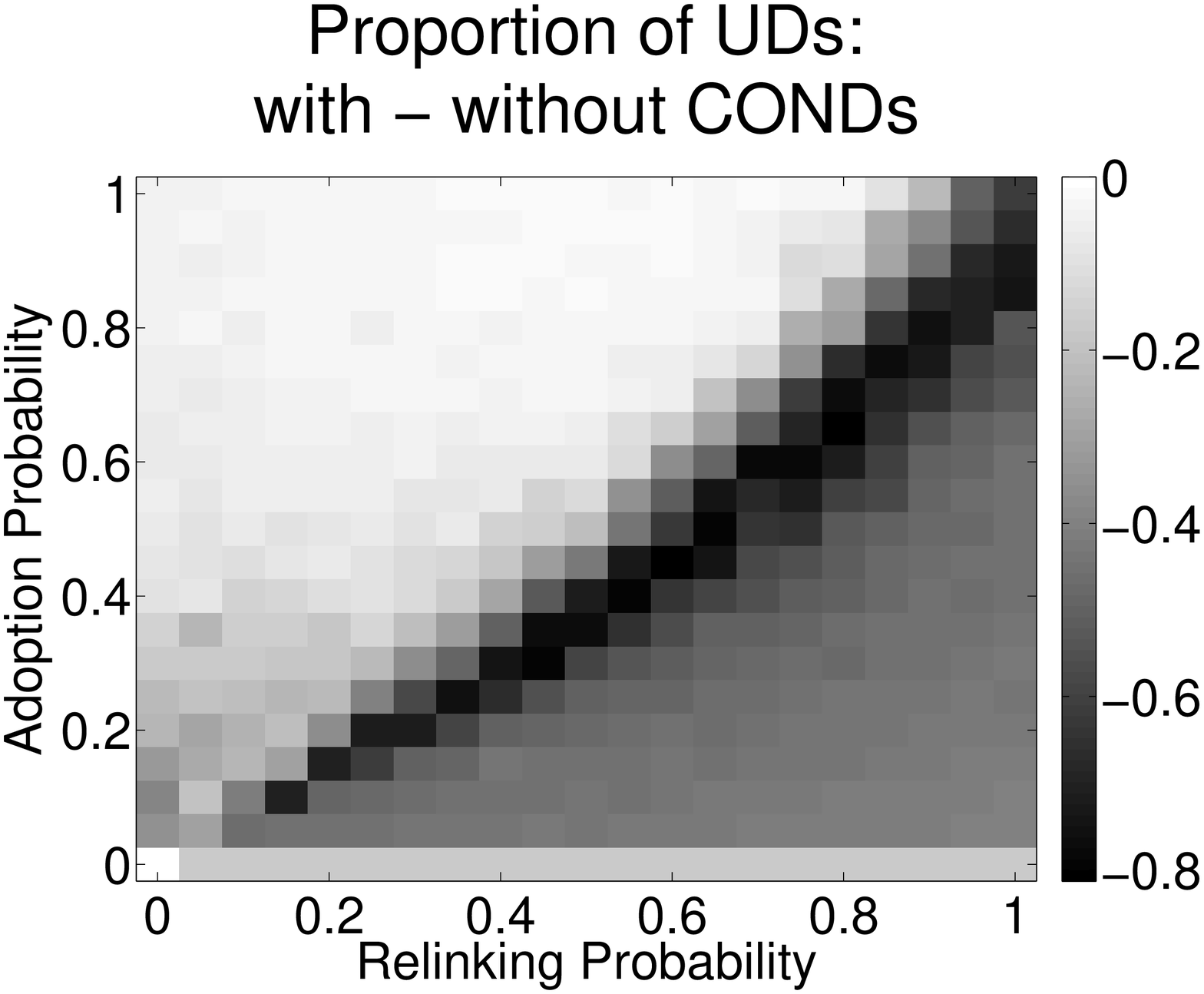}
\includegraphics[width=0.32\textwidth]{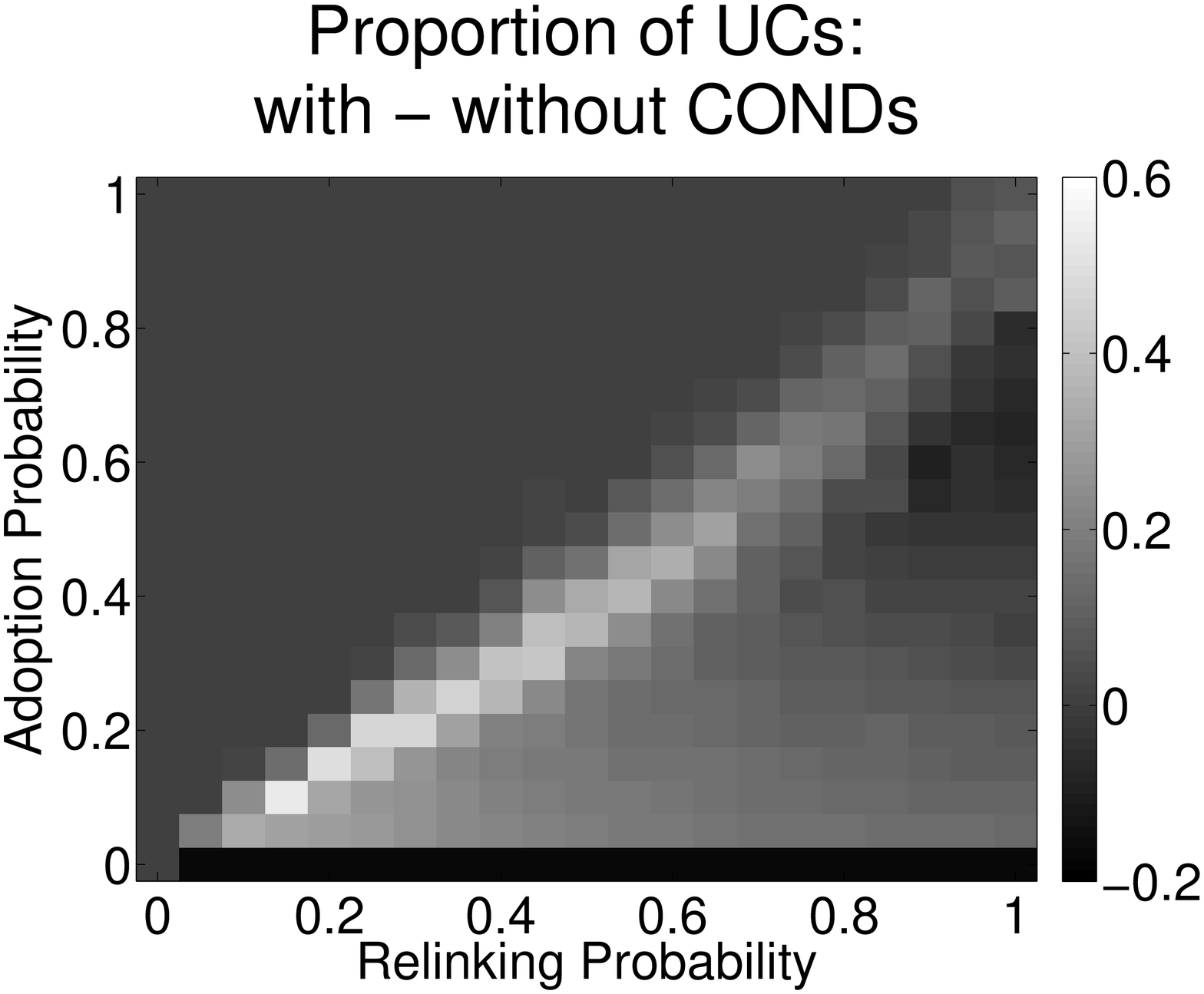}
\caption{The effect of the competing dynamics of strategy adoption (vertical axis) and rewiring of stressed links (horizontal axis) on the final proportion of negative ties in the network (Left Panels), of UDs (Central Panels) and of UCs (Right Panels). The Top Panels show the results in the absence of Conditional Players. The Middle Panels show the results in the case of populations that initially included one third of each strategy. The Lower Panels show the difference in observed proportions with respect to the case in which no CONDs are present at the onset. In all simulations N=200, $P_{neg}=0.2$, $P_{neg}=0.1$ and $P_{rand}=0.01$.. Network signs are initially randomly positive or negative with equal probability. The probability of the existence of a tie is $P_{link}=0.05$.}
\label{FixedPFlip}
\end{figure}

Our model integrates two main dynamic forces that drive the evolution of cooperation and that can be compared with similar mechanisms in the literature. First, the agents tend to adopt those strategies in their social neighborhood that perform better. Second, stressed relationships can be rewired. To analyze the joint influence of these two dynamics, we systematically changed their relative strength (measured as their probability to happen at each time step and interaction). Figure \ref{FixedPFlip} shows the results for $P_{adopt}\in[0,1]$ and $P_{rew}\in[0,1]$, where the values of both variables were changed progressively in steps of $0.05$. For each combination of parameters, we ran 50 simulations. The averages are reported in Figure \ref{FixedPFlip}. In addition, the relative standard deviations are reported in the Appendix in Figure \ref{stds}.

Figure \ref{FixedPFlip} shows that cooperation is sustained for relatively low invasion and high rewiring probabilities. This is in line with the findings of the literature for positive networks. The rewiring of stressed connections enhances the segregation of the agents by type of strategies for any adoption probability. In the context of this model, this decreases the chances of observing negative signs and thus limits the spread of UDs in the population, fostering instead the emergence of a cooperative synergy between CONDs and UCs. 

The results also confirm that, for any rewiring probability, when the likelihood of adoption of more fitting strategies is higher than a threshold, cooperative behavior disappears. Indeed, the speed of update positively affects the proportion of those agents in the population that adopt the strategy with a higher payoff in the dyadic PD (i.e., unconditional defection), thus enhancing its diffusion in the population.

In the absence of CONDs (Top Panels), the two dynamic forces have similar strengths, with a slight advantage for strategy adoption (in the sense that it is required that $P_{rew}>P_{adopt}$ in order for the cooperation to be sustained).   
Moreover, the proportion of cooperators who survive at the end of the simulation increases with the size of the difference $P_{rew}-P_{adopt}$. A similar reasoning applies to the progressive decrease of the proportion of negative ties. Since the proportion of the unconditional cooperators is never above its initial value (50\% of the population) in the absence of emotional strategies, we can conclude that the prevelance of cooperation mainly depends on the speed with which cooperators manage to get isolated from defectors (as in \cite{santos2006cooperation}).

Introducing CONDs in the population produces subtle but important changes. To highlight them, we need to study the last two rows of Figure \ref{FixedPFlip}. First, the area that guarantees the survival of cooperative strategies is extended and now encompasses the values $P_{rew} \approx P_{adopt}$ too. 
Second, when cooperation is supported, the proportion of UCs can rise above its initial value, and in some cases even above the initial sum of conditional players and unconditional cooperators. This is the case when the adoption probability is sufficiently smaller than the rewiring probability. Under other parameter conditions in which cooperation survives, the presence of the emotional strategy can also lower the proportion of unconditional cooperators. This, however, gives advantage to the CONDs themselves and not to the unconditional defectors. A dynamic environment with a high probability of rewiring associated with a high probability of adoption makes the emotional strategy powerful. CONDs tend to diffuse since they obtain systematically higher payoffs when they find themselves in mixed environments of both UCs and UDs. The intensity of the adoption rate makes it possible for CONDs to spread both in the direction of UDs and of UCs. Under these conditions, since the network rewires relatively fast, the conditional agents tend to dominate the population when the equilibrium is reached.

The Lower Panels of Figure \ref{FixedPFlip} compare cases with and without the presence of emotional strategies. These Panels are calculated by taking the difference between the averages in the two cases. The negative values thus indicate a decrease in the proportion of the variable observed as a consequence of the introduction of conditional players.
Overall, across the whole spectrum of the parameters studied, the introduction of emotional strategies reduces both the proportion of unconditional defectors and the proportion of negative ties in the network. The size of the difference is larger in the transitional area in which the two competing dynamics studied have almost the same strength, since the introduction of CONDs there allows for the survival of cooperation where it was impossible before.

We noted in connection with Figure \ref{examples_and_distros} that the dynamics of our model can lead to two possible outcomes, one dominated by cooperative behavior, and the other dominated by defection. The extensive analysis performed here allows us to understand which type of equilibrium dominates. When rewiring is the strongest dynamic force, the cooperative-dominated equilibrium is the only outcome. The distribution of outcomes across 50 simulations is reported for the illustrative case $P_{rew}=0.3$ and $P_{adopt}=0.1$ in the insets in the second row of Figure \ref{FixedPFlip}. When the adoption probability is too strong, only defection-dominated configurations are found. 
Between the two areas a sharp phase transition occurs where we can observe the bimodal distribution of Figure \ref{examples_and_distros}.
The general validity of this observation is shown in Figure \ref{stds} in the Appendix, where the standard deviations of the results are presented. 

\subsection{Robustness checks} 
In the previous section we studied the impact of adoption and rewiring probabilities, by fixing the amount of sign-switches respectively at $P_{neg}=0.2$ and $P_{pos}=0.1$.  However, a complete analysis of our model requires some understanding of the impact of these variables. We repeated the simulations presented in Figure \ref{FixedPFlip} by progressively increasing the values of $P_{neg}$ and $P_{pos}$, while maintaining their ratio. The results are reported in Figure \ref{differentppos} (in the Appendix) for the case $P_{pos}=0.2$ and $P_{neg}=0.4$ as well as for the case $P_{pos}=0.4$ and $P_{neg}=0.8$. The structure of outcomes for different adoption and rewiring probabilities remains qualitatively similar to the one just discussed. The most noticeable change occurs in the area around the phase transition between the parameter combinations where defection dominates and where cooperation prevails. Here, an increase in the sign-switching probability significantly decreases the number of negative ties and that of UDs, while increasing the proportion of UCs. This result indicates that, during the phase transition when the two outcomes are possible, the proportion of outcomes dominated by cooperation increases with the amounts of sign-switch. Above the phase transition ($P_{adopt}>P_{cond}$), the number of unconditional defectors decreases, but this variation does not result in changes in the proportion of UCs or of negative signs. This hints at the fact that the missing UDs are defecting CONDs, who are functionally equivalent to pure defectors. Finally, below the phase transition ($P_{adopt}<P_{cond}$), UC domination increases with the probability of sign changes. Once again, changing the sign-switch probabilities does not have an impact on the measure of cooperative behavior, only on the balance between conditional and unconditional cooperation.
We can therefore conclude that while some quantitative differences occur as we change the speed of sign-switching relative to other dynamic forces, our results are qualitatively resilient to these changes.

Another piece of criticism one could level at our model is that our rewiring mechanism endogenously creates clusters of cooperators thus increasing the cooperation when enough rewiring is introduced. In the attempt of modeling the rewiring process in line with the sociological literature, we assumed that new links can only be created among friends of friends.  An alternative model specification is to assume that new links are created at random. To analyze the impact of this alternative specification, we performed two sets of simulations setting $P_{rand}=1$ and keeping all other relevant parameters equal. The first set reproduces the test of the impact of the proportion of CONDs on cooperation, while the second reproduces the analysis of the joint effect of adoption and rewiring probability. Both are reported, in Figure \ref{prop_cond}  and in Figure \ref{Pflip_rand1} (in the Appendix), respectively. The first set of results shows that there are still two types of equilibrium, and that their type is the same as when $P_{rand}<<1$. We underline, however, that at each level of COND, the proportion of simulations in which the population converges to the defection-dominated equilibrium is consistently higher with random rewiring. The reason for the decrease in the share of cooperative type of equilibrium is that a purely random rewiring reduces the segregation effect produced by our baseline setup, making it less likely for cooperation to emerge. Figure \ref{Pflip_rand1} also shows, nevertheless, that a sharp decrease in cooperation is limited to the area around the phase transition. In the area where unconditional cooperation dominates with transitive update, it does even more so with the random rewiring procedure.
This is a surprising result and indicates that rewiring based on transitive closure provides an advantage for cooperation in the critical parameter domain, while a random rewiring procedure is a more efficient engine for dissemination when the conditions for cooperation are highly favorable. This is in line with the literature on the diffusion of innovations, which highlights the limitations for the spread in the case of transitive closure and demonstrates the efficiency of bridging (or random) ties in the process \cite {burt1992social,granovetter1973strength,watts1998collective}.
Furthermore, we point out that with random rewiring, the proportion of UCs only depends on the adoption probability and not on the degree of rewiring. This is a consequence of the fact that the random procedure has a neutral effect on the segregation of different strategies, therefore, even when the rewiring happens fast, the slow-down in the diffusion of cooperation does not happen. Summing up, the introduction of a random rewiring mechanism as a substitute of the one discussed before brings quantitative changes into our model but preserves our main results, concerning the effects of negative ties and emotional strategies on cooperation.

\subsection{The effect of density and network size}\label{ConnandSize}

Following the literature on the evolution of cooperation in networks \cite{ohtsuki2006simple}, we studied the consequences of varying the connectivity. We increased the density of the Erd\H{o}s-R\'{e}nyi random network progressively. In the experiments reported here, both the probability of rewiring and that of adoption were equal to $0.1$. This meant that we selected parameter conditions that were on the frontier between the area in which cooperation was sustained and the area in which it disappeared. It follows from this that the results for cases in which $P_{rew}>P_{adopt}$ would be even more pronounced.

\begin{figure}[ht!]
\centering
\includegraphics[width=0.49\textwidth]{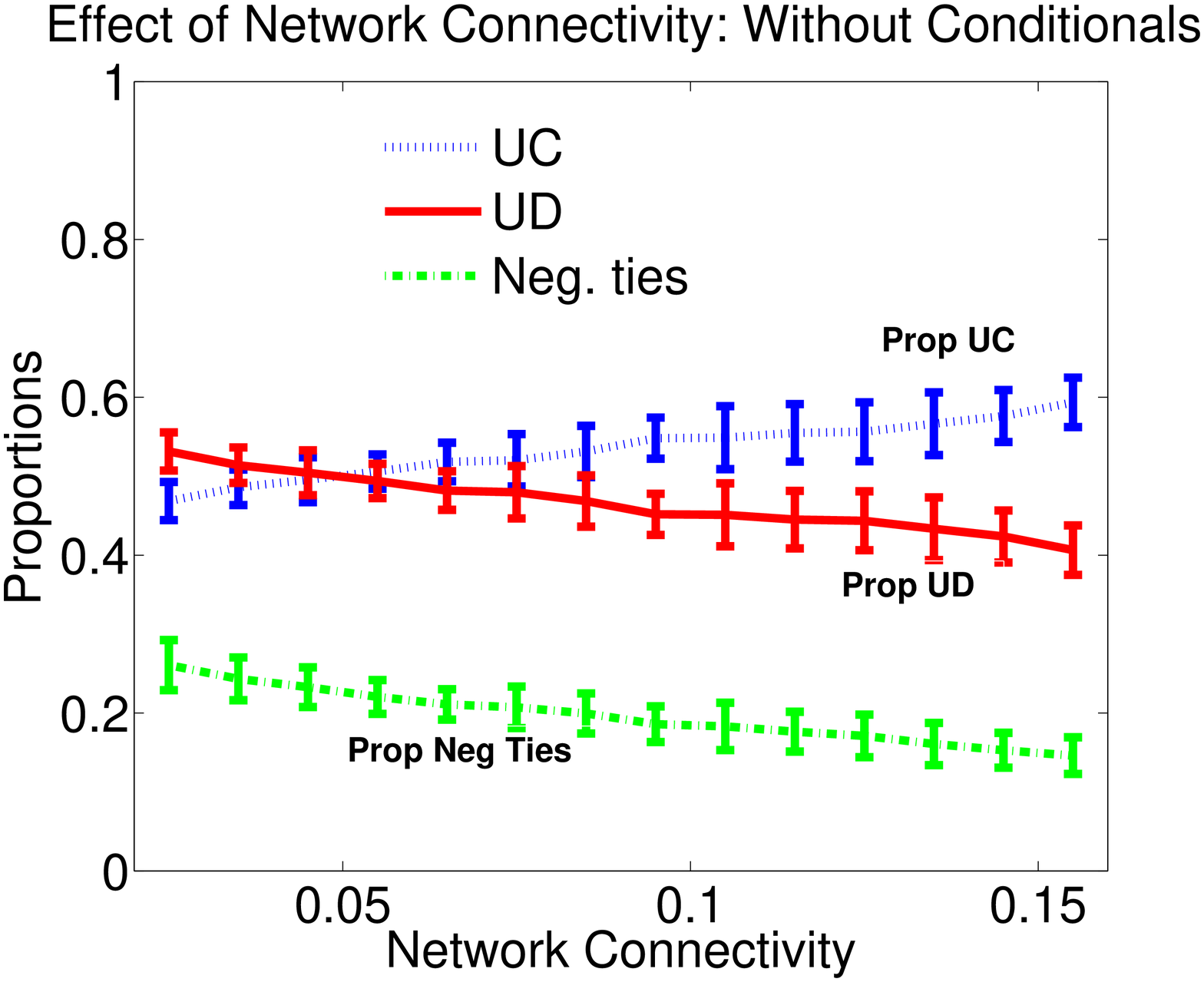}
\includegraphics[width=0.49\textwidth]{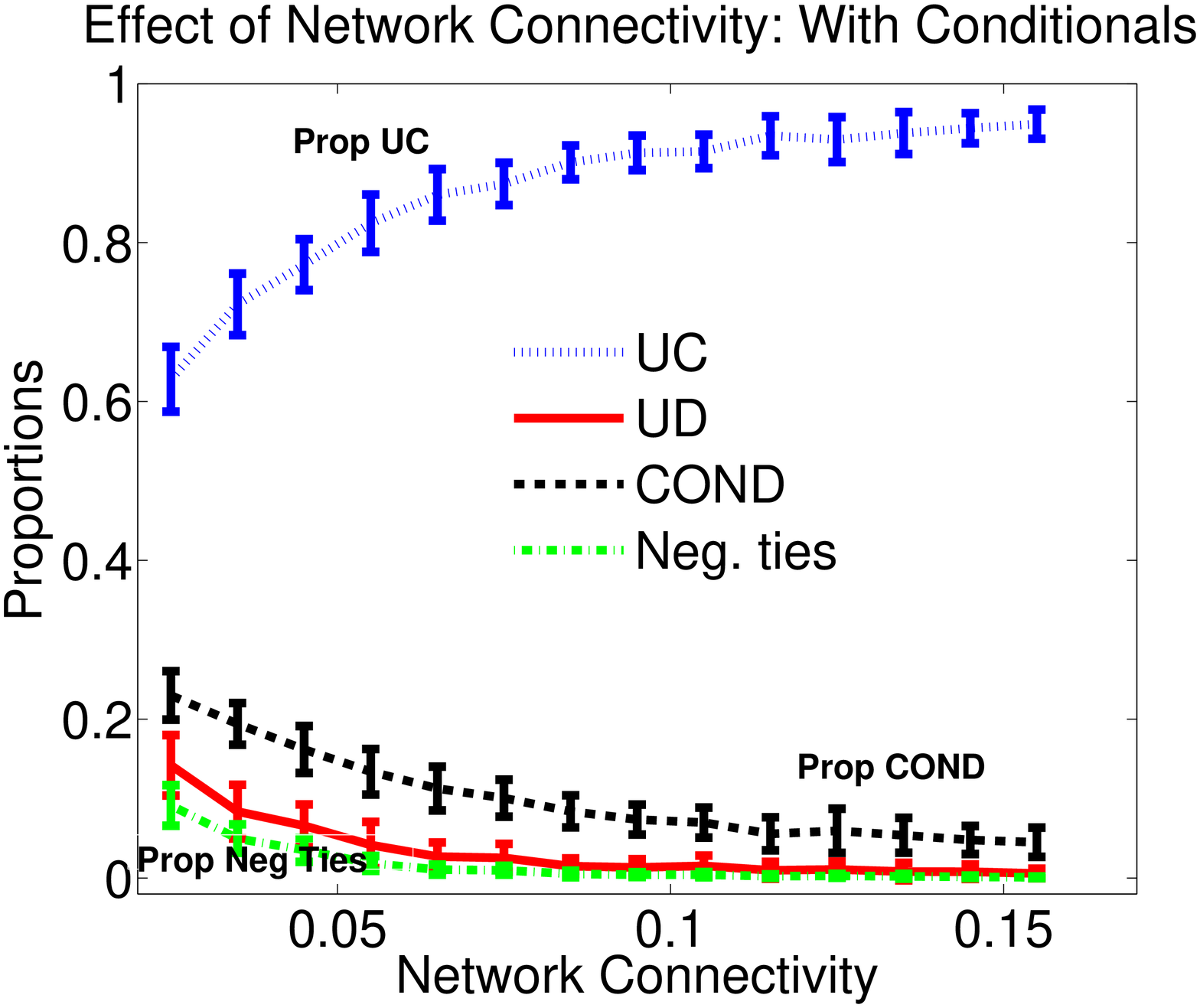}
\caption{The effect of the probability of existence of each tie ($P_{link}$) on the final proportions of agent types and on the final proportion of negative signs. Left Panel: results in the absence of conditional players. Right Panel: results in the presence of conditional players. In all simulations N=200, $P_{rew}=P_{adopt}=0.1$, $P_{neg}=0.2$, $P_{neg}=0.1$ and $P_{rand}=0.01$. The network signs are initially randomly positive or negative with equal probability.}
\label{Connectivity_TypesANDSigns}
\end{figure}

As Figure \ref{Connectivity_TypesANDSigns} shows, in the absence of emotional strategies, the proportion of the agents of each type remains very similar to the initial value of $0.5$, regardless of network density. 
In contrast, when CONDs are added, increasing the density progressively increases the proportion of UCs in the final population, up to the point where they come to constitute the whole population. At the same time, the final proportion of conditional players, defectors and negative signs becomes negligibly small. 
Therefore, in our model when emotional strategies are introduced, higher density leads to a significant increase in cooperative behavior. This intriguing result is in contrast with the one obtained for unsigned networks by \cite{ohtsuki2006simple}, which shows how an increase in density leads to a decrease in cooperation.

This result follows from the combination of some rewiring (without which UD dominates) and the presence of conditional players (without whom the UC strategy does not diffuse). As the density increases, the average agent becomes more connected and thus it takes more rewiring (i.e., more time) to create disconnected clusters of strategies. This provides more time for the combination of COND and UC strategies to locally outperform UD, thus diffusing cooperation.

\begin{figure}[ht!]
\centering
\includegraphics[width=0.49\textwidth]{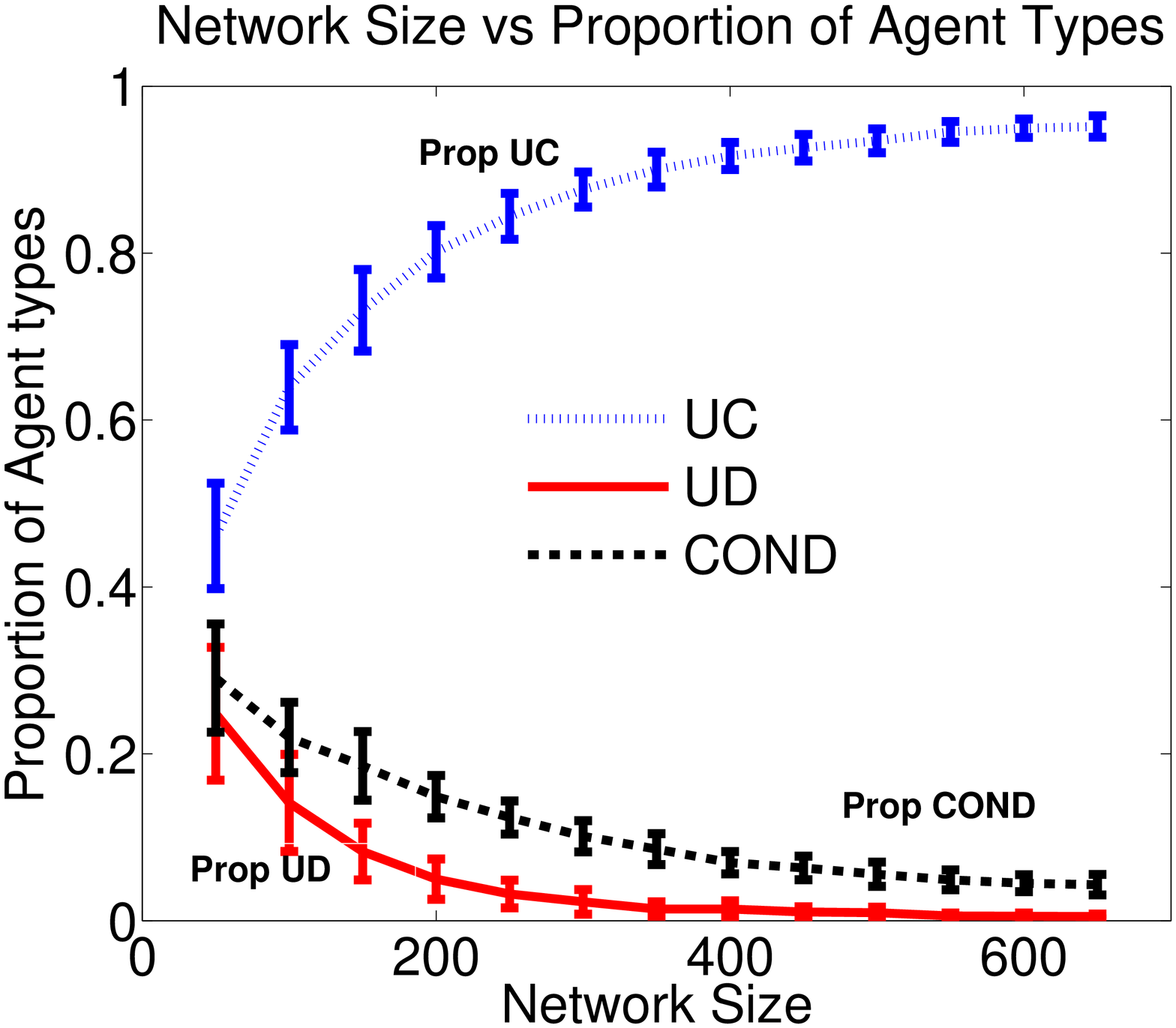}
\includegraphics[width=0.49\textwidth]{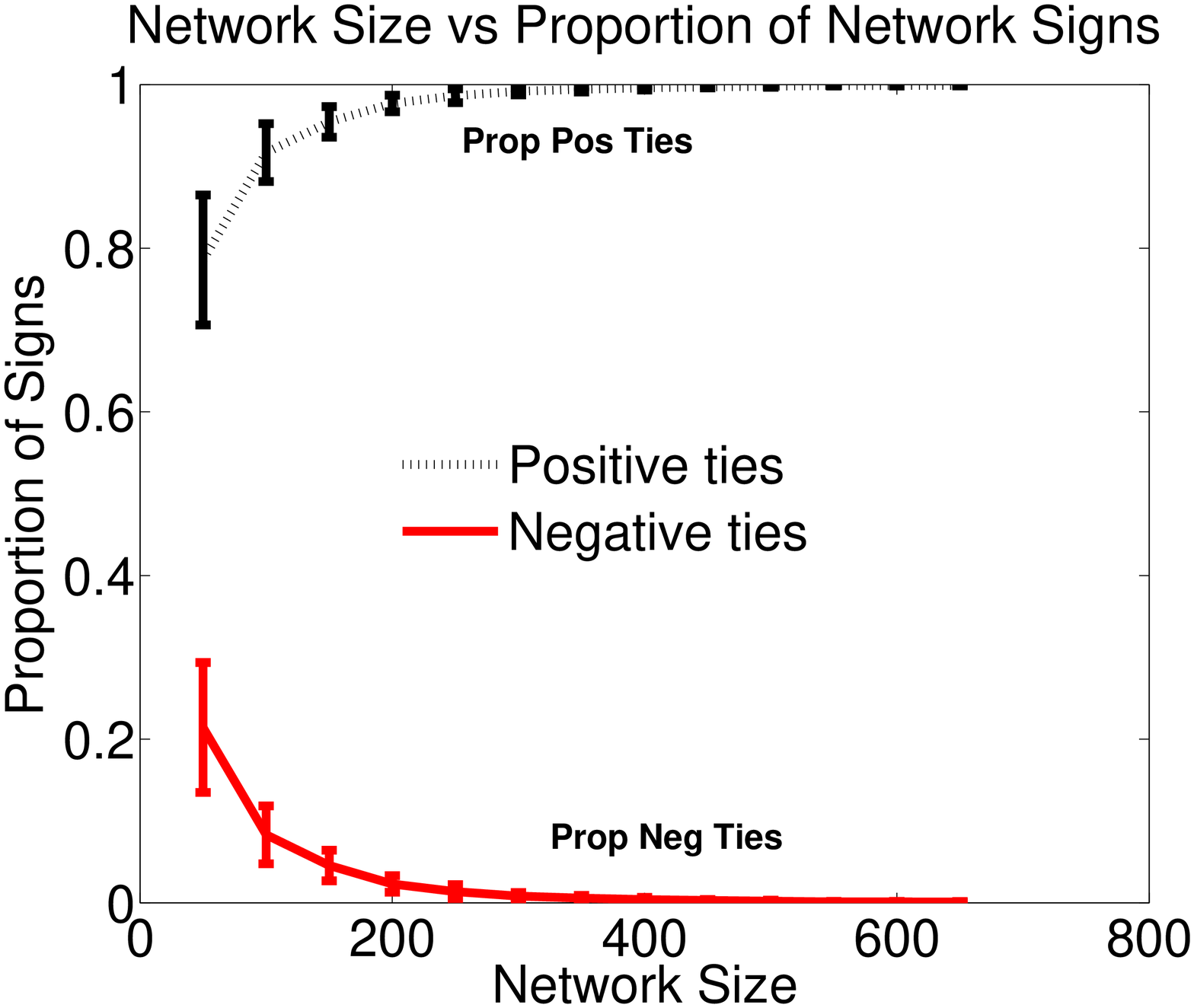}

\caption{The effect of the network size on the final proportion of agent types (Left Panel) and on the final proportion of negative and positive tie signs (Right Panel). In all simulations $P_{rew}=P_{adopt}=0.1$, $P_{neg}=0.2$, $P_{neg}=0.1$ and $P_{rand}=0.01$. The network ties are initially randomly positive or negative with equal probability and the probability of existence for each of them is $P_{link}=0.05$.}
\label{Size_TypesANDSigns}
\end{figure}

To conclude our analyses, as a further robustness check of our results, we studied the influence of network size on the evolution of cooperation. Once again, we fixed all parameters at their baseline values (the rewiring and adoption probabilities were fixed at $0.1$, while the network density was $0.05$) and we increased network size progressively.
Increasing the size (Figure \ref{Size_TypesANDSigns}) simply reinforced our results, making them more pronounced in the sense that the proportion of UCs becomes even more dominant than in smaller networks. This reassures us about the fact that the results observed are not artifacts of small-network noise effects.

\section{Conclusions}\label{Conclusions}
In this paper, we have suggested a new solution to the puzzle of the evolution of unconditional cooperation. We have shown that strategies that make use of the emotional content of social relationships can act as catalysts, creating favorable conditions for the spread of unconditional cooperation, while they do not gain dominance. This is a similar idea to \cite{Jain16012001}, who developed a simple mathematical model for the evolution of an idealized chemical system to study how a signed network of cooperative molecular species arises and evolves to become more complex and structured. In that paper, however, they used network signs to represent catalytic and inhibitory interactions among the molecular species. 

We have built a model in which the simple strategies of the single-shot PD are adopted locally and network ties and signs co-evolve with interactions. The update of network ties and of relational signs, however, are not part of the agents' strategies and thus subject to probabilistic updates and not to evolution. We have chosen this simple setup because we wanted to model situations in which strategies are simple and agents have no memory of interactions. Alternative interpretations and assumptions have been widely studied in the network game literature \cite{jackson2010social,jackson1996strategic,takacs2008collective}.

We interpreted sign-dependent strategies as emotional strategies in relation to everyday observations in that humans react with affection, liking, disliking, frustration, anger, commitment, and gratitude to their various experiences in social dilemma interactions. In line with the suggestion of \cite{trivers1971evolution}, we highlighted the key role of emotions in the evolution of cooperation. A possible limitation of our study is that we did not let strategies mutate in order to develop more fine-tuned emotional responses to their interaction experience.

The main result of this paper is that the emotional strategies can act as catalysts for cooperation but they rarely gain dominance themselves, which resembles that of \cite{nowak2006five} about the role of TFT in structured populations. \cite{nowak2006five} observed that this strategy stimulates the diffusion of cooperation when many defectors are present, but it is then taken over by other strategies such as the {\it Win Stay, Lose Shift}.

Besides our main finding, we have determined key parametric conditions under which this evolutionary catalysis takes place in populations similar to those typically assumed for human ancestral communities \cite{dunbar1992neocortex}. The most important conditions that favor cooperation are high rewiring probability, low adoption probability of strategies that perform better, and high network density. This latter finding contradicts earlier observations  \cite{nowak2006five,ohtsuki2006simple,santos2006cooperation} on the evolution of cooperation in graphs, and thus confirms the importance of studying the co-evolution of networks and cooperation on signed graphs in this context. The result, however, is in line with the sociological literature that claims that a cohesive structure of communities supports better cooperation than a sparse one \cite{coleman1994foundations,hechter1988principles}.

Another novelty of this study concerns the interaction between emotional strategies and rewiring. In line with the literature on non-signed networks, we have found that the possibility of rewiring tense relationships enhances the chances of diffusion of cooperative behavior. However, we have refined this result, showing that when rewiring happens too fast, the segregation induced can also inhibit the diffusion of cooperation, as negative relations that could potentially become positive do not have enough time to do so.

Finally, while we assumed an Erd\H{o}s-R\'{e}nyi random initial distribution of degrees in the network, a natural extension of this paper would be to study more realistic topologies. In particular, scale-free, small world, and core-periphery structures could be analyzed to test the resilience of the results obtained in this work with the introduction of sign-dependent emotional strategies. While some studies find that certain structures are more efficient in sustaining cooperation \cite{masuda2003spatial}, other theoretical and experimental studies do not find any advantages due to topology in positive networks \cite{gargiulo2012influence,gracia2012human,grujic2010social,gracia2012heterogeneous,suri2011cooperation}. Which result applies to signed networks remains an open question. We make a first step in the direction of answering it in \cite{righi2014degree}.

\section*{Acknowledgments}
The authors wish to thank the "Lend\"{u}let" program of the Hungarian Academy of Sciences for financial support, two anonymous reviewers, Gabriella Mezei and Elena Franchini for their useful comments.

\section*{Appendix}
\begin{figure}[ht!]
\centering
\includegraphics[width=0.32\textwidth]{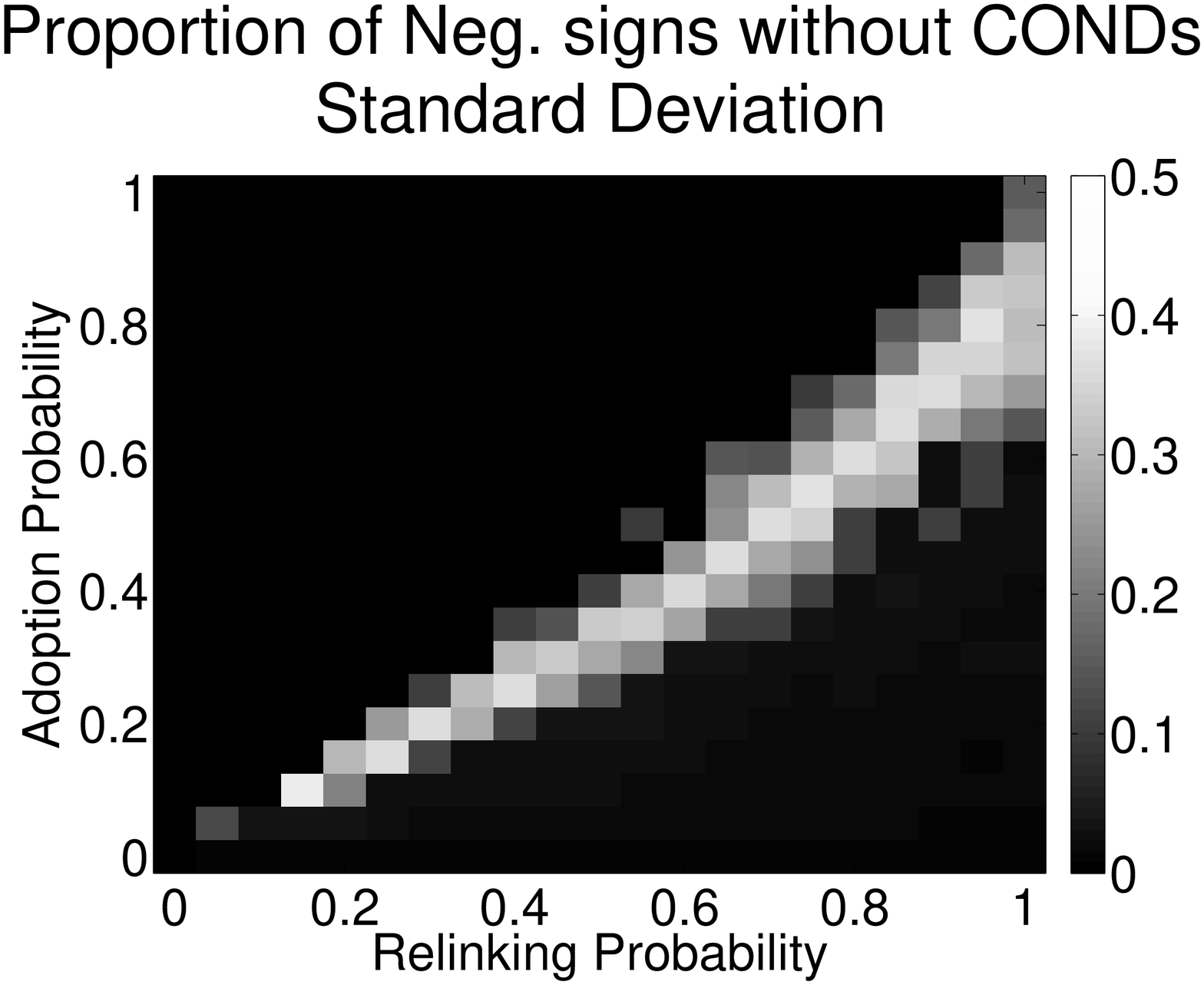}
\includegraphics[width=0.32\textwidth]{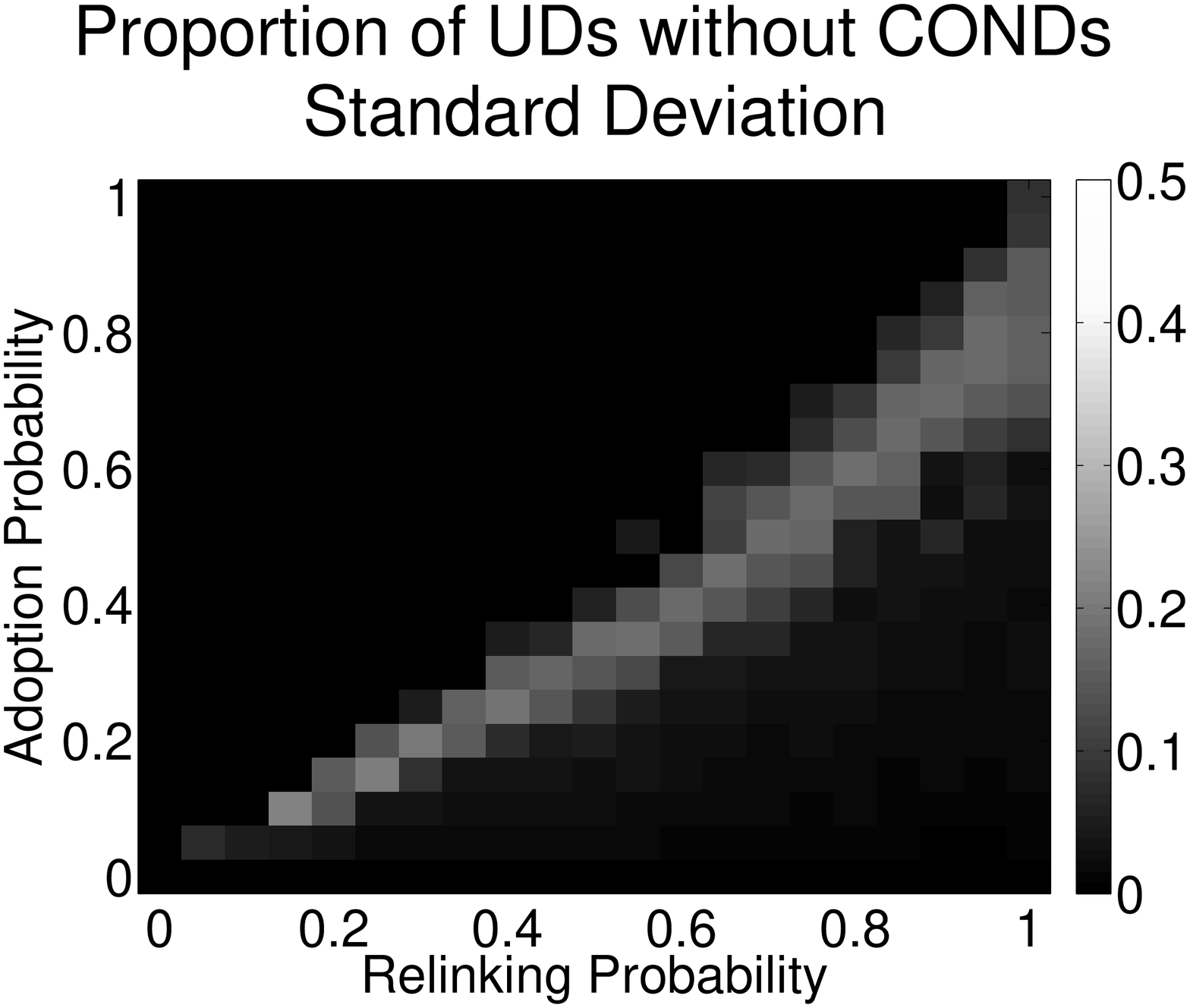}
\includegraphics[width=0.32\textwidth]{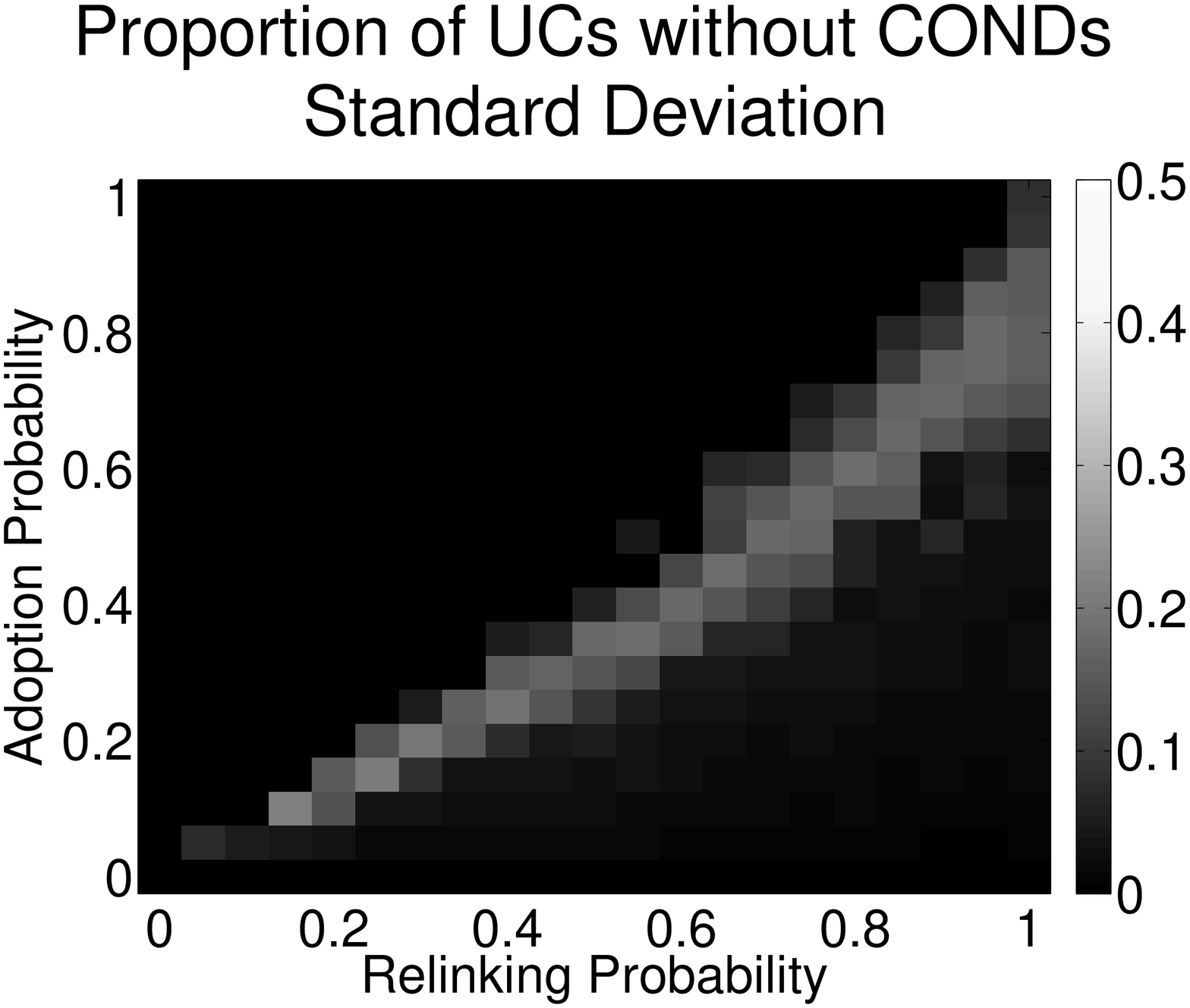}
\includegraphics[width=0.32\textwidth]{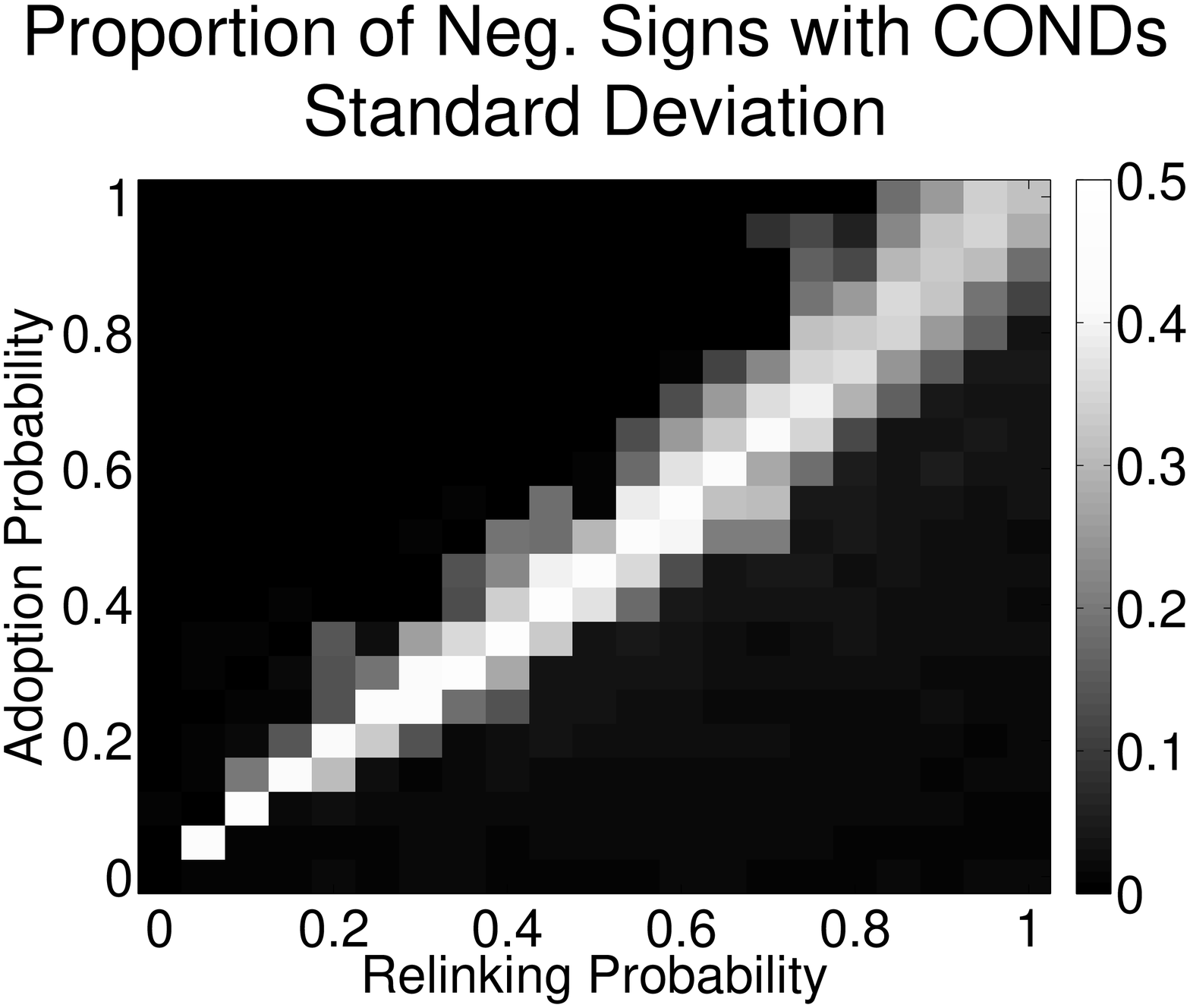}
\includegraphics[width=0.32\textwidth]{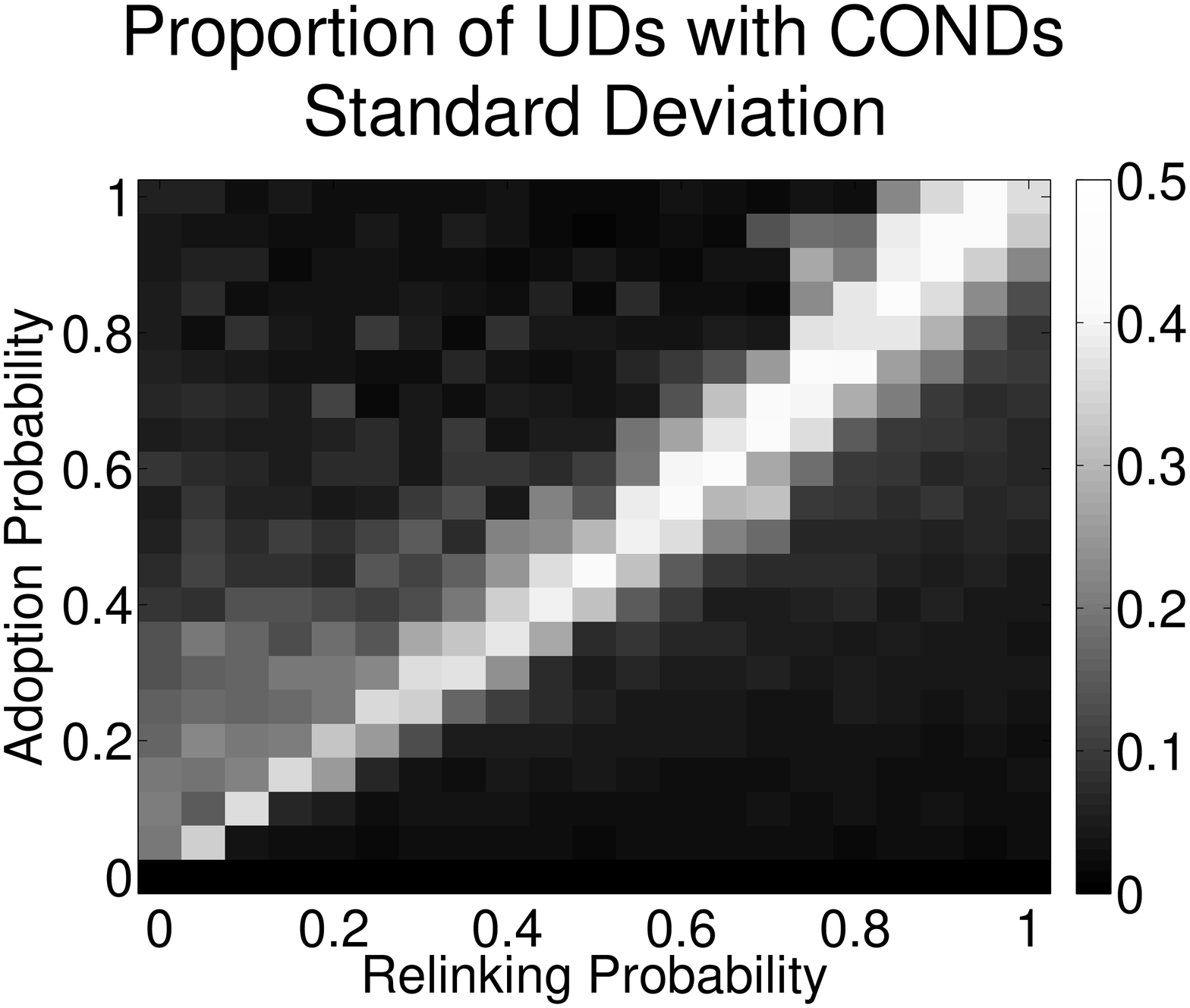}
\includegraphics[width=0.32\textwidth]{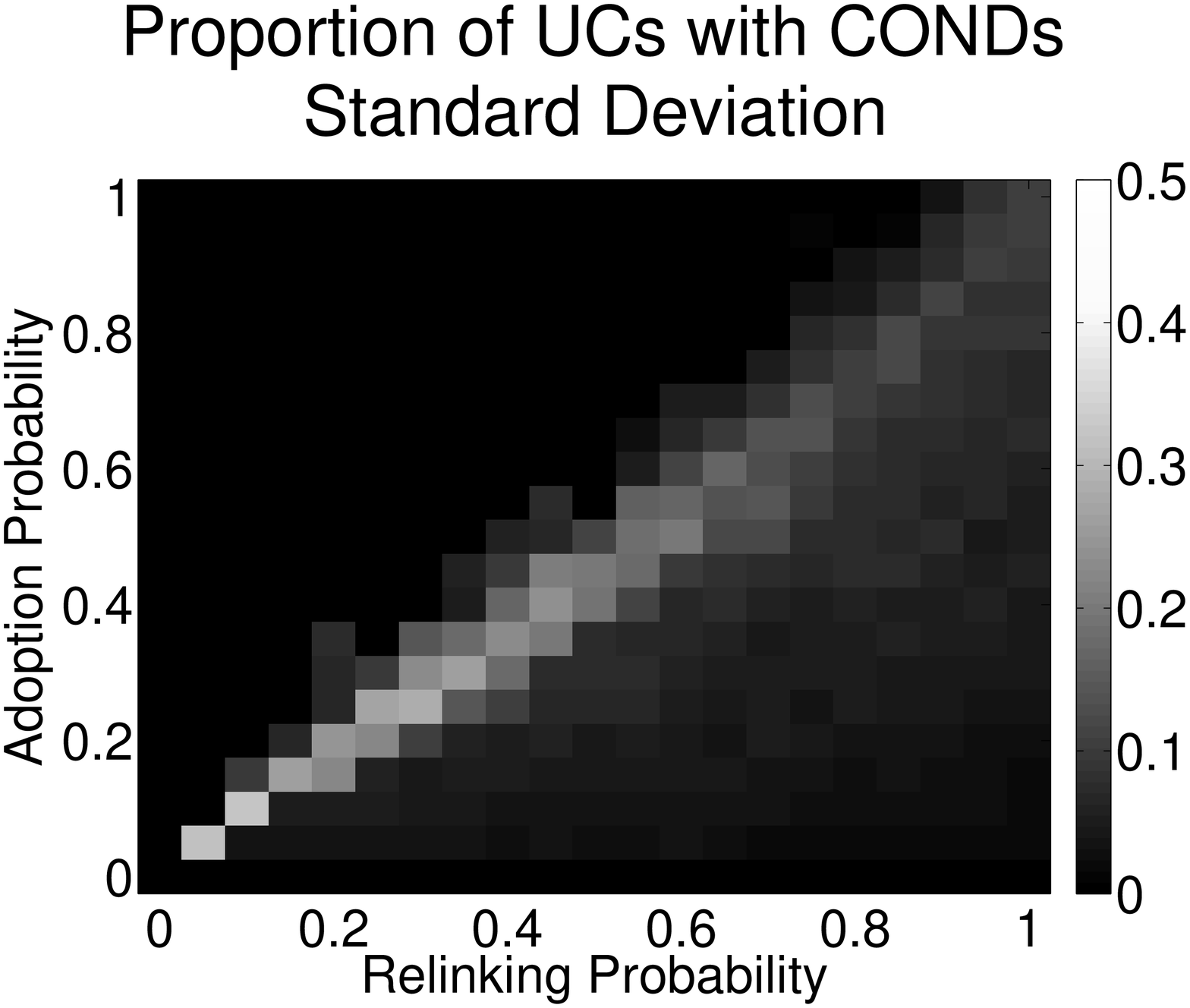}\\
\caption{Effect of the competing dynamics of adoption of strategies with higher payoffs (vertical axis) and of rewiring of stressed links (horizontal axis) on the {\it Standard Deviations} of the final proportion of negative ties in the network (Left Panels), of UDs (Central Panels) and of UCs (Right Panels). Top Panels show the results in the absence of Conditional Players. Lower Panels show results for populations initially including one third of each strategy. In all simulations N=200, $P_{neg}=0.2$, $P_{neg}=0.1$ and $P_{rand}=0.01$. Network signs are randomly initialized positive or negative with equal probability. The probability of existence for each tie is $P_{link}=0.05$.}
\label{stds}
\end{figure}

\begin{figure}[ht!]
\centering
\includegraphics[width=0.32\textwidth]{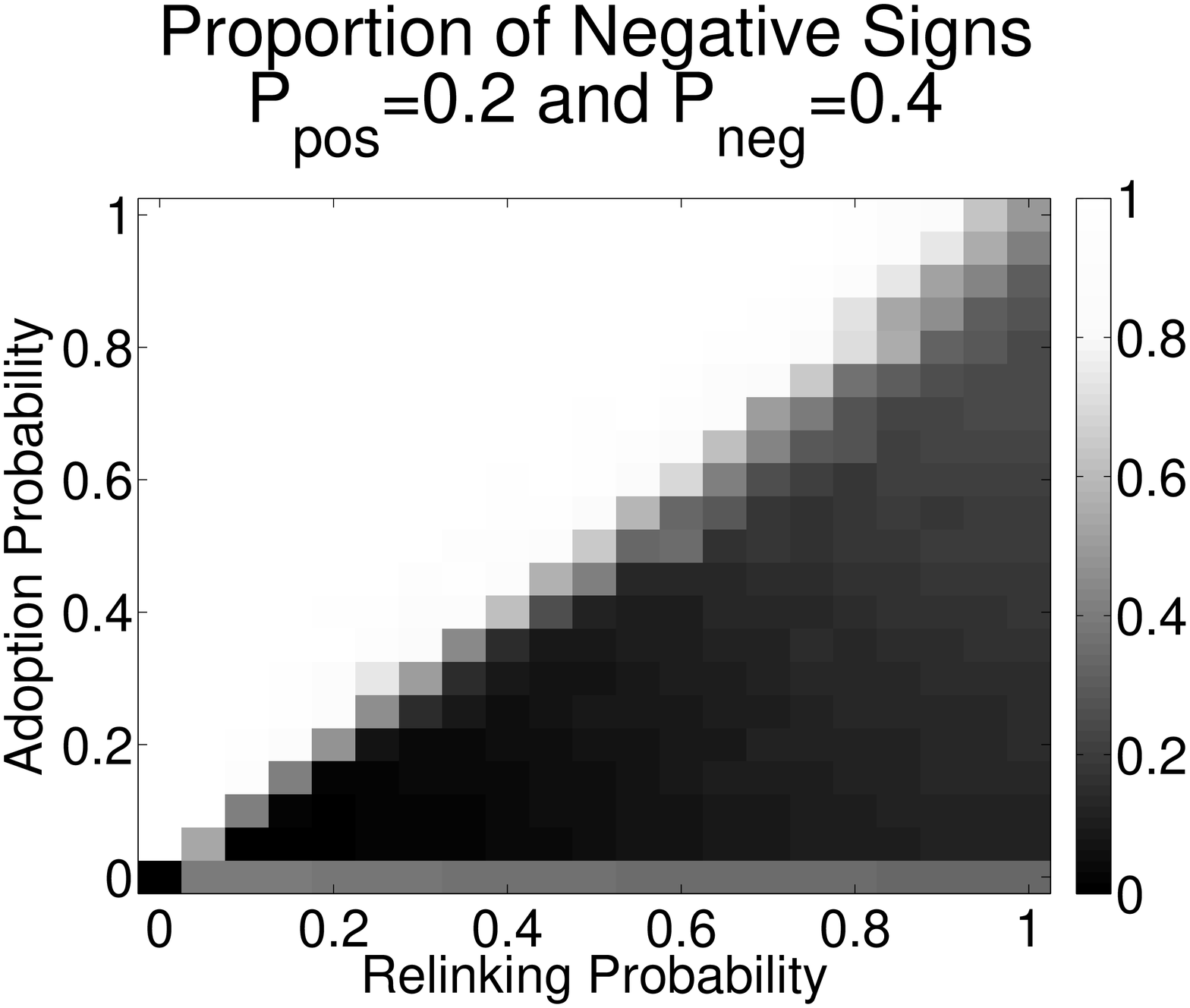}
\includegraphics[width=0.32\textwidth]{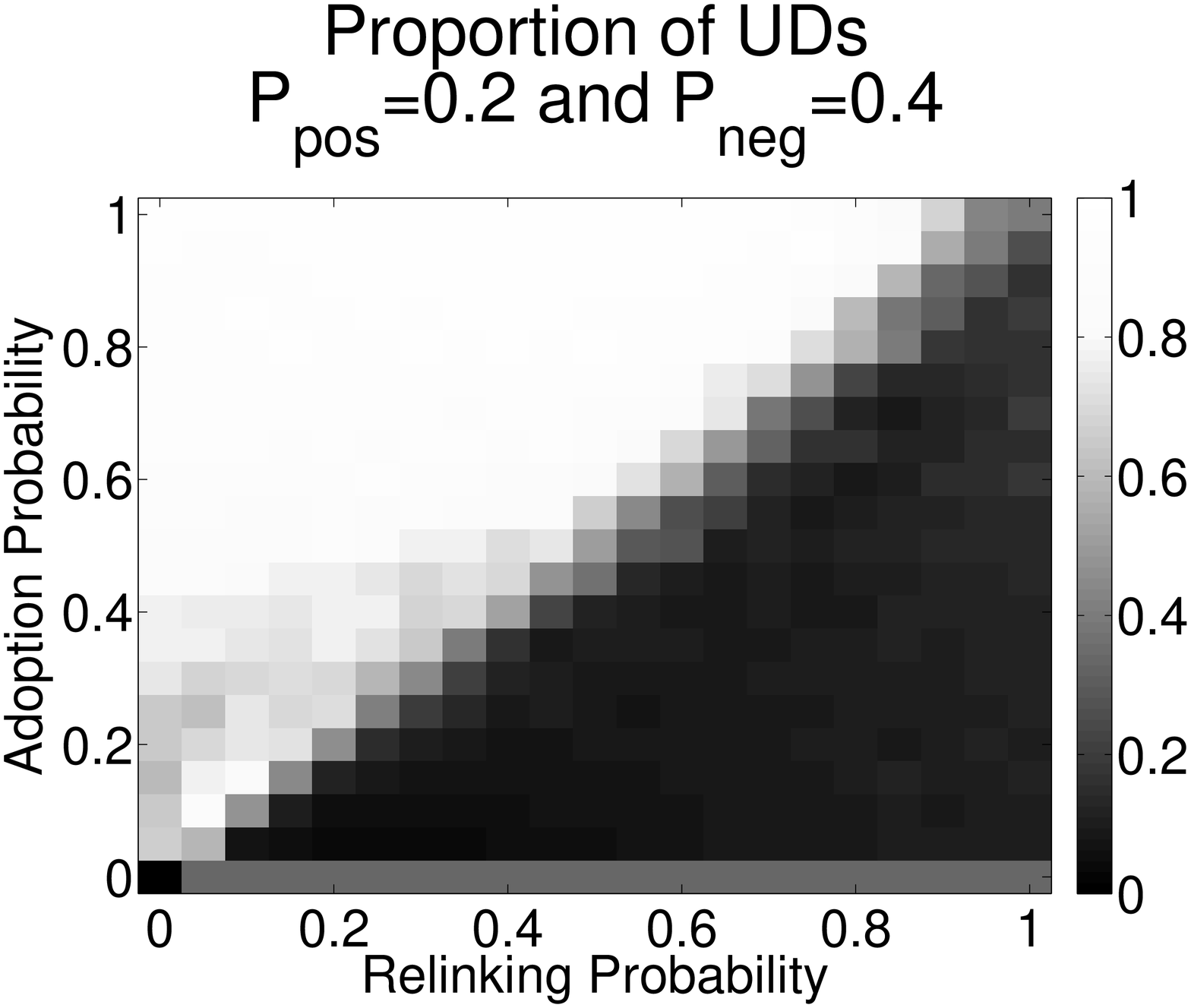}
\includegraphics[width=0.32\textwidth]{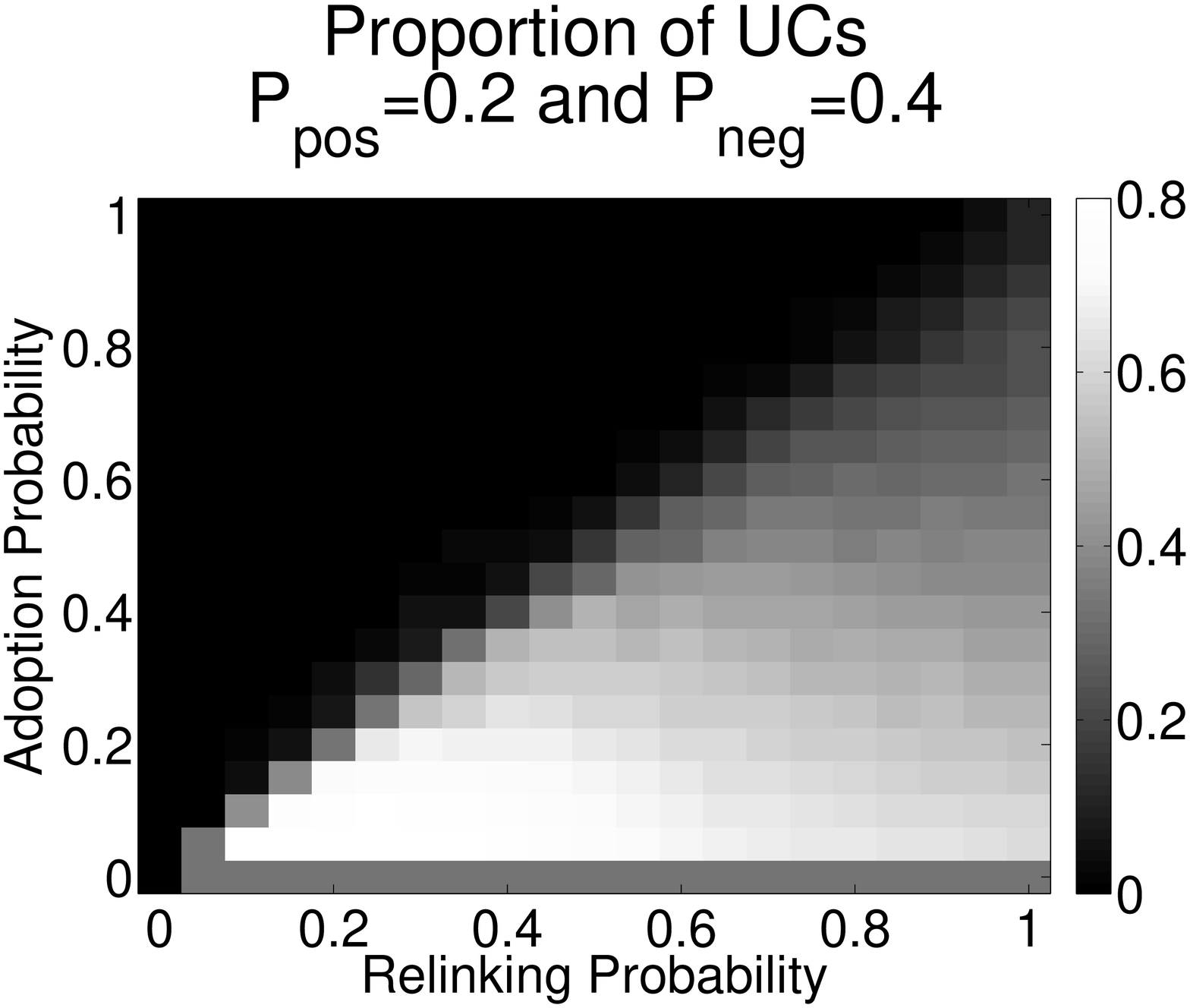}\\
\includegraphics[width=0.32\textwidth]{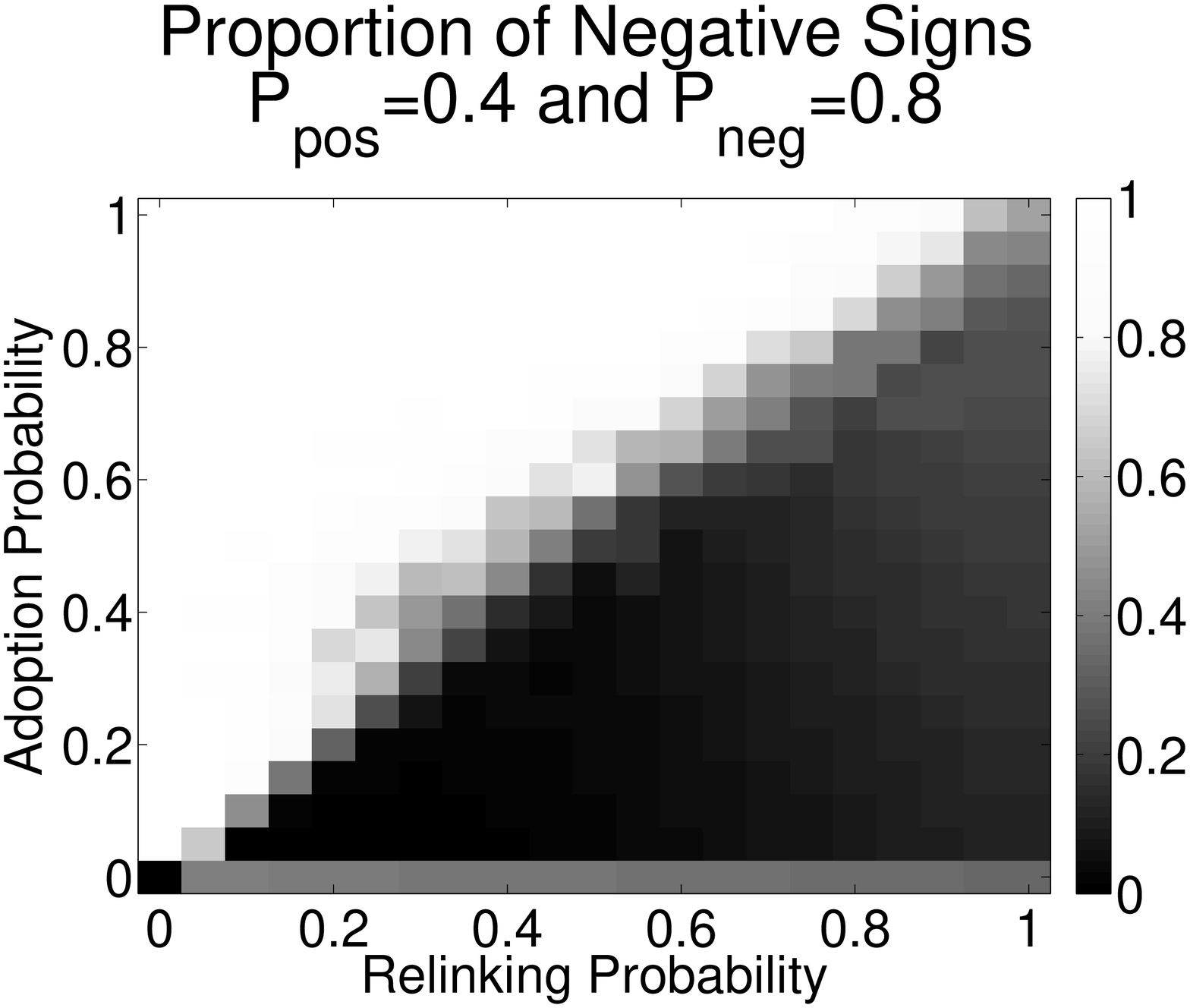}
\includegraphics[width=0.32\textwidth]{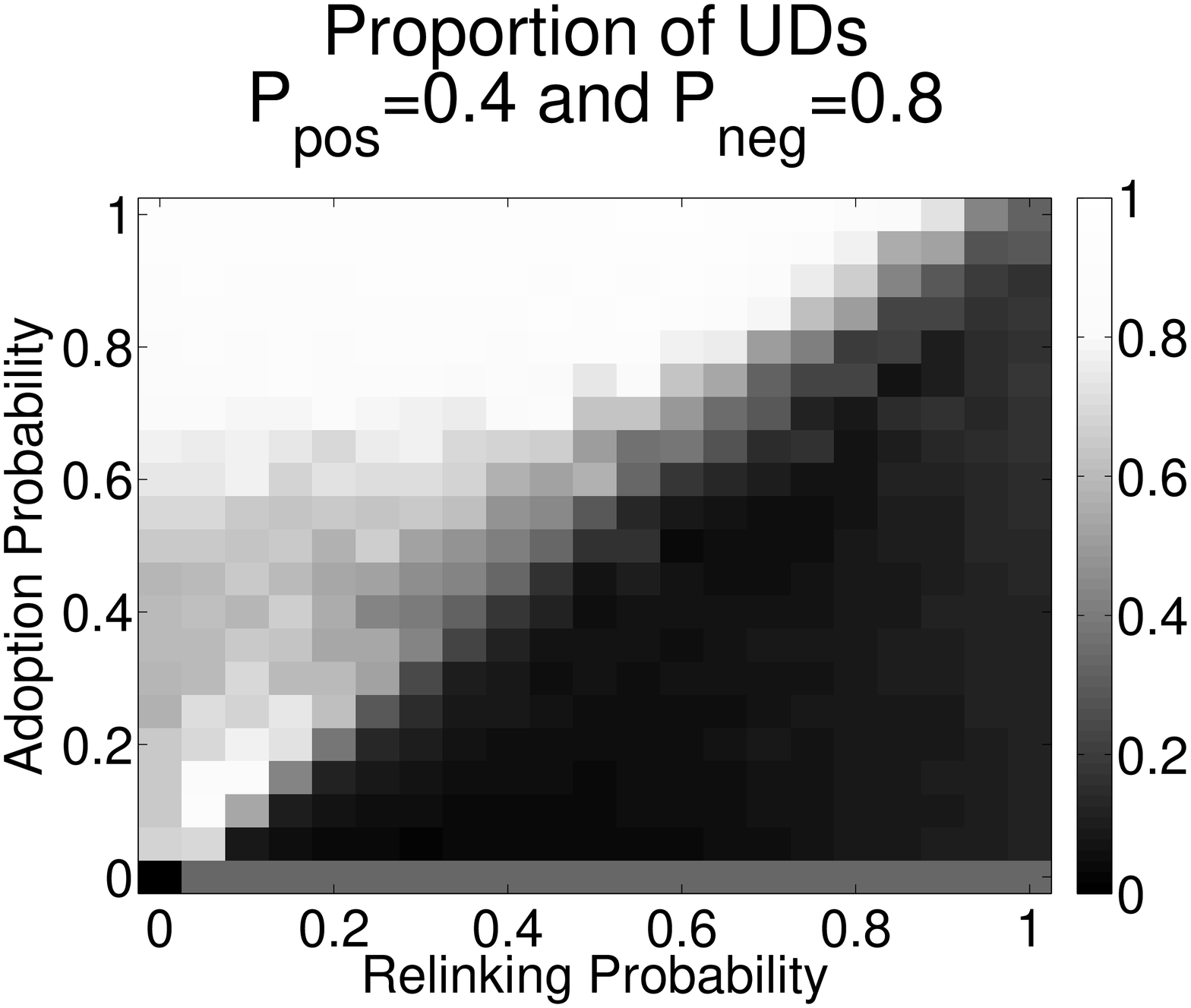}
\includegraphics[width=0.32\textwidth]{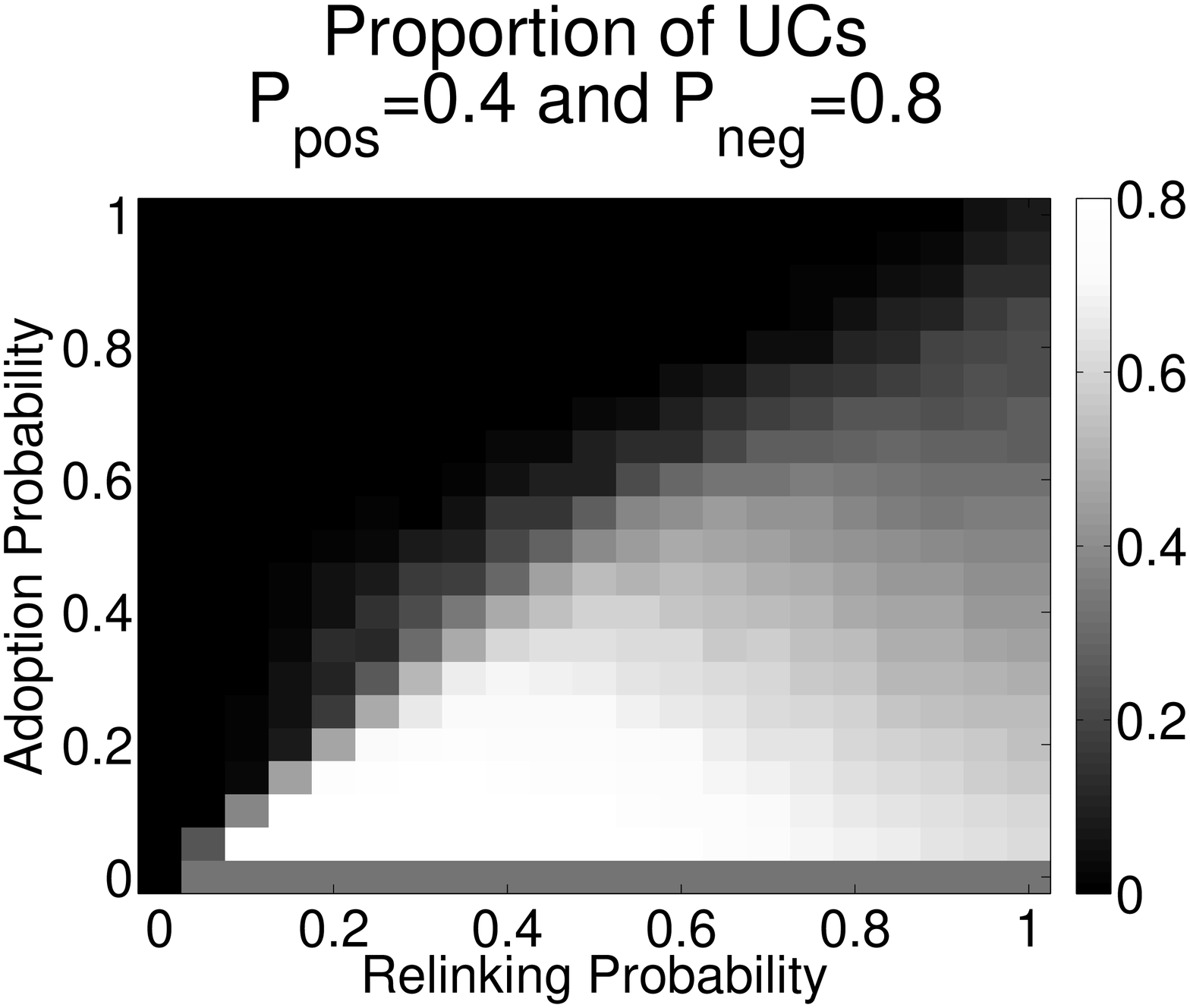}\\
\includegraphics[width=0.32\textwidth]{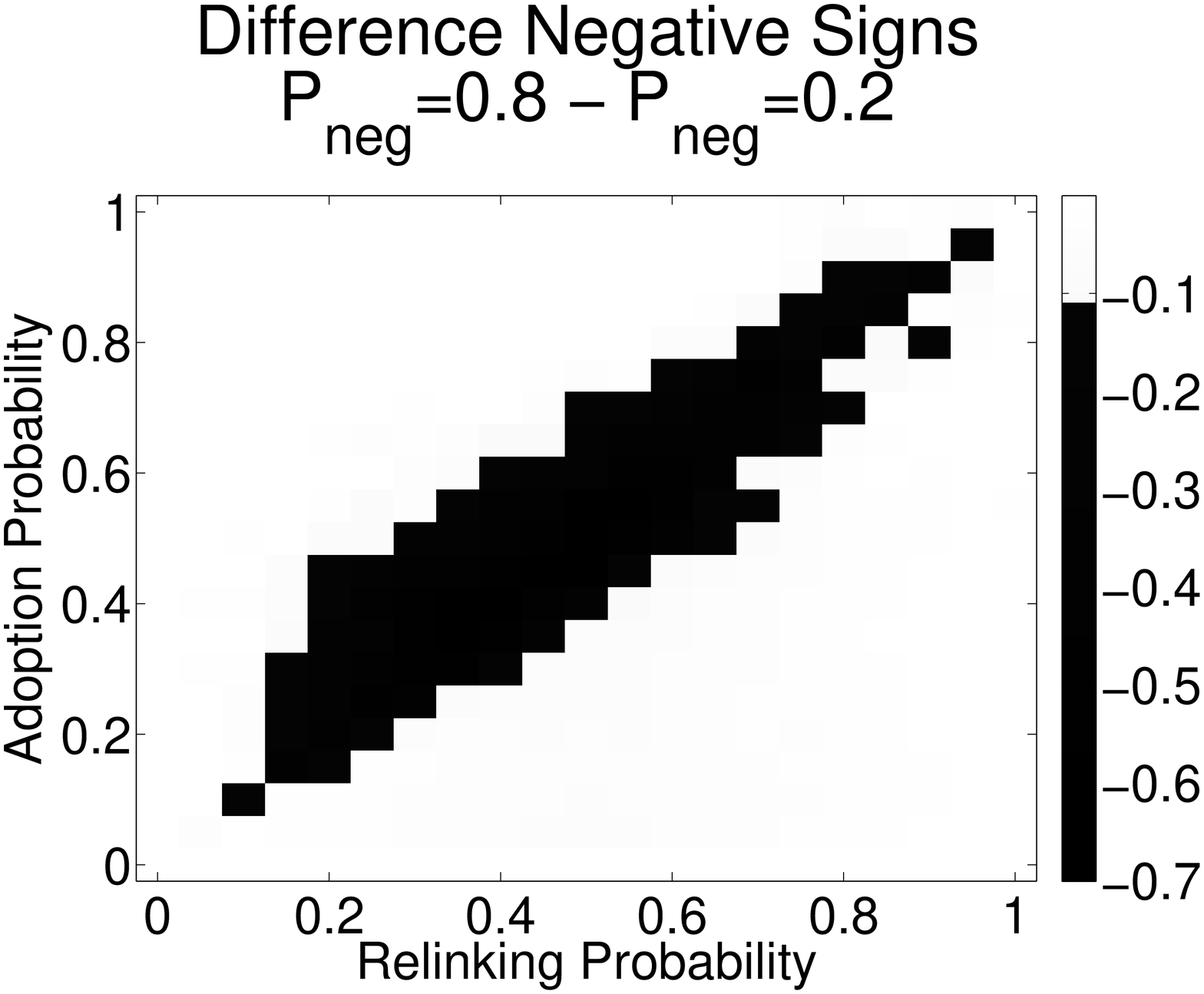}
\includegraphics[width=0.32\textwidth]{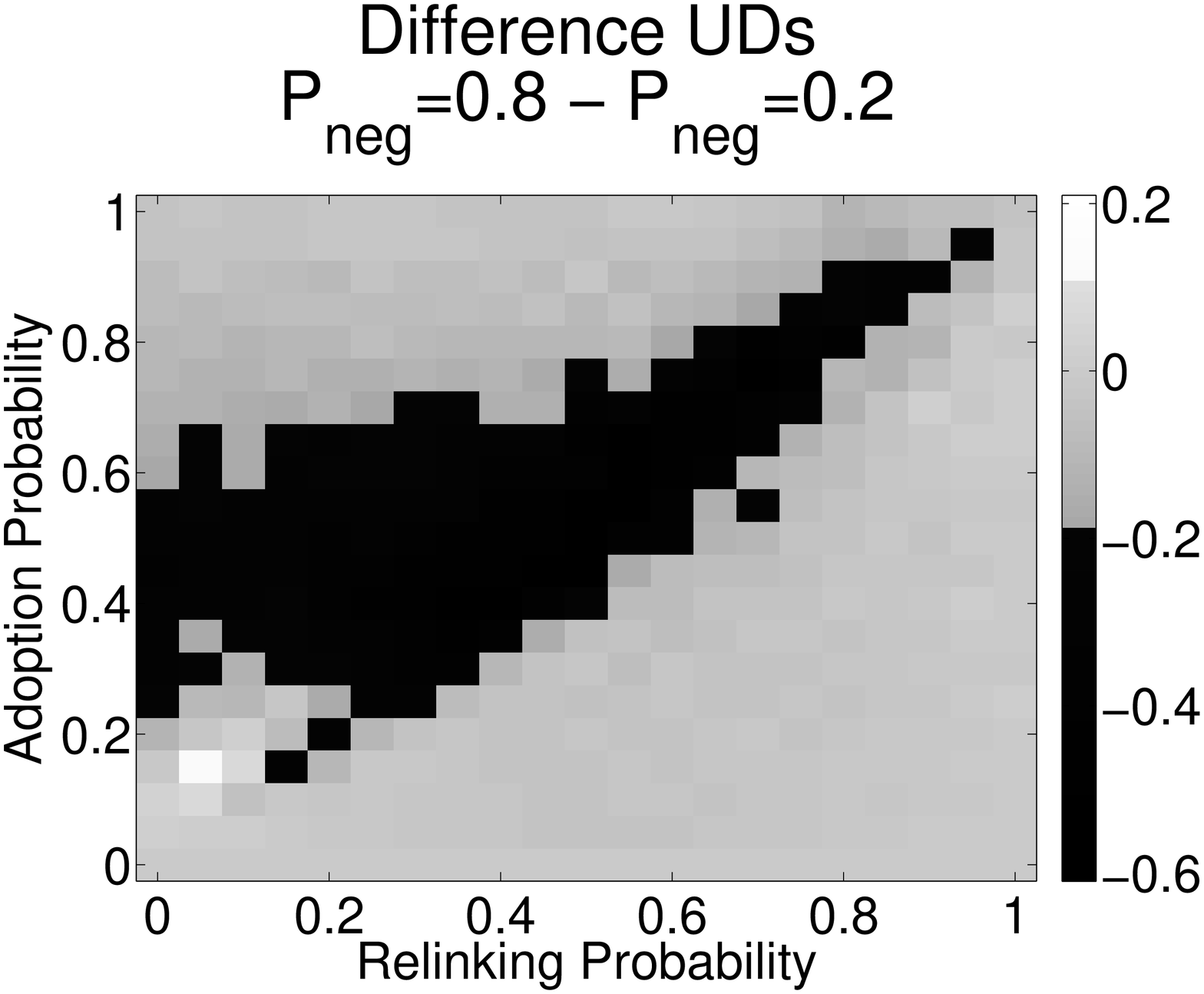}
\includegraphics[width=0.32\textwidth]{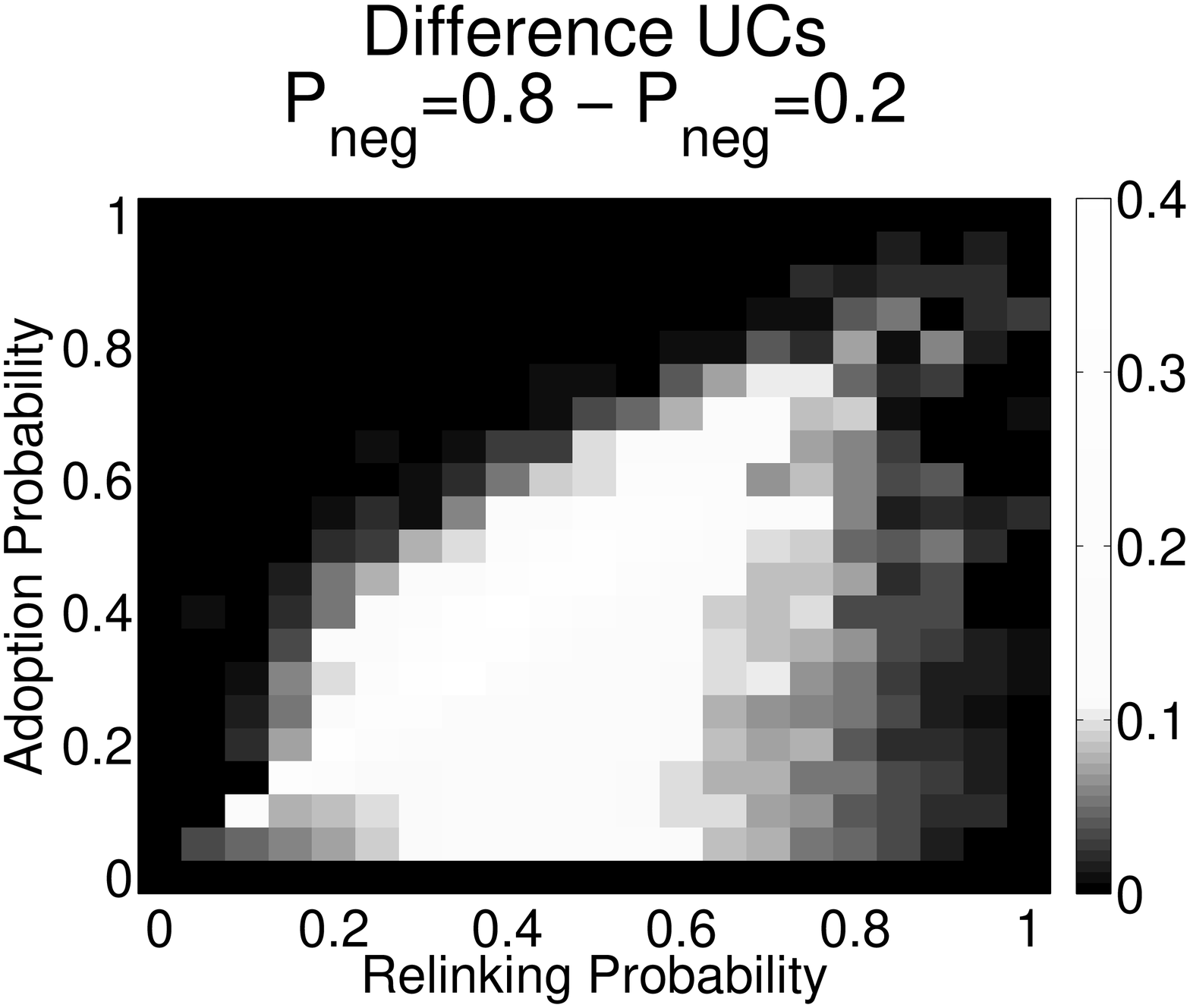}
\caption{Effect of the competing dynamics of adoption of strategies with higher payoffs (vertical axis) and of rewiring of stressed links (horizontal axis) on the final proportion of negative ties in the network (Left Panels), of UDs (Central Panels) and of UCs (Right Panels) for {\it different probabilities of switching signs}. Top Panels show the results for the case $P_{pos}=0.2$ and $P_{neg}=0.4$. Middle Panels display results for the case  $P_{pos}=0.4$ and $P_{neg}=0.8$. Lower Panels display the difference between the case in the Middle Panels and the baseline of the Middle row of Figure \ref{FixedPFlip}. The color map is tweaked to highlight the most significant changes in proportions. In all simulations the populations initially includes each strategy in proportion one third and N=200. Network signs are randomly initialized positive or negative with equal probability. The probability of existence for each tie is $P_{link}=0.05$ while $P_{rand}=0.01$.}
\label{differentppos}
\end{figure}
\begin{figure}[ht!]
\centering
\includegraphics[width=0.32\textwidth]{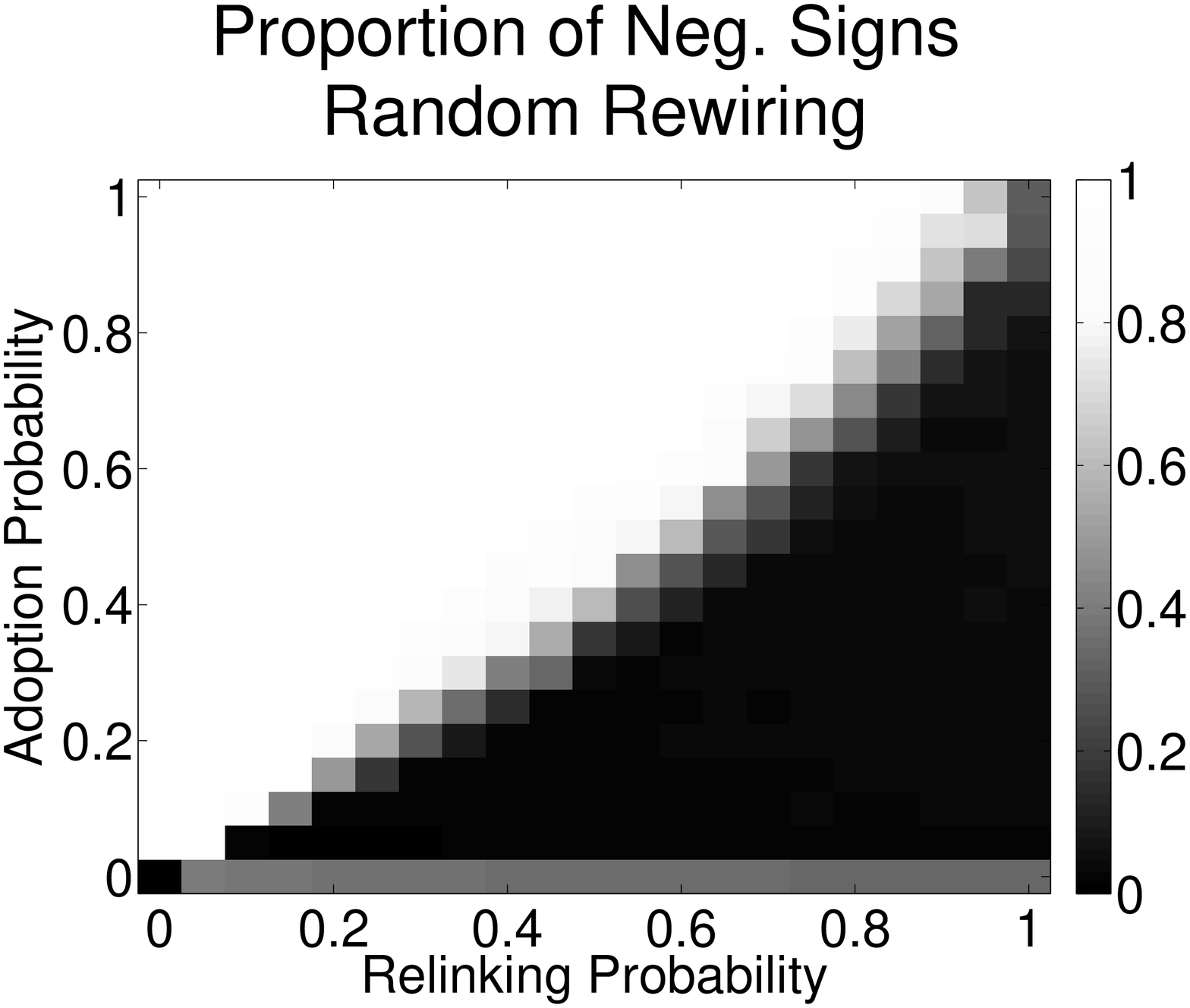}
\includegraphics[width=0.32\textwidth]{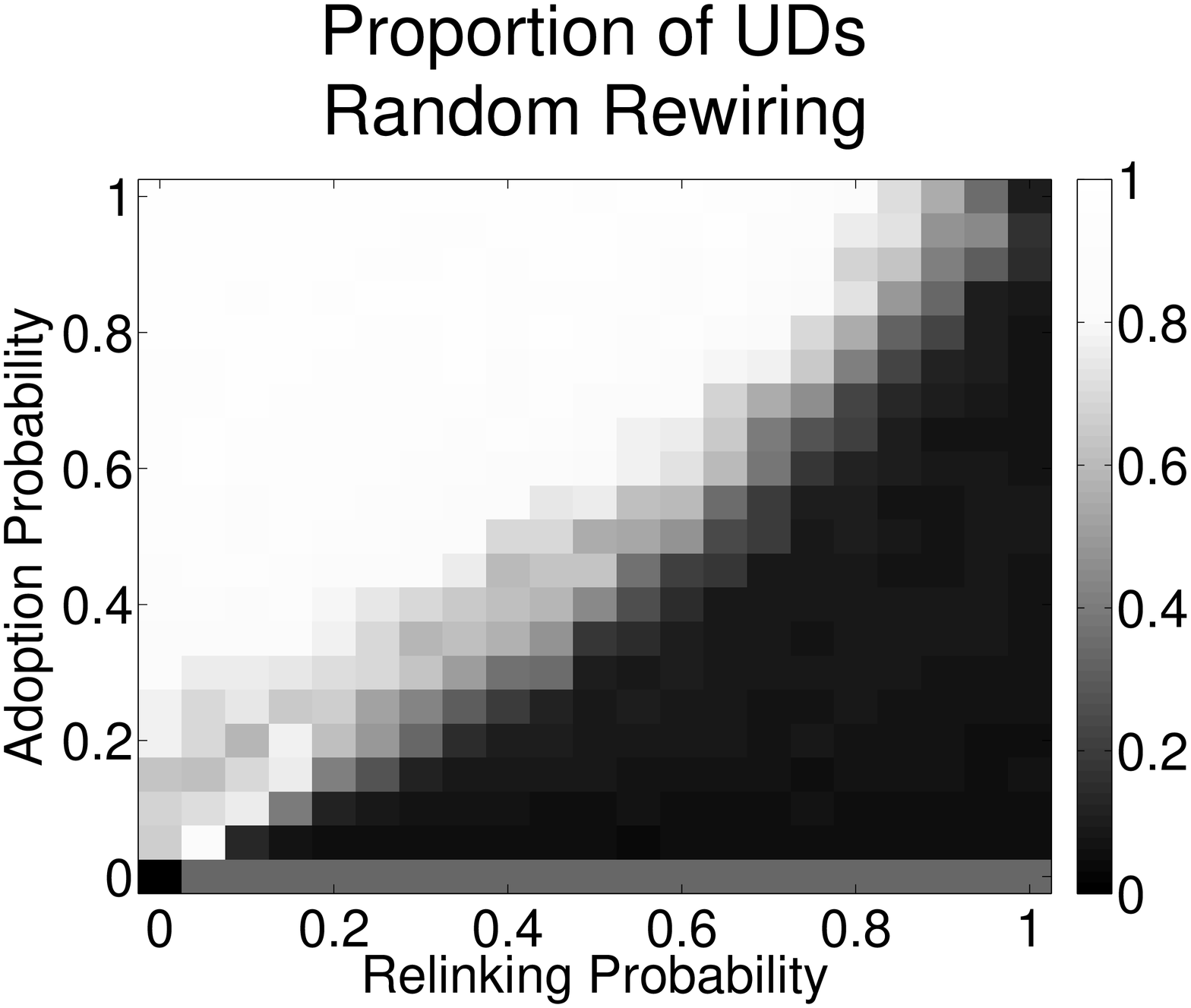}
\includegraphics[width=0.32\textwidth]{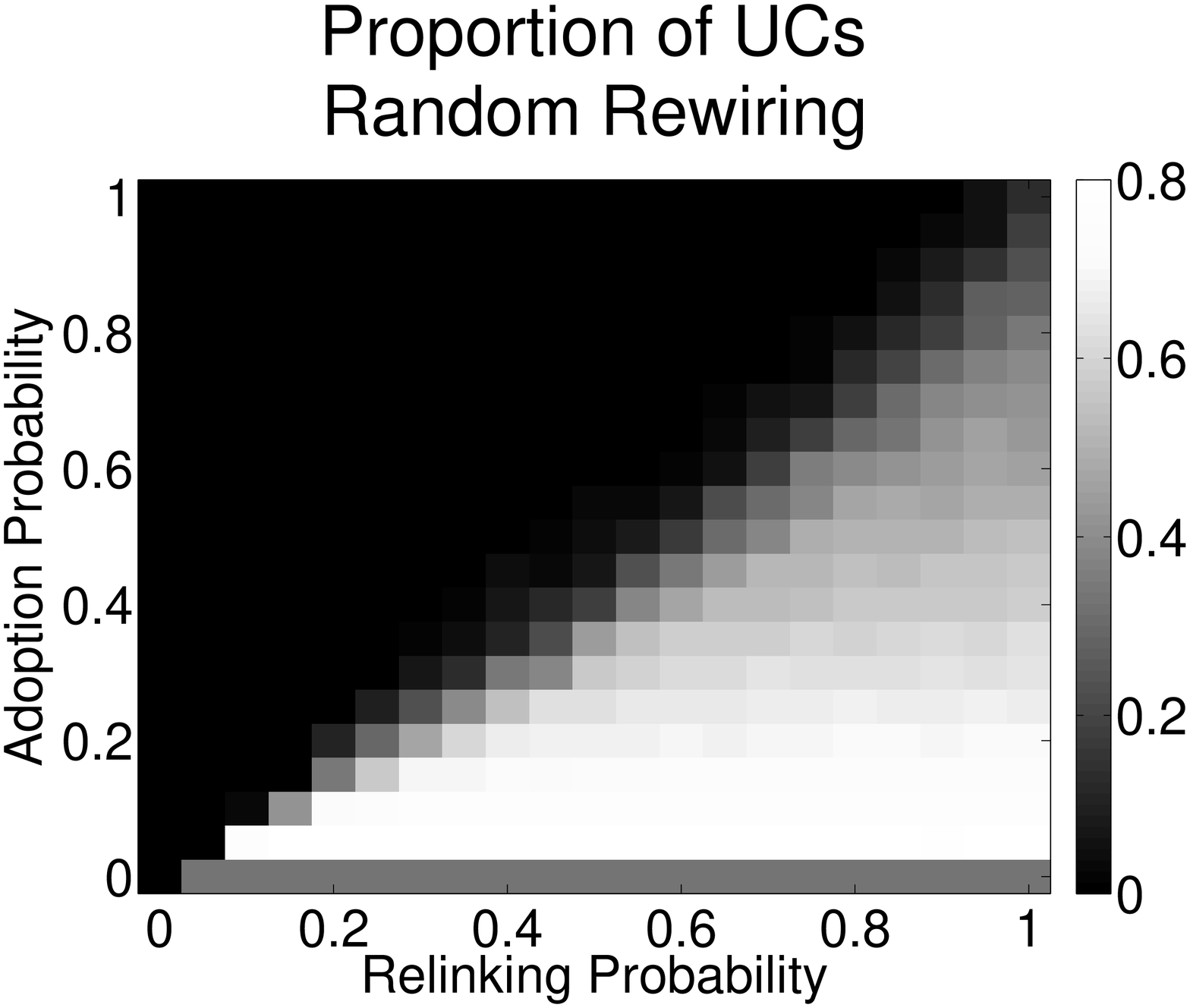}\\
\includegraphics[width=0.32\textwidth]{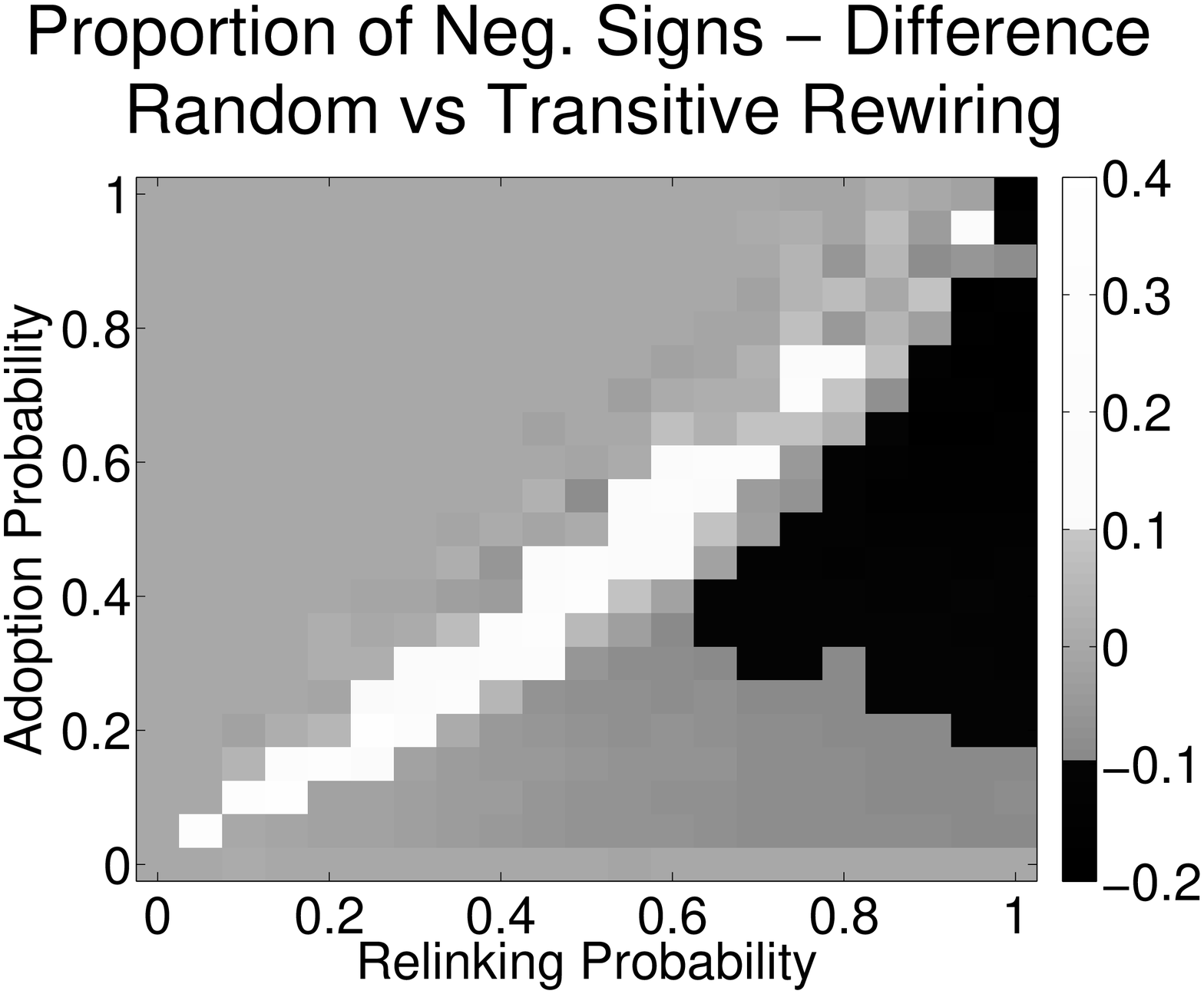}
\includegraphics[width=0.32\textwidth]{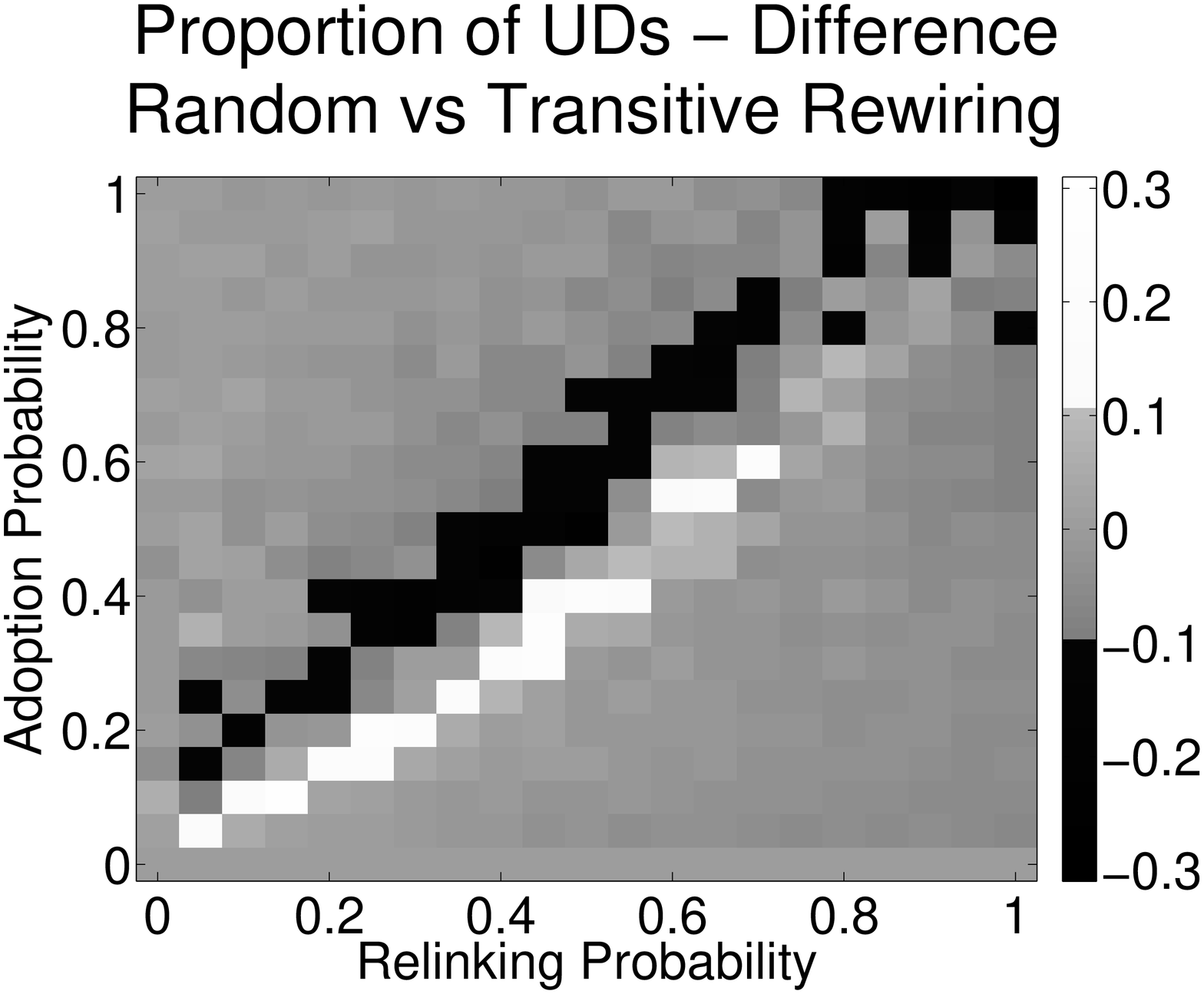}
\includegraphics[width=0.32\textwidth]{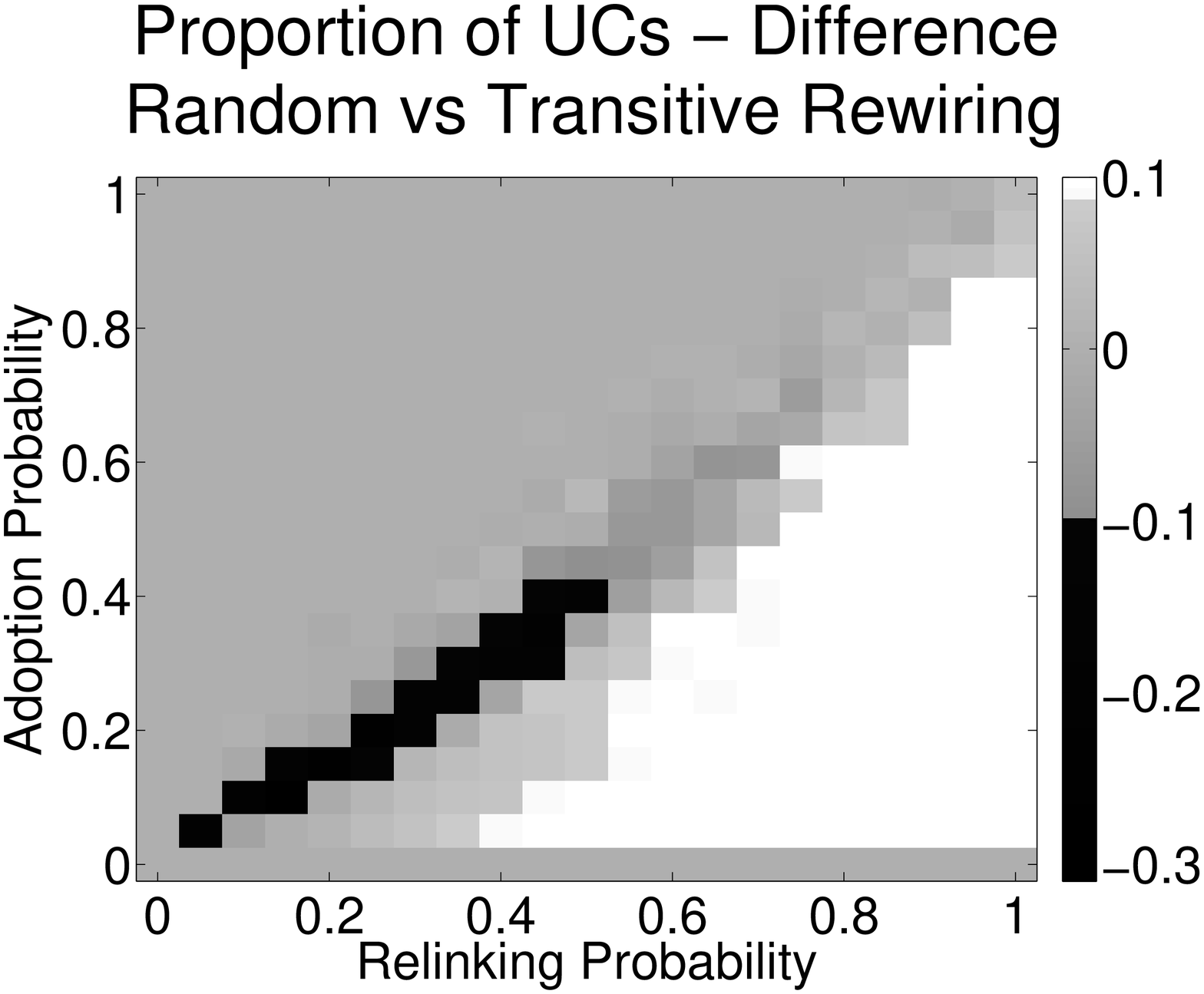}
\caption{Top Panels: Effect of the competing dynamics of adoption of strategies with higher payoffs (vertical axis) and of rewiring of stressed links (horizontal axis) on the final proportion of negative ties in the network (Left Panels), of UDs (Central Panels) and of UCs (Right Panels) for the case of {\it completely random rewiring} ($P_{rand}=1$). Lower Panels show the difference between the case of random rewiring and the baseline in figure \ref{FixedPFlip}. The color map is tweaked to highlight the most significant changes in proportions. In both cases and in all simulations N=200, $P_{neg}=0.2$ and $P_{neg}=0.1$. The population at the outset is equally divided among the three agents' types. Network signs are randomly initialized positive or negative with equal probability. The probability of existence for each tie is $P_{link}=0.05$. }
\label{Pflip_rand1}
\end{figure}

\bibliographystyle{ws-acs}
\bibliography{RighiTakacsACS}

\begin{thebibliography}{10}
\providecommand{\urlprefix}{}
\expandafter\ifx\csname urlstyle\endcsname\relax
  \providecommand{\doi}[1]{doi:\discretionary{}{}{}#1}\else
  \providecommand{\doi}{doi:\discretionary{}{}{}\begingroup
  \urlstyle{rm}\Url}\fi

\bibitem{axelrod2006evolution}
Axelrod, R., \emph{The Evolution of Cooperation} (Basic Books, 1984).

\bibitem{axelrod1997complexity}
Axelrod, R., \emph{The Complexity of Cooperation: Agent-based Models of
  Competition and Collaboration}, Princeton studies in complexity (Princeton
  University Press, 1997).

\bibitem{axelrod1981evolution}
Axelrod, R. and Hamilton, W.~D., The evolution of cooperation, \emph{Science}
  \textbf{211} (1981) 1390--1396.

\bibitem{bowles2004evolution}
Bowles, S. and Gintis, H., The evolution of strong reciprocity: cooperation in
  heterogeneous populations, \emph{Theoretical Population Biology} \textbf{65}
  (2004) 17--28.

\bibitem{burt1992social}
Burt, R.~S., The social structure of competition, \emph{Networks and
  organizations: Structure, form, and action} \textbf{57} (1992) 91.

\bibitem{burtsev2006evolution}
Burtsev, M. and Turchin, P., Evolution of cooperative strategies from first
  principles, \emph{Nature} \textbf{440} (2006) 1041--1044.

\bibitem{coleman1994foundations}
Coleman, J.~S., \emph{Foundations of social theory} (Harvard University Press,
  1990).

\bibitem{darwin1965expression}
Darwin, C., \emph{The expression of the emotions in man and animals}, Vol. 526
  (University of Chicago Press, 1965).

\bibitem{dreber2008winners}
Dreber, A., Rand, D.~G., Fudenberg, D., and Nowak, M.~A., Winners don’t punish,
  \emph{Nature} \textbf{452} (2008) 348--351.

\bibitem{dunbar1992neocortex}
Dunbar, R.~I., Neocortex size as a constraint on group size in primates,
  \emph{Journal of Human Evolution} \textbf{22} (1992) 469--493.

\bibitem{erdHos1959random}
Erd{\H{o}}s, P. and R{\'e}nyi, A., On random graphs, \emph{Publicationes
  Mathematicae Debrecen} \textbf{6} (1959) 290--297.

\bibitem{ernst2005human}
Ernst, F. and Gaechter, S., Human behaviour: Egalitarian motive and altruistic
  punishment (reply), \emph{Nature} \textbf{433} (2005).

\bibitem{fehr2003nature}
Fehr, E. and Fischbacher, U., The nature of human altruism, \emph{Nature}
  \textbf{425} (2003) 785--791.

\bibitem{flood1958some}
Flood, M.~M., Some experimental games, \emph{Management Science} \textbf{5}
  (1958) 5--26.

\bibitem{fowler2005altruistic}
Fowler, J.~H., Altruistic punishment and the origin of cooperation,
  \emph{Proceedings of the National Academy of Sciences of the United States of
  America} \textbf{102} (2005) 7047--7049.

\bibitem{fowler2005egalitarian}
Fowler, J.~H., Johnson, T., and Smirnov, O., Egalitarian motive and altruistic
  punishment, \emph{Nature} \textbf{433} (2005).

\bibitem{frank1988passions}
Frank, R.~H., \emph{Passions within reason: The strategic role of the
  emotions.} (WW Norton \& Co, 1988).

\bibitem{gargiulo2012influence}
Gargiulo, F. and Ramasco, J.~J., Influence of opinion dynamics on the evolution
  of games, \emph{PloS one} \textbf{7} (2012) e48916.

\bibitem{gracia2012human}
Gracia-L{\'a}zaro, C., Cuesta, J.~A., S{\'a}nchez, A., and Moreno, Y., Human
  behavior in prisoner's dilemma experiments suppresses network reciprocity,
  \emph{Scientific reports} \textbf{2} (2012).

\bibitem{gracia2012heterogeneous}
Gracia-L{\'a}zaro, C., Ferrer, A., Ruiz, G., Taranc{\'o}n, A., Cuesta, J.~A.,
  S{\'a}nchez, A., and Moreno, Y., Heterogeneous networks do not promote
  cooperation when humans play a prisoner’s dilemma, \emph{Proceedings of the
  National Academy of Sciences} \textbf{109} (2012) 12922--12926.

\bibitem{granovetter1973strength}
Granovetter, M.~S., The strength of weak ties, \emph{American Journal of
  Sociology}  (1973) 1360--1380.

\bibitem{grujic2010social}
Gruji{\'c}, J., Fosco, C., Araujo, L., Cuesta, J.~A., and S{\'a}nchez, A.,
  Social experiments in the mesoscale: Humans playing a spatial prisoner's
  dilemma, \emph{PloS one} \textbf{5} (2010) e13749.

\bibitem{Virtuallabs}
Hauert, C., Virtuallabs, \url{http://www.univie.ac.at/virtuallabs/Moran/}
  (2004), accessed: 2010-09-30.

\bibitem{hauert2004spatial}
Hauert, C. and Doebeli, M., Spatial structure often inhibits the evolution of
  cooperation in the snowdrift game, \emph{Nature} \textbf{428} (2004)
  643--646.

\bibitem{hechter1988principles}
Hechter, M., \emph{Principles of group solidarity}, Vol.~11 (Univ of California
  Press, 1988).

\bibitem{jackson2010social}
Jackson, M., \emph{Social and Economic Networks}, Princeton University Press
  (Princeton University Press, 2010).

\bibitem{jackson1996strategic}
Jackson, M.~O. and Wolinsky, A., A strategic model of social and economic
  networks, \emph{Journal of Economic Theory} \textbf{71} (1996) 44--74.

\bibitem{Jain16012001}
Jain, S. and Krishna, S., A model for the emergence of cooperation,
  interdependence, and structure in evolving networks, \emph{Proceedings of the
  National Academy of Sciences} \textbf{98} (2001) 543--547.

\bibitem{keltner2006social}
Keltner, D., Haidt, J., and Shiota, M.~N., Social functionalism and the
  evolution of emotions, in \emph{Evolution and social psychology}, eds.
  Schaller, M., Simpson, J.~A., and Kenrick, D.~T. (Psychology Press, 2013),
  pp. 115--142.

\bibitem{kerr2008detection}
Kerr, N.~L. and Levine, J.~M., The detection of social exclusion: Evolution and
  beyond., \emph{Group Dynamics: Theory, Research, and Practice} \textbf{12}
  (2008) 39.

\bibitem{kurzban2001evolutionary}
Kurzban, R. and Leary, M.~R., Evolutionary origins of stigmatization: the
  functions of social exclusion, \emph{Psychological Bulletin} \textbf{127}
  (2001) 187.

\bibitem{lieberman2005evolutionary}
Lieberman, E., Hauert, C., and Nowak, M.~A., Evolutionary dynamics on graphs,
  \emph{Nature} \textbf{433} (2005) 312--316.

\bibitem{masuda2003spatial}
Masuda, N. and Aihara, K., Spatial prisoner's dilemma optimally played in
  small-world networks, \emph{Physics Letters A} \textbf{313} (2003) 55--61.

\bibitem{mcnamara2008coevolution}
McNamara, J.~M., Barta, Z., Fromhage, L., and Houston, A.~I., The coevolution
  of choosiness and cooperation, \emph{Nature} \textbf{451} (2008) 189--192.

\bibitem{nowak2006evolutionary}
Nowak, M., \emph{Evolutionary Dynamics: Exploring the Equations of Life}
  (Belknap Press of Harvard University Press, 2006).

\bibitem{nowak2006five}
Nowak, M.~A., Five rules for the evolution of cooperation, \emph{Science}
  \textbf{314} (2006) 1560--1563.

\bibitem{ohtsuki2006simple}
Ohtsuki, H., Hauert, C., Lieberman, E., and Nowak, M.~A., A simple rule for the
  evolution of cooperation on graphs and social networks, \emph{Nature}
  \textbf{441} (2006) 502--505.

\bibitem{poundstone1992prisoner}
Poundstone, W., \emph{Prisoner's dilemma} (Doubleday, 1992).

\bibitem{righi2014degree}
Righi, S. and Tak{\'a}cs, K., Degree variance and emotional strategies as
  catalysts of cooperation in signed networks, \emph{Proceedings of the
  European Conference on Modeling and Simulation}  (2014).

\bibitem{righi2014parallel}
Righi, S. and Tak{\'a}cs, K., Parallel versus sequential update and the
  evolution of cooperation with the assistance of emotional strategies,
  \emph{arXiv preprint arXiv:1401.4672}  (2014).

\bibitem{santos2006cooperation}
Santos, F.~C., Pacheco, J.~M., and Lenaerts, T., Cooperation prevails when
  individuals adjust their social ties, \emph{PLoS Computational Biology}
  \textbf{2} (2006) e140.

\bibitem{schuessler1989exit}
Schuessler, R., Exit threats and cooperation under anonymity, \emph{Journal of
  Conflict Resolution} \textbf{33} (1989) 728--749.

\bibitem{suri2011cooperation}
Suri, S. and Watts, D.~J., Cooperation and contagion in web-based, networked
  public goods experiments, \emph{PLoS One} \textbf{6} (2011) e16836.

\bibitem{takacs2008collective}
Tak{\'a}cs, K., Janky, B., and Flache, A., Collective action and network
  change, \emph{Social Networks} \textbf{30} (2008) 177--189.

\bibitem{trivers1971evolution}
Trivers, R.~L., The evolution of reciprocal altruism, \emph{Quarterly Review of
  Biology}  (1971) 35--57.

\bibitem{vanberg1992rationality}
Vanberg, V.~J. and Congleton, R.~D., Rationality, morality, and exit, \emph{The
  American Political Science Review}  (1992) 418--431.

\bibitem{vukov}
Vukov, J., Szab\'o, G., and Szolnoki, A., Cooperation in the noisy case:
  Prisoner's dilemma game on two types of regular random graphs, \emph{Phys.
  Rev. E} \textbf{73} (2006) 067103.

\bibitem{wang2008learning}
Wang, S., Szalay, M.~S., Zhang, C., and Csermely, P., Learning and innovative
  elements of strategy adoption rules expand cooperative network topologies,
  \emph{PLoS One} \textbf{3} (2008) e1917.

\bibitem{wang2013int}
Wang, Z., Szolnoki, A., and Perc, M., Interdependent network reciprocity in
  evolutionary games, \emph{Scientific Reports} \textbf{3} (2013).

\bibitem{watts1998collective}
Watts, D.~J. and Strogatz, S.~H., Collective dynamics of ‘small-world’networks,
  \emph{nature} \textbf{393} (1998) 440--442.

\bibitem{wedekind2000cooperation}
Wedekind, C. and Milinski, M., Cooperation through image scoring in humans,
  \emph{Science} \textbf{288} (2000) 850--852.

\bibitem{yamagishi1996selective}
Yamagishi, T. and Hayashi, N., Selective play: Social embeddedness of social
  dilemmas, in \emph{Frontiers in social dilemmas research} (Springer, 1996),
  pp. 363--384.

\bibitem{yamagishi1994prisoner}
Yamagishi, T., Hayashi, N., and Jin, N., Prisoner’s dilemma networks: selection
  strategy versus action strategy, in \emph{Social dilemmas and cooperation}
  (Springer, 1994), pp. 233--250.

\end{thebibliography}

\end{document}